\documentclass[acmlarge,screen,nonacm]{acmart}

\AtBeginDocument{%
  \providecommand\BibTeX{{%
    \normalfont B\kern-0.5em{\scshape i\kern-0.25em b}\kern-0.8em\TeX}}}

\usepackage{algorithmic}
\usepackage{graphicx}
\usepackage{textcomp}
\usepackage{xcolor}
\usepackage{booktabs}
\usepackage{multirow}
\usepackage{multicol}
\usepackage{graphicx}
\usepackage{subfigure}
\usepackage{listings}
\usepackage{color}
\usepackage{cleveref}

\setcopyright{none}
\begin{document}

\title{ A Comparison of Code Embeddings and Beyond
}
\author{Siqi Han}
\email{sqhan@stu.ecnu.edu.cn}
\author{DongXia Wang}
\email{dxwang@stu.ecnu.edu.cn}
\author{Wanting Li}
\email{wtli@stu.ecnu.edu.cn}
\author{Xuesong Lu}
\email{xslu@dase.ecnu.edu.cn}
\affiliation{
  \institution{School of Data Science and Engineering, East China Normal University}
   \country{China}
}


\begin{abstract}
Program representation learning is a fundamental task in software engineering applications. With the availability of ``big code'' and the development of deep learning techniques, various program representation learning models have been proposed to understand the semantic properties of programs and applied on different software engineering tasks. However, no previous study has comprehensively assessed the generalizability of these deep models on different tasks, so that the pros and cons of the models are unclear. In this experience paper, we try to bridge this gap by systemically evaluating the performance of eight program representation learning models on three common tasks, where six models are based on abstract syntax trees and two models are based on plain text of source code. We kindly explain the criteria for selecting the models and tasks, as well as the method for enabling end-to-end learning in each task. The results of performance evaluation show that they perform diversely in each task and the performance of the AST-based models is generally unstable over different tasks. In order to further explain the results, we apply a prediction attribution technique to find what elements are captured by the models and responsible for the predictions in each task. Based on the findings, we discuss some general principles for better capturing the information in the source code, and hope to inspire researchers to improve program representation learning methods for software engineering tasks.
\end{abstract}



\maketitle

\section{Introduction}\label{sec:intro}
The task of program representation learning is to learn continuous vectors for representing code snippets, such that semantically-similar snippets are mapped to close vectors in the continuous space. The learnt representations, or the program embeddings, can be further used for downstream tasks in programming language processing and software engineering~\cite{wei2017supervised,gupta2017deepfix,wan2018improving,huo2020control,svyatkovskiy2020intellicode,cambronero2019deep,brody2020structural}. In the past few years, the availability of massive source code from public repositories has brought a boost to the use of deep learning techniques for program representation learning~\cite{ahmad2020transformer,yu2019neural,zhang2019novel,bui2021infercode,rabinovich2017abstract,yin2018tranx,liu2020atom,alon2020structural}. A large fraction of such studies leverage the abstract syntax trees (ASTs) of programs and use deep learning models to learn programs' representations from the ASTs~\cite{yu2019neural,zhang2019novel,alon2019code2vec,rabinovich2017abstract,yin2018tranx,liu2020atom,alon2020structural}. The assumption is that compared to the plain text of source code, an AST reflects better the structural information of a program so that the learnt representation can capture more precise semantic property of the program. The AST-based deep models have shown the state-of-the-art performance on tasks such as code classification and code clone detection~\cite{zhang2019novel,wang2020detecting}.

Despite the great success, we observe that there is a lack of comprehensive comparison of these models and in particular the evaluation of the generalizability of the program embeddings. While each model shows the advantage on one or two selected tasks in the original paper, it is unclear how they generalize to other tasks. Moreover, no previous work has investigated what information these embeddings capture from the source code W.R.T a particular task. Therefore it is hard to make suggestions on the use and improvement of these embedding models on subsequent tasks. To bridge this gap, first, we systemically evaluate the performance of six AST-based program embedding models on three common tasks in programming language processing (see Section~\ref{sec:experiments}). We kindly select the evaluated models based on four criteria to ensure that they represent the state of the art in the field (see Section~\ref{sec:ast-models}), and select the representative evaluation tasks that require the models to use as input the embedding of an entire program (see Section~\ref{sec:design}). Second, we attempt to explain the performance generated by these models using a prediction attribution technique (see Section~\ref{sec:attr}). With the attribution technique, we may visualize what elements in the original source code are responsible for the predictions, which helps to reveal why the models perform diversely on the tasks. As a baseline, we additionally train two classic neural networks in natural language processing to learn directly from the plain text of source code and evaluate the corresponding performance as well.

A related but different study is conducted by Kang et al.~\cite{kang2019assessing}, which assesses the generalizability of \emph{code2vec} token embeddings~\cite{alon2019code2vec}. They use code2vec as an algorithm (like word2vec in NLP) to embed the tokens in the source code and use the token embeddings as input to the models for downstream tasks. In other words, they assess whether code2vec produces robust word embeddings for downstream tasks. In contrast, we want to evaluate the generalizability of the embeddings that represent an entire program (like doc2vec in NLP). The reason is twofold. First, many software engineering models require the embedding of an entire program as input~\cite{allamanis2018survey,chen2019literature,zhang2019novel,wang2020detecting}. As such it is worth investigating which program embedding models are suitable for what kind of tasks. Second and more importantly, program embedding models focus on how to handle the structural information in the program rather than training the embeddings of individual tokens. Therefore the resulted token embeddings may be sensitive to the structural context they are trained in. Yet all program embedding models attempt to capture the semantic property of source code in general, therefore it is fairer to directly compare the generalizability of program embeddings.

To summarize, we mainly ask two questions: \emph{how do the representative AST-based program embedding models perform on different tasks} and \emph{why do they perform well or not on each task?} We attempt to answer the questions by the following contribution:
\begin{itemize}
    \item We select six representative AST-based program embedding models and evaluate their performance on three common software engineering tasks. We also implement two classic NLP models as baselines. We explain the performance discrepancy of the models by analyzing how they extract the structural information from the ASTs. To the best of our knowledge, this is the first effort in literature to systemically evaluate these program embedding models.
    \item We propose to use the prediction attribution technique to explain the model performance. The technique finds what features in the input data are responsible for the prediction by comparing with the prediction of the baseline input. We design a proper baseline input for each task so that the prediction can be attributed to the input tokens in the original programs. We visualize the attribution results and discuss what features in the source code are important for each task.
    \item We provide a discussion on the key lessons learned from the experimental study and draw some general principles for better capturing the information in the source code. The insights may motivate further research on program representation learning models that can either generalize better to various downstream tasks or achieve the new state-of-the-art performance on specific tasks.
\end{itemize}

The rest of the paper is organized as follows. In Section~\ref{sec:preliminairs}, we describe the evaluated program embedding models and explain our implementations to the models for fair comparison. In Section~\ref{sec:design}, we describe the three selected tasks that are suitable to evaluate the generalizability of program embeddings and the network architecture for each task. We report and analyze the performance of the embedding models in Section~\ref{sec:experiments}, followed by the prediction attribution analysis in Section~\ref{sec:attr}. We discuss the lesson learned from the study in Section~\ref{sec:disscusion} and threats to validity in Section~\ref{sec:threat}. Finally, we conclude in Section~\ref{sec:conclusion}.

\section{The Models for Program Embedding}\label{sec:preliminairs}
In this section, we briefly review the models evaluated in the experiments. We mainly investigate the representative models that learn program representations from abstract syntax trees. We refer to this type of models as the AST-based models. In comparison, we also apply two classic models in natural language processing, namely, LSTM and Transformer (only the encoder), to learn from the source code text. We refer to the two models as the token-based models. We describe the modifications in our implementation for a fair comparison of the models. 

\subsection{The Token-based Models}
\subsubsection{LSTM}
LSTM can capture the token dependencies in a sequence and is suitable for encoding the source code text. At each step, LSTM computes a hidden state that represents the information in the source code until the current input token. We follow previous work~\cite{dam2016deep,liang2019seml} and simply use the last hidden state to represent the entire program.

\subsubsection{Transformer}
The Transformer model encodes an input sequence using self-attentions, where each input token is encoded by attending to all the tokens in the same sequence. The Transformer encoder is typically constructed using several self-attention layers, and the encoding at the first position of the last layer is used to represent the entire program.

\subsection{The AST-based Models}\label{sec:ast-models}
The AST-based models evaluated in this work are selected based on four criteria. First, the model should learn a single embedding which represents the entire program. Second, for fair comparison, it should only use the static structural information of an AST and not incorporate any dynamic information such as concrete execution traces. Third, the underlying work is published on a top-ranked conference or journal, so that the model represents the state of the art in the field. Finally, the authors should have made the stable source code publicly available so that we may refer to their implementation details and re-implement the models correctly using a unified deep learning framework (PyTorch).

Based on the criteria, we finally identify six representative deep learning models with high citations. According to the methods of manipulating the structure of an AST, we roughly divide them into three categories. The first category recursively aggregates the structural information from the leaves to the root in a bottom-up approach. The representative models are TBCNN~\cite{mou2016convolutional} and AutoenCODE~\cite{white2016deep}, where the former proposes tree-based convolutional neural networks and the latter adopts recursive neural networks, respectively, for the bottom-up aggregation. The second category of models aggregate the structural information from the selected paths in the ASTs. The representative models are code2vec~\cite{alon2019code2vec} and code2seq~\cite{alon2018code2seq}. Both of them embed the AST paths and aggregate the path embeddings using the attention mechanism. The last category considers the control/data flow in the original source code while extracting the structural information in the ASTs. The representative models are GGNN~\cite{allamanis2018learning} and ASTNN~\cite{zhang2019novel}. GGNN constructs a graph by adding semantic edges into the AST and then adopts gated graph neural networks to learn from the graph, whereas ASTNN splits an AST into a sequence of statement subtrees and adopts a bidirectional GRU network to learn from the sequence.

\subsubsection{TBCNN}
TBCNN~\cite{mou2016convolutional} applies tree-based convolution kernels on an AST to gather the information in each subtree. The subtree features are aggregated using dynamic pooling to obtain the program embedding. In particular, TBCNN pretrains the embeddings of AST nodes and encodes each node as a linear combination of its own embedding and the aggregated embedding of its children. Then it uses tree-based convolution kernels to slide over the entire tree, and obtains a ``feature tree''. To cope with arbitrary sizes of feature trees, TBCNN applies dynamic pooling to aggregate all the features into one vector representing the entire program.

In their original implementation\footnote{https://sites.google.com/site/treebasedcnn/home}, the leaf nodes in the AST, which correspond to the tokens in the source code, are discarded before the convolutional layer. However, we found that incorporating the features of the leaf nodes improved the performance on the downstream tasks. Therefore, we keep the leaf nodes in our implementation. 

\subsubsection{AutoenCODE} 
AutoenCODE~\cite{white2016deep} transforms an AST into a binary tree and trains an autoencoder to learn the program embedding from it. In particular, AutoenCODE recursively combines the embeddings of every pair of children in a bottom-up approach, and then reconstructs the embeddings of the two children. The training loss is computed as a weighted sum of all the reconstruction losses over the entire binary tree. The trained embedding of the root represents the entire program.

In practice, the authors~\cite{white2016deep} found rather than combining two sibling nodes in each step, the greedy algorithm that minimizes each reconstruction loss yields better performance. Therefore in our implementation, we also adopt the greedy variant.

\subsubsection{Code2vec}
Code2vec~\cite{alon2019code2vec} learns to represent a program with its AST paths. An AST path connects two leaf nodes in the AST. For each AST path, code2vec concatenates the representations of the two leaf nodes and the representation of the internal path, and forms a context vector. A fixed number of sampled context vectors are transformed and aggregated using attention to form the program embedding.

Code2vec maintains two vocabularies for the leaf nodes and the paths, respectively. The internal nodes on paths are inseparable in the path representation.

\subsubsection{Code2seq}
Code2seq~\cite{alon2018code2seq} follows the general idea of code2vec, and extracts more fine-grained information from the AST paths. To incorporate the information of internal nodes, code2seq feeds the corresponding node embeddings sequentially into an LSTM network and obtains the encoding for the entire path. 

In the original paper, the program embedding is the average of the vectors of the selected AST paths. However, we notice that using attentions as code2vec to aggregate the AST paths produces better results. Therefore we adopt the attention mechanism in our implementation.

\subsubsection{GGNN}
GGNN~\cite{allamanis2018learning} adds semantic edges into an AST and learns from the resulted graph. The edges represent the flow of control and data in the program. For instance, a \texttt{NextToken} edge connects a leaf node to its successor in the source code. Once the program graph is obtained, GGNN uses the gated graph neural networks to learn the node representations. Following~\cite{wang2020detecting}, we use a global attention layer to aggregate all node representations and obtain the program embedding.

\subsubsection{ASTNN}
ASTNN~\cite{zhang2019novel} splits an AST into a set of statement subtrees corresponding to the statements in the source code. The subtrees are organized into a sequence in accordance with the order of the statements in the source code. Therefore the flow of control in the program is preserved. The sequence of the encoded subtrees is fed into a bidirectional GRU network. Finally, ASTNN uses max pooling to aggregate all hidden states and forms the program embedding.

\subsection{The Software Used}
For fair comparison, we re-implement all the models using PyTorch. For the token-based models, we split each code snippet into a sequence of subtokens according to camel rules\footnote{https://en.wikipedia.org/wiki/camelcase.}. For the AST-based models, following previous work~\cite{mou2016convolutional,zhang2019novel}, we use \emph{pycparser}\footnote{https://github.com/eliben/pycparser} and \emph{javalang}\footnote{https://github.com/c2nes/javalang} to convert C and Java programs into ASTs, respectively. For code2vec and code2seq, we use \emph{astminer}\footnote{https://github.com/JetBrains-Research/astminer}~\cite{kovalenko2019pathminer} to directly obtain the AST paths, which is recommended in its source code repository. To construct the graphs used by GGNN, we follow the work~\cite{allamanis2018learning} and use \emph{graph-ast}\footnote{https://github.com/bdqnghi/graph-ast} to add semantic edges between AST nodes.

\subsection{The Model Detail}
The efficiency of a model is often an useful aspect when assessing it.
For the selected eight models, we calculated the model parameters based on our re-implementation with the PyTorch framework.
Combined with the default settings of each model, We set the embedding size to 128, and the vocab size is assumed to be fixed at 1000. 
The result is shown in Tab.~\ref{tab:params}.

\begin{table}[h]
    \caption{The Parameter Size of Models.}  \label{tab:params}
    \begin{center}
    \begin{tabular}{c|cc}
    \toprule
    \textbf{Group}      & \textbf{Model} & \textbf{Parameters(k)} \\ \midrule
    Token-based         & LSTM           & 656                \\
                        & Transformer    & 1924                \\ \midrule
                        & TBCNN          & 231                \\
                        & AutoenCODE     & 66                \\ \cmidrule(l){2-3} 
    AST-based           & code2vec       & 404                \\
                        & code2seq       & 938                \\ \cmidrule(l){2-3} 
                        & GGNN           & 294 \\
                        & ASTNN          & 329                \\ \bottomrule
    \end{tabular}
    \end{center}
\end{table}

Among the eight models, LSTM and Transformer have larger number of parameters owing to it not only considers each token but the model structure is also complicated.
Both TBCNN and AutoenCODE recursively learn the program representation on ASTs.
TBCNN has a small number of parameters due to the shared parameters in convolution on different kernels, and AutoenCODE is a recursive auto-encoder which only need two encoding-decoding matrices so that it contains the smallest parameters.
Except for these two models, other AST-based models transform the information on the AST via focusing on a path or a statement, so there will produce more parameters.
Between code2vec and code2seq, the number of code2seq's parameters is larger due to the richer input information and the more complicated model structure.
However, in the real situation, code2seq is lighter than code2vec because it splits the token of leaf nodes so that the vocab size is reduced.
For example, in the code clone detection task, the number of tokens included in code2seq is reduced from 43343 to 20659, and the number of paths is reduced from 28763 to 178 in contrast to code2vec.
GGNN abstracts AST as a hypergraph, which combines some nodes and adds edges to the AST.
Therefore, the number of parameters of this model with graph neural network is relatively small.
For ASTNN, it extracts the statement sequences to simplify the representation of the AST.

Above all, the token-based models have more parameters because they pay attention to every token in the programs, containing redundant information.
The parameters of the AST-based models are less because the structural representation can refine input features and filter redundant information.

\section{The Selected Tasks}\label{sec:design}
In this section, we describe the selected tasks for evaluation. For each task, we also describe the subsequent component to connect the embedding model and the output. For the pretrained model AutoenCODE, we continue to fine-tune the parameters during the subsequent training.

Since the purpose of the current work is to evaluate the generalizability of program embedding, we decide to select the tasks that directly utilize the embedding of an entire program. We first select the \emph{code classification} task, which is a very common task in literature~\cite{frantzeskou2008examining,mou2016convolutional,zhang2019novel}. It aims at classifying programs by their functionality. We also refer to the CodeXGlue~\cite{lu2021codexglue} benchmark tasks for programming language, and select two other tasks, namely, \emph{code clone detection} and \emph{code search}. The former task detects whether two programs are similar with respect to a clone type~\cite{roy2007survey}, and the latter retrieves the most relevant code given a natural language query.

\subsection{Code Classification} 
Classifying programs by functionalities is important for program understanding and maintenance~\cite{kawaguchi2006mudablue,linares2014using,chihada2015source,bui2018cross}. For example, in a large software repository, automatically tagging programs with their functionality benefits the reuse and maintenance of the source code during the development process. Among the models under evaluation, TBCNN, code2vec and ASTNN also apply their program embeddings on code classification.

\begin{figure}[h]
    \centering
    \includegraphics[width=0.68\columnwidth]{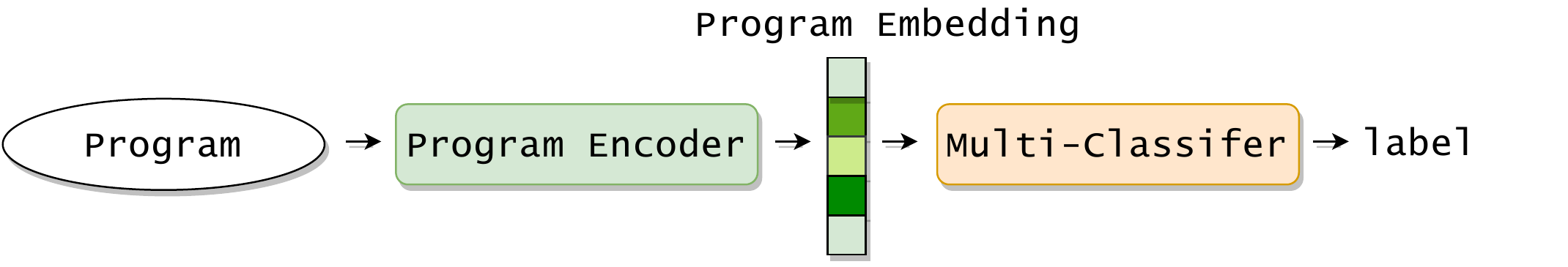}
    \caption{The model structure for code classification.}
    \label{fig:Classification}
\end{figure}
Fig.~\ref{fig:Classification} displays the framework of the models for code classification, which follows the settings in~\cite{zhang2019novel}. On the left-hand side, the program encoder can be implemented using any of the aforementioned embedding models. The resulted program embedding is directly input to a multi-classifier, which predicts the class of the program. We implement the classifier using a fully-connected layer with the Softmax activation. We use the cross-entropy loss function and Adamax for optimization.

\subsection{Code Clone Detection} 
Code clone detection is also widely studied in software engineering research~\cite{kamiya2002ccfinder,wei2017supervised,baxter1998clone,zhang2019novel,wang2020detecting}. During software development, programmers usually copy and paste common code. It may cause inefficiency in developing and affect the stability of systems if duplicated code snippets are not properly used. In programming education, detecting similar code can help teachers group student solutions or find code plagiarism during programming tests. Among the models, AutoenCODE and ASTNN also apply their program embeddings on code clone detection.

\begin{figure}[h]
    \centering
    \includegraphics[width=0.7\columnwidth]{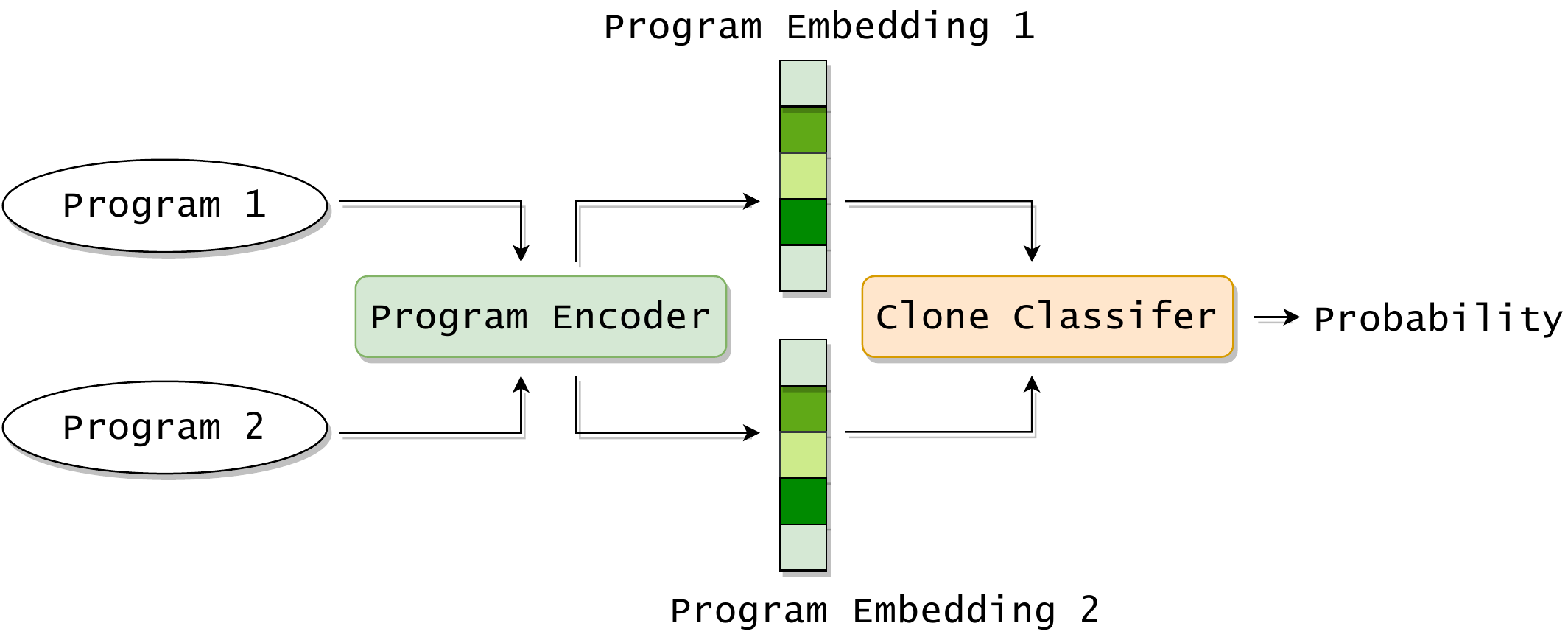}
    \caption{The model structure for code clone detection.}
    \label{fig:Clone}
\end{figure}
Fig.~\ref{fig:Clone} displays the framework of the models for code clone detection, following the settings in~\cite{zhang2019novel}. On the left-hand side, two programs are separately encoded using the embedding models. Then the two embeddings are input into a classifier, which predicts whether the two programs are clones. We implement the clone classifier by subtracting the two embeddings and transforming the resulted embedding using a fully-connected layer. The output is activated using the Sigmoid function. During training, we use the cross-entropy loss function and Adamax to optimize the parameters.

\subsection{Code Search} 
Programmers often search code fragments with natural language queries. The code search task automatically returns the most relevant code snippets from a prepared code collection according to a query~\cite{gu2018deep,liu2019neural,sun2020pscs,sivaraman2019active,cambronero2019deep,yan2020code}. None of the aforementioned models has been evaluated on this task in the original paper. 

\begin{figure}[h]
    \centering
    \includegraphics[width=0.6\columnwidth]{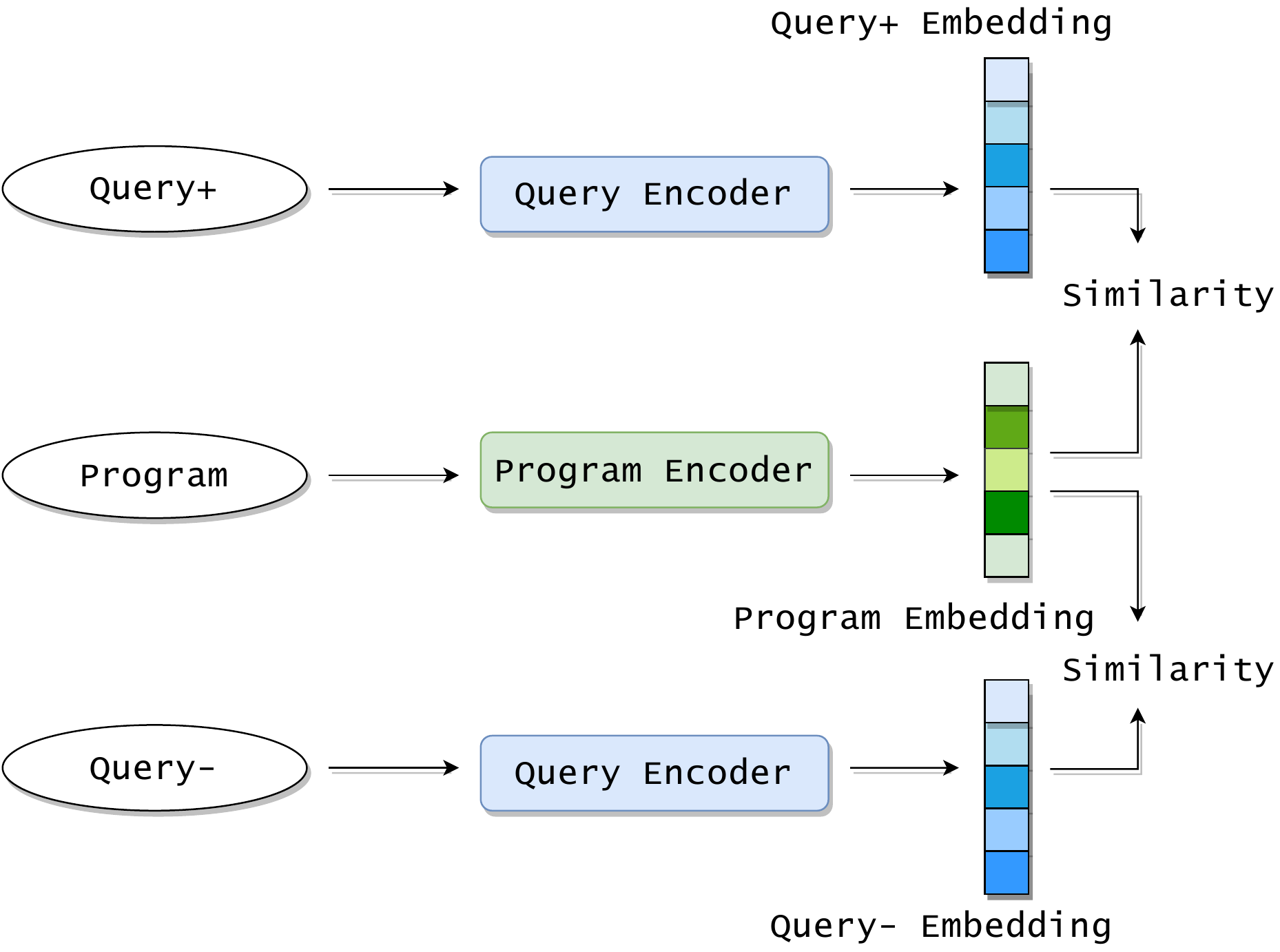}
    \caption{The model structure for code search.}
    \label{fig:Search}
\end{figure}

Fig.~\ref{fig:Search} displays the framework of the models for code search, which follows the approach in~\cite{gu2018deep,sun2020pscs}. During training, we first select a pair of query and code snippet $<Q^+,C^+>$, where $Q^+$ denotes a query and $C^+$ denotes a relevant code. We also select a random query $Q^-$ as the negative sample. The queries and the code are encoded using the query encoder and the program encoder, respectively. With the query and program embeddings $\textbf{v}^+_Q$, $\textbf{v}^-_Q$ and $\textbf{v}^+_C$, the training objective is to simultaneously maximize the cosine similarity between $\textbf{v}^+_Q$ and $\textbf{v}^+_C$ and minimize the cosine similarity between $\textbf{v}^-_Q$ and $\textbf{v}^+_C$. We use Adam for optimization. To boost the performance, we split the composed identifiers and words into subwords. For instance, "fastPathOrderedEmit" is split into "fast path ordered emit" and then encoded separately. We build a shared vocabulary for the tokens in the source code and the ASTs.







\section{Performance Evaluation}\label{sec:experiments}
For each task, we describe the dataset used in the experiments and the evaluation metric, and report the performance.

\subsection{The statistics of the used dataset}
\begin{table}[t]
    \caption{The Statistics of Three Dataset.}  \label{tab:data-oj}
    \begin{center}
    \setlength{\tabcolsep}{0.5mm}{
    \begin{tabular}{c|cccc}
    \toprule
                           &                  & POJ-104 & BigCloneBench & CodeSearchNet \\ \midrule
                           & Max code tokens  & 24395   & 65091         & 346763        \\
    Token                       & Avg. code tokens & 569.25  & 1115.68       & 449.25        \\
                           & Max doc tokens   & -       & -             & 813           \\
                           & Avg. doc tokens  & -       & -             & 47.36         \\ \midrule
                               & Max AST depth    & 76      & 62            & 470           \\
    AST                     & Avg. AST depth   & 13.32   & 10.57         & 8.81          \\
                           & Max AST nodes    & 7027    & 9602          & 79940         \\
                          & Avg. AST nodes   & 189.57  & 241.46        & 103.77        \\ \bottomrule 
    \end{tabular}}
    \end{center}
\end{table}
Table~\ref{tab:data-oj} is the specific statistics of the dataset we employed under different input structures on three tasks.
Since AST is an abstract structure of the program, it extracts effective syntactical structures so that it can shorten the input size from redundant token sequences.

\subsection{Code Classification}\label{sec:cc}
\textbf{Dataset}:
We use \textbf{POJ-104}\footnote{https://sites.google.com/site/treebasedcnn/} for the task, which is a widely used public dataset in literature~\cite{mou2016convolutional, zhang2019novel,ben2018neural,zhang2020generating,liu2020modeling}. The dataset is collected from an online judge system and contains 104 programming problems. Each problem contains 500 C programs written by the students, which are considered to belong to the same class. The total $52,000$ programs are randomly divided into training, validation and testing set with proportion 3:1:1.
The detailed dataset information is displayed in the second column in Fig.~\ref{tab:data-oj} with statistics of tokens, ASTs, and graphs. Since AST is an abstract structure of the program, it extracts effective syntactical structures so that it can shorten the input size from redundant token sequences.

\textbf{Metric}:
We select the class with the largest score as the predicted class. The evaluation metric is the multi-class accuracy, which is calculated as the proportion of correctly classified programs to all programs.

\begin{table}[t]
    \caption{The Results of Code Classification.} \label{tab:cla-res}
    \begin{center}
    \setlength{\tabcolsep}{6mm}{
    \begin{tabular}{@{}l|ll@{}}
    \toprule
    Groups      & Model       & Accuracy         \\ \midrule
    Token-based & LSTM        & 94.31\%          \\
                & Transformer & 90.74\%          \\ \midrule
                & TBCNN       & 94.00\%          \\
                & AutoenCODE  & 80.83\%          \\ 
                \cmidrule(l){2-3} 
    AST-based   & code2vec    & 91.32\%          \\
                & code2seq    & 84.85\%          \\ \cmidrule(l){2-3} 
                & GGNN        & 96.06\%          \\
                & ASTNN       & \textbf{98.09\%} \\
    \bottomrule
    \end{tabular}}
    \end{center}
\end{table}
\textbf{Results}:
The results on the testing set are shown in Table~\ref{tab:cla-res}. We observe that all models except AutoenCODE and code2seq have the test accuracy above $90\%$, indicating that current program embedding models can effectively predict the functionality of a program. Among them, ASTNN achieves the highest accuracy $98.09\%$.

According to the taxonomy in Section~\ref{sec:preliminairs}, first we observe that GGNN and ASTNN have the best performance. This may be because the two models capture the flow of control and data in a program in addition to the structural information in the corresponding AST. Second, the two token-based models perform overall better than the two categories of AST-based models that solely extract the structural information from a program's AST (i.e., TBCNN, AutoenCODE, code2vec and code2seq). Given that the token-based models learn from the plain text of source code, the results may indicate that for functionality prediction the information of control and data flow in a program is more important than the syntactic structure information in the AST. Third, for the two models that aggregate the node information in an AST using a bottom-up approach, TBCNN performs much better than AutoenCODE. Apart from using the different ways for bottom-up aggregation, TBCNN preserves the original structure of the AST, whereas AutoenCODE ignores the original structure and aggregates the nodes using a greedy algorithm. The results indicate that the greedy algorithm may destroy the AST's structure, leading to the failure of capturing the semantic property of a code. Finally, the two models leveraging AST paths perform worse than TBCNN and better than AutoenCODE. This may justify the previous conjecture because using AST paths partially ignores the original tree structure. 

In short, when predicting the functionality of a program, both the information of control and data flow and the structural information are important for an accurate prediction. Between the two types of information, the former might be a stronger signal to reflect the program's functionality. Nevertheless, learning jointly the two types of information can achieve the best prediction performance.

\subsection{Code Clone Detection}\label{sec:ccd}
\textbf{Dataset}:
We use the \textbf{BigCloneBench}\footnote{https://jeffsvajlenko.weebly.com/bigcloneeval.html} dataset for this task, which is also used in~\cite{wei2017supervised, sajnani2016sourcerercc, wang2020detecting}. The dataset is collected from a Java repository~\cite{svajlenko2014towards}. Each program in BigCloneBench is a single method. According to the functionality of a method, BigCloneBench constructs in total $6,000,000$ true clone pairs and $260,000$ false clone pairs. Following~\cite{wei2017supervised,wang2020detecting}, we select 9,134 programs to generate the pairs and randomly divide the dataset into training, validation and testing set with proportion 3:1:1.The detailed dataset information is displayed in the third column in Fig.~\ref{tab:data-oj} with statistics of tokens, ASTs, and graphs.

In~\cite{roy2007survey}, four types of code clones are defined based on the degree of similarity between two programs. \textsc{Type-1}: The two programs are identical except for variations in comments and layout. \textsc{Type-2}: The two programs are identical except for variations in identifier names and literal values in addition to \textsc{Type-1} differences. \textsc{Type-3}: The two programs are syntactically similar and have statements added, modified, or removed with respect to each other, in addition to \textsc{Type-2} differences. \textsc{Type-4}: The two programs are syntactically dissimilar but implement the similar functionality. In BigCloneBench, more than $98\%$ clone pairs belong to \textsc{Type-4}. Therefore in this task we mainly evaluate whether the program embeddings can capture functional similarity between programs.

We select another common C-language dataset on code clone detection task for test better generalizability. OJClone is a dataset for detecting clones extracted from the POJ-104~\cite{mou2016convolutional}. Code files submitted to the same programming problem can be regarded as similar code snippets. Consistent with the previous work~\cite{zhang2019novel}, we choose the first 15 problems out of 104 program problems to construct the clone dataset. Fifty thousand clone pairs are randomly selected into OJClone for time-saving reasons. These code pairs are mainly among the clone types of \textsc{Type-3} or \textsc{Type-4}.

\textbf{Metric}:
At testing time, the clone classifier predicts the probability that the two programs are clones. We set probability 0.5 as the threshold to determine whether they are clones. The evaluation metric is thereby precision, recall and F1-score. We do not distinguish among the clone types and report the overall results.

\begin{table*}[]
    \caption{The Results of Code Clone Detection.} \label{tab:clo-res}
    \begin{center}
\begin{tabular}{c|cccc|ccc}
 \toprule
Groups      & Models & \multicolumn{3}{c|}{BigCloneBench}  &   \multicolumn{3}{c}{OJClone}  \\ \cline{3-8} 
 &            & Precision      & Recall & F1    & Precision & Recall         & F1    \\ \midrule
Token-based & LSTM   & 0.882     & 0.86  & 0.872    & 0.892     & 0.825          & 0.857          \\
         & Transformer    & 0.914     & \textbf{0.968}   & \textbf{0.940}    & 0.945     & 0.262          & 0.410 \\ \midrule
AST-based    & TBCNN      & 0.891     & 0.950  & 0.920   &0.665     & 0.395          & 0.496          \\
 & AutoenCODE & 0.154   &  0.759  & 0.256 & 0.382 &0.241 &0.274 \\ \cmidrule(l){2-8} 
 & Code2Vec   & 0.833     & 0.911  & 0.870 & 0.721     & 0.623          & 0.668  \\
 & Code2Seq   & 0.858     & 0.938          & 0.896 & 0.852     & 0.146         & 0.249  \\ \cmidrule(l){2-8} 
 & GGNN       & 0.935 & 0.879         & 0.906      & 0.801     & \textbf{0.981}          & 0.880   \\
 & ASTNN      & \textbf{0.947}     & 0.930 & 0.939 & \textbf{0.974}     & 0.900  & \textbf{0.935}   \\ \bottomrule
\end{tabular}\end{center}
\end{table*}

\textbf{Results}:
The results are shown in Table~\ref{tab:clo-res}. We observe that all the models except AutoenCODE have similar performance. GGNN, TBCNN and Transformer perform better than other models.

The interesting observation is that the two token-based models perform very comparative to the AST-based models, and Transformer performs overall the best. The rationale is that in practice most clone programs are generated when programmers copy and reuse code snippets, therefore even the \textsc{Type-4} clone programs are likely to have very similar text and statements, which are easily captured by the token-based models. To verify this, we obtain the mean of \emph{word2vec}~\cite{mikolov2013efficient,mikolov2013distributed} embeddings of each program and use it to calculate the average cosine similarity between clone programs~\cite{kenter2015short,jatnika2019word2vec}. The result is as high as $0.866$. We therefore argue that on the currently used benchmark dataset for code clone detection, token-based models are sufficient for detecting most clones. However, this does not mean the AST-based models are not necessary for code clone detection. Indeed, TBCNN, GGNN and ASTNN can reach very high values of precision or recall. As such, one may choose different models to cater for different requirements for clone detection in real applications. Also, we expect the contribution from the relevant area to the construction of the dataset that contains more functionally similar but textually different clone pairs, which may better examine the advantage of learning program embeddings with the ASTs. 

The performance of code2vec and code2seq is slightly lower than other models, indicating that the AST paths convey less useful feature for clone detection. Finally, AutoenCODE performs bad in the current experimental setting. Although the model is proposed for the clone detection task, the clones are detected via clustering using coarse-grained distance metrics in the original paper~\cite{white2016deep}. Therefore it is worth thinking what is the proper way to determine clone pairs based on the program embeddings. We leave this for future work.

We then discuss the performance of eight program representation models on OJClone shown in the right part of Table~\ref{tab:clo-res}, which can be concluded into two situations. The first case is that the models have comparable precision and recall. ASTNN and GGNN achieve the best, with ASTNN reaching 0.974 in precision and GGNN having a peak recall of 0.981. In general, these two models take advantage of control/data flow information and extract more sophisticated structural features. LSTM performs moderately on detecting clones, indicating that functionally-similar programs are very likely to have very similar statements, which are easily captured by the token-based models. What's more, code2vec and AutoenCODE aren't as effective as the previous models for insufficient syntactical information. Code2vec embeds each path as a whole separately, thus hiding a lot of structural information in the path. AutoenCODE reconstructs the tree in a recursive way so that only leaf nodes in ASTs are utilized. The second case is that there is a high precision from the models' results, but the recall is pretty low. The high precision indicates that most of the code pairs that are predicted to be positive are true positive samples.  However, a large proportion of positive samples are incorrectly predicted as non-cloned pairs, resulting in a low recall. We inspect that Transformer, TBCNN and code2seq are more complex than other models from the model construction. In addition, OJClone is unbalanced that positive samples in OJClone only account for 6\%, so that the difficulty of training rises, and the requirements for the model to detect clones in OJClone also increase. Their low recall illustrates that they are very cautious in the unbalanced training process because they incline to classify uncertain samples to negative class. 
We inspect that they are less capable of detecting programs with complicated array calculations and multiple functions. Thus, they misjudge more positive samples to negative ones. 

In short, for the programs collected from a real code repository, the token-based embedding models may be the sufficiently good choice for code clone detection. If precision or recall is the metric of interest, one may also apply AST-based models such as ASTNN or TBCNN.

\subsection{Code Search}
\textbf{Dataset}:
We use the \textbf{CodeSearchNet}~\cite{husain2019codesearchnet} dataset for this task. The dataset is collected via the Bing search engine, where each query is paired with a returned code snippet. The queries and programs are combined with intent rewrites in StaQC~\cite{yao2018staqc}. Following the work in~\cite{sun2020pscs}, we use the Java code for the experiments, and obtain a training set of $330,000$ query-code pairs and a testing set of $19,000$ query-code pairs.

\textbf{Metric}: At testing time, for a pair of query and program, the model computes a similarity score, indicating how much the program matches the query. Based on the similarity scores, we calculate two commonly used metrics for code search~\cite{li2016relationship,lv2015codehow,raghothaman2016swim,ye2014learning}, namely SuccessRate@k and Mean Reciprocal Rank (MRR). For a given query, both the metrics evaluate whether the paired program is among the top candidates returned by the model. The set of candidate programs are formed by the paired program plus 998 randomly selected programs from the dataset. Then the model computes the similarity scores between the query and each candidate program, and sorts the scores in descending order. The metric SuccessRate@k considers a search to be successful if the score of the paired program is among the top-k in the sorted list. Then SuccessRate@k is calculated as the proportion of successful searches to all searches. We set $k=10$ in the experiments~\cite{sun2020pscs,gu2018deep}. MRR takes a step forward and considers the rank of the paired program in the list. It is calculated as the average of the inverse of ranks for all the paired programs. For both metrics, a higher value indicates better performance.

\begin{table}[t]
    \caption{The Results of Code Search}\label{tab:se-res}
    \begin{center}
    \begin{tabular}{c|ccc}
    \toprule
    Groups & Model &    SuccessRate@10 & MRR       \\ \midrule
    Token-based & LSTM        & 60.27\% & 0.3946 \\
                & Transformer & 53.99\% & 0.3376 \\ \midrule
                & TBCNN       & 0.75\%  & 0.0088 \\
                & AutoenCODE  & 1.32\% & 0.0088 \\ \cmidrule(l){2-4} 
    AST-based   & code2vec    & 39.48\%  & 0.2170\\
                & code2seq    & \textbf{72.73\%}  &  \textbf{0.5391} \\ \cmidrule(l){2-4} 
                & GGNN        & 44.37\%  & 0.2583 \\ 
                & ASTNN       & 40.72\%  & 0.2284\\ \bottomrule
    \end{tabular}
    \end{center}
\end{table}
\textbf{Results}:
The results for code search are shown in the Table~\ref{tab:se-res}. We observe that code2seq has much better performance than all the other models on both metrics, where SuccessRate@10 is $72.73\%$ and MRR is $0.5391$, and the two token-based models perform better than the remaining AST-based models except code2seq.

For the two models extracting features from the AST paths, code2seq has a drastic performance improvement over code2vec. Since the difference between them is that code2seq uses more fine-grained information of the internal nodes, we may conclude that the internal nodes of the AST paths convey vital features to match the program with the corresponding query and therefore cannot be neglected. The moderate performance of the two token-based models again indicates that the plain text of the source code already contains essential features for code search. To see this, we apply the method described in Section~\ref{sec:ccd} and calculate the average cosine similarities between a program and the corresponding query and between a program and the negative sample queries. The results are 0.784 and 0.677, respectively. Therefore, the token-based models can benefit from the textual similarity between a program and the corresponding query. TBCNN and AutoenCODE perform very bad on this task, which may indicate that the key features for matching with a query are not captured with the bottom-up aggregation approach. GGNN and ASTNN have similar performance with code2vec, although they integrate the information of internal nodes into the program embedding. Therefore we may conclude that with the precise information of the internal nodes, the structural information of the AST paths is the strongest signal for matching a program with the corresponding natural language query.

In short, for tasks such as code search that need to match the content of natural language (NL) and programming language (PL), using the token-based models to embed the programs yields overall better results than using the AST-based models. However, the AST-paths may contain the key structural features that can drastically boost the performance of such tasks. Indeed, code2seq outperforms all baseline models on another NL-PL task, code summarization, in the original paper~\cite{alon2018code2seq}, which aims to automatically generate a piece of natural language comment for a program. This is another evidence for our conjecture.

 \section{Explanatory Study}\label{sec:attr}
In this section, we study why the models demonstrate the performance reported in Section~\ref{sec:experiments}. We attempt to answer the question by investigating what elements in the source code are mostly relevant to the model prediction, which is a widely studied problem in the field of explainable deep learning in recent years~\cite{serrano2019attention,wiegreffe2019attention,jacovi2020towards}. Note that although the method does not directly manipulate the structure of an AST, it tells what each AST-based model captures from the source code via the corresponding structural information. As such, the results could be viewed as a proxy of how well each particular AST structure captures the semantic information.

To this end, a practical solution is to calculate the attention weights in the models and see which part of the source code is mostly attended~\cite{vig2019multiscale,clark2019does,tao2018get,chen2019deep,tang2018analysis}. However, not all the models under evaluation take advantage of the attention mechanism, e.g., LSTM, TBCNN, GGNN, which makes the approach infeasible for our experiments.

\subsection{Explaining with Integrated Gradients}
An alternative solution is to attribute the predictions of the models to their input features, i.e., to investigate what elements in the source code are responsible for the predictions. The underlying techniques are referred to as prediction attribution techniques~\cite{baehrens2010explain,simonyan2014deep, shrikumar2017learning,gupta2019neural}. A simple yet effective prediction attribution technique is proposed by Sundararajan et al.~\cite{sundararajan2017axiomatic} and called \emph{integrated gradients}. The intuition of the technique is that, when attributing a prediction to some features in the input, one also needs to know the ``neutral'' output of the model given the absence of the features in the input. Such an input is called the baseline input. For example, when attributing a sentiment prediction of a sentence to the individual words, the sequence of all-zero vectors could be used as a baseline input~\cite{sundararajan2017axiomatic, murdoch2018beyond,wallace2019allennlp,he2019towards,du2019attribution}. Then the integrated gradients technique distributes the difference between the prediction of the original input and the prediction of the baseline input to the individual input features.

For the models under evaluation, the individual input features are the tokens in the source code (LSTM and Transformer), the tokens of the leaf/internal nodes (all AST-based models), and the tokens of the paths (code2vec and code2seq). Given a proper baseline input for each model on each task, integrated gradients technique could attribute the prediction to the tokens in the source code. Then we visualize the attribution results by highlighting the elements in the source code with high attribution scores. In particular, we rank the input tokens by their attribution score in the descending order and pick the top $60\%$ tokens. We further equally divide the picked tokens into three groups, where the group with the highest scores are highlighted using the red color, the group with the medium scores are highlighted using the orange color, and the group with the lowest scores are highlighted using the yellow color. For each task, we choose three programs for visualization and render the programs for each individual model. Due to the page limit, we demonstrate the results of one program for each task and leave the results of the other two programs in the supplementary material. In the end, we also try to systematically interpret the attribution results by grouping different types of input tokens in all code.

\subsection{The Baseline Input}

\begin{table*}[t]
    \caption{The neutral characteristic of the all-zero baseline.} \label{tab:baseline}
    \begin{center}
    \setlength{\tabcolsep}{0.5mm}{
    \begin{tabular}{c|c|cc|cc}
    \toprule
 & \multicolumn{1}{c|}{Classification Task} & \multicolumn{2}{c|}{Clone Task} & \multicolumn{2}{c}{Search Task} \\
            & p-value of chisquare test   & Mean                          & Standard Deviation              & Mean                          & Standard Deviation              \\ \midrule
LSTM        & 1.0000  & 0.2466                        & 3.77e-01    & 0.3596        & 2.03e-01                        \\
Transformer & 0.9999  & 0.0200                        & 5.86e-02    & 0.1727        & 1.58e-01                        \\ \midrule
TBCNN       & 1.0000  & 0.0000 & 3.30e-09 & 0.0489                        & 5.02E-02                        \\
AutoenCODE  & 1.0000  & 0.1065    & 2.02e-02    & 0.3149        & 1.61E-02                        \\
code2vec    & 1.0000  & 0.1679    & 1.99e-01    & 0.0937        & 1.65e-01                        \\
code2seq    & 1.0000  & 0.0143    & 1.05e-01    & -0.0235       & 1.96e-01                        \\
GGNN        & 1.0000  & 0.0004    & 1.70e-02    & 0.0000        & 0.00e+00                       \\
ASTNN       & 1.0000  & 0.0272    & 1.32e-01    & 0.0742        & 7.59e-02       \\ \bottomrule
    \end{tabular}}
    \end{center}
\end{table*}

A baseline input should produce ``neutral'' prediction for each task~\cite{sundararajan2017axiomatic}. For code classification, neutral prediction means the model cannot identify which class a given input belongs to, i.e., the model should output a uniform distribution over all classes for the input. For code clone detection, neutral prediction means the output should be close to zero when the input pair consists of a reference code and a baseline code. Then once we replace the baseline with a code to be detected, we could blame the clone probability to the input features of the detected code. Similarly, for code search, neutral prediction means the output cosine distance is close to zero given a query and a baseline code. Then once we replace the baseline with a candidate code, we could blame the cosine distance to the candidate code, no matter the distance (i.e., correlation between the query and the code) is positive or negative.

Borrowing from the idea in NLP, we use all-zero embedding vectors to constitute the baseline code and study the corresponding output for each task. For code classification, neutral prediction means the model cannot identify which class a given input belongs to, that is, the output distribution is uniform over all classes. To verify this, we calculate the average output of all baseline code in each class and then compute the mean of the output for all classes. We assume the mean output is a uniform distribution over all classes, and conduct a goodness-of-fit test with chi-square statistics. From the second column in Table~{\ref{tab:baseline}}, with the default setting on confidence ($0.95$) and degree of freedom ($n-1, n=104$), we cannot reject the uniformity hypothesis for every model since the p-value of chi-square test is greater than the confidence.
For code clone detection, neutral prediction means the model output should be close to zero when the input pair consists of a reference code and a baseline code. We calculate the mean and standard deviation of the clone probabilities between all reference code and the baseline code, and demonstrate that most of the program embedding models get a close-to-zero output, which are displayed in the third column of Table~{\ref{tab:baseline}}.
Similarly, for code search, neutral prediction means the output cosine distance is close to zero when the input pair consists of a query and a baseline code. From the last column in Table~{\ref{tab:baseline}}, we calculated the mean and standard deviation of the cosine distances between all queries and the baseline code, and demonstrate that the baseline input can produce neutral prediction for all models.
The results show that the all-zero input generally satisfies the requirement of neutral prediction in all three tasks for all the program embedding models. 
\begin{figure*}[t]
    \centering
	\subfigure[Attribution score on LSTM]{
      \includegraphics[width=0.37\columnwidth]{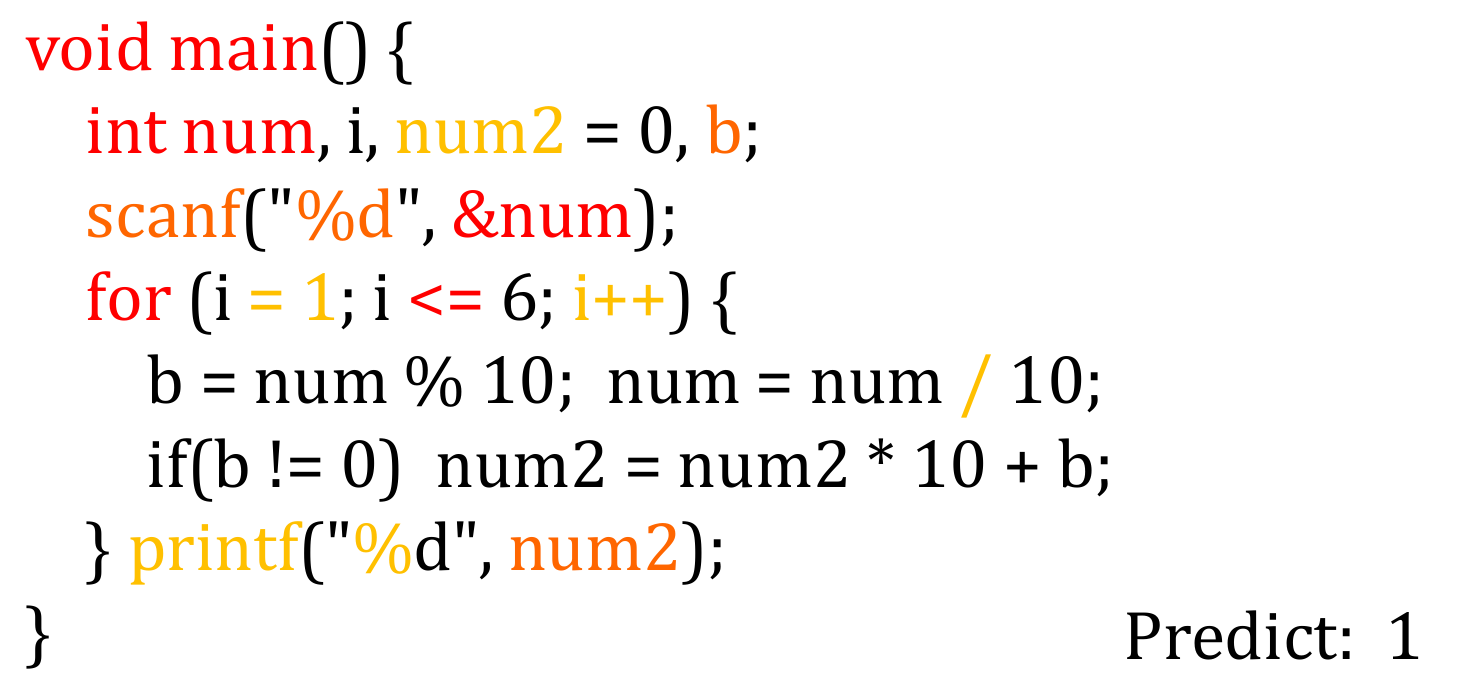}
      \label{fig:cla-32212-LSTM}
    } 
    \subfigure[Attribution score on Transformer]{
      \includegraphics[width=0.37\columnwidth]{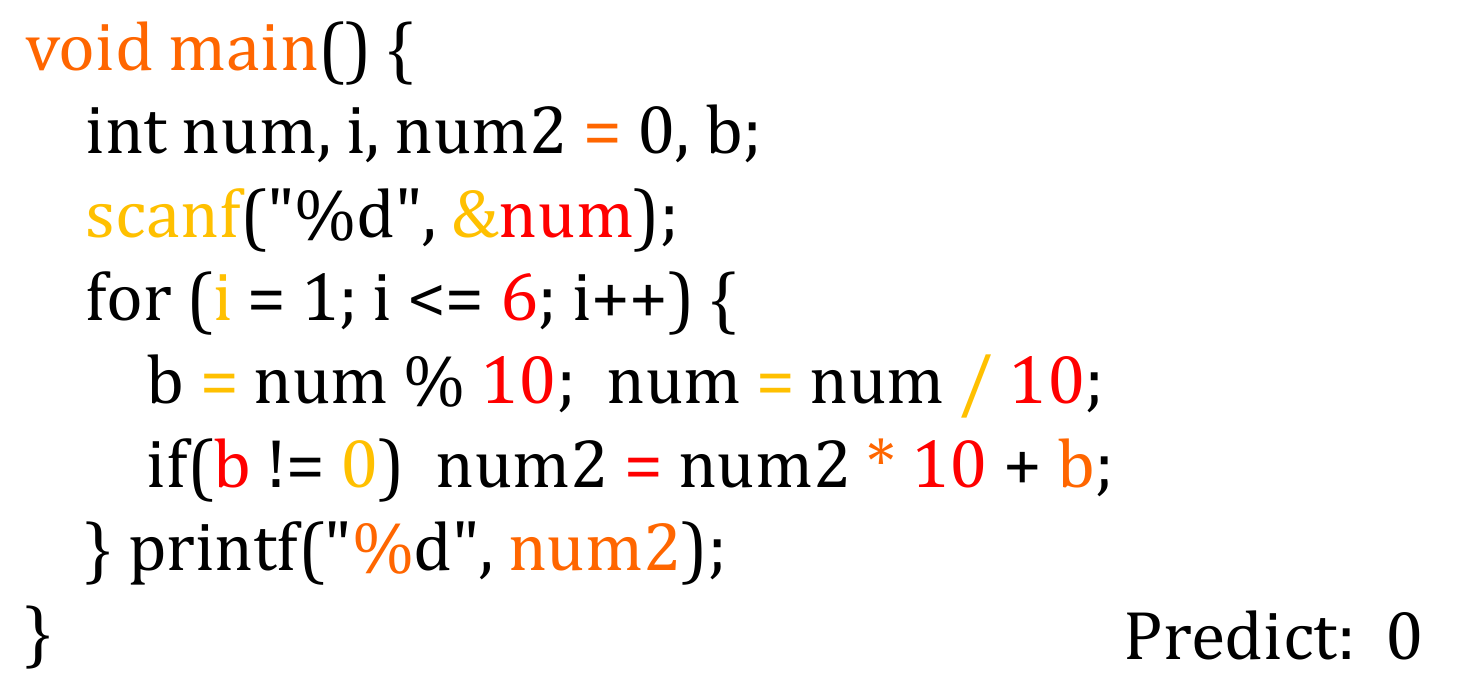}
      \label{fig:cla-32212-Transformer}
    } \\
    \subfigure[Attribution score on TBCNN]{
      \includegraphics[width=0.37\columnwidth]{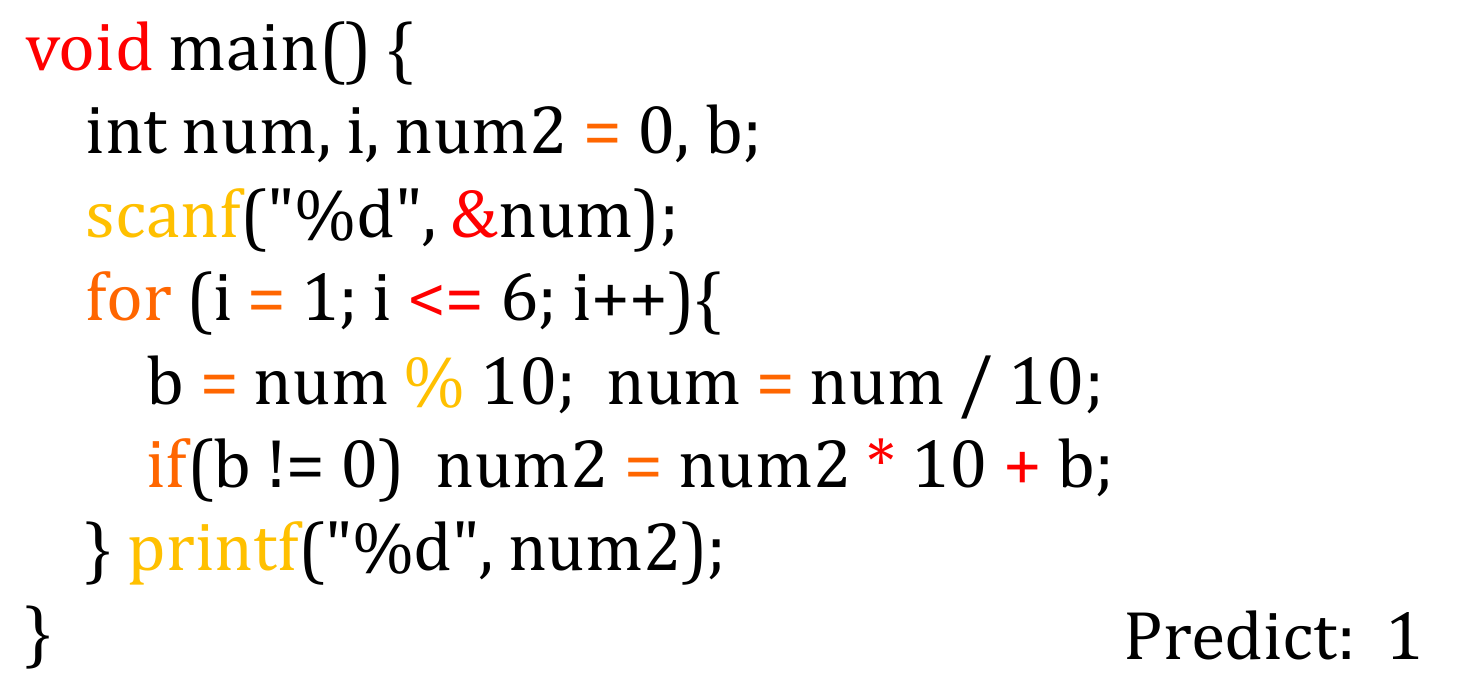}
      \label{fig:cla-32212-TBCNN}
    } 
    \subfigure[Attribution score on AutoenCODE]{
      \includegraphics[width=0.37\columnwidth]{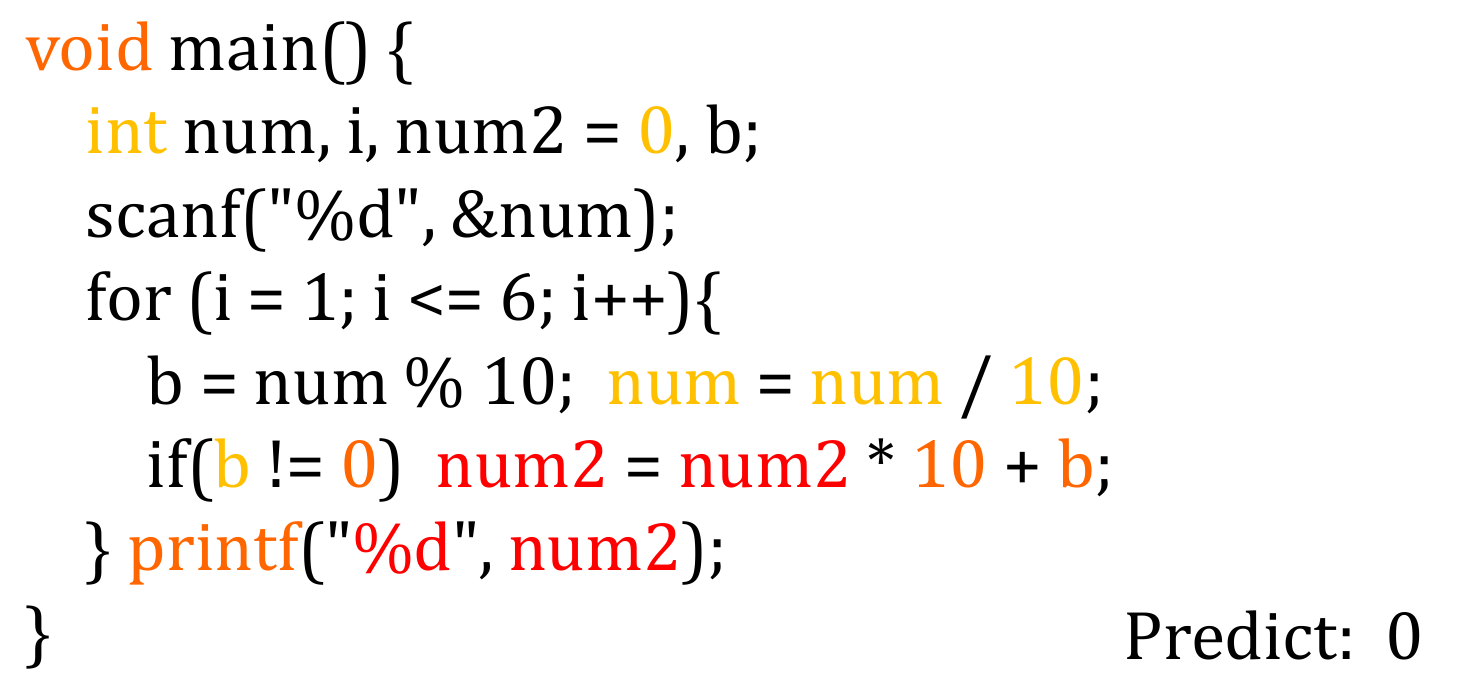}
      \label{fig:cla-32212-autoencoder}
    } \\
    \subfigure[Attribution score on code2vec]{
      \includegraphics[width=0.37\columnwidth]{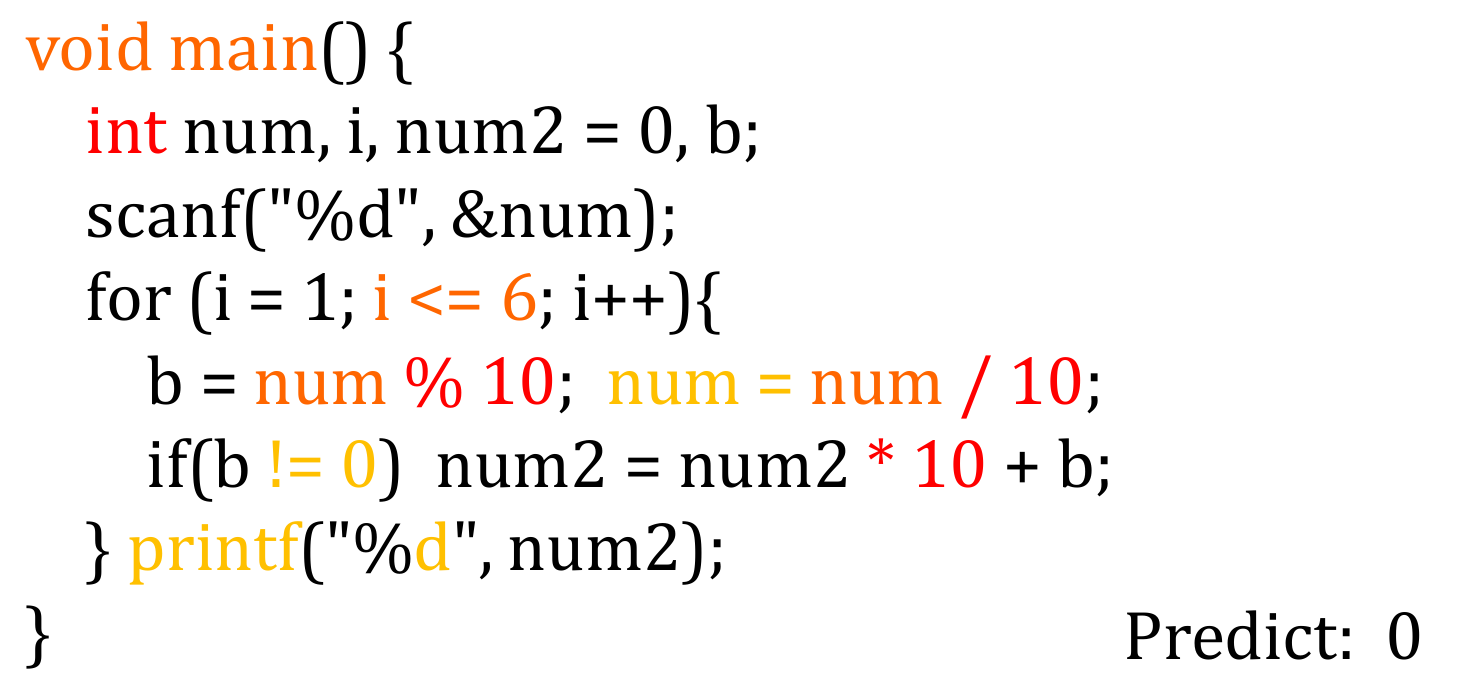}
      \label{fig:cla-32212-code2vec}
    }
    \subfigure[Attribution score on code2seq]{
      \includegraphics[width=0.37\columnwidth]{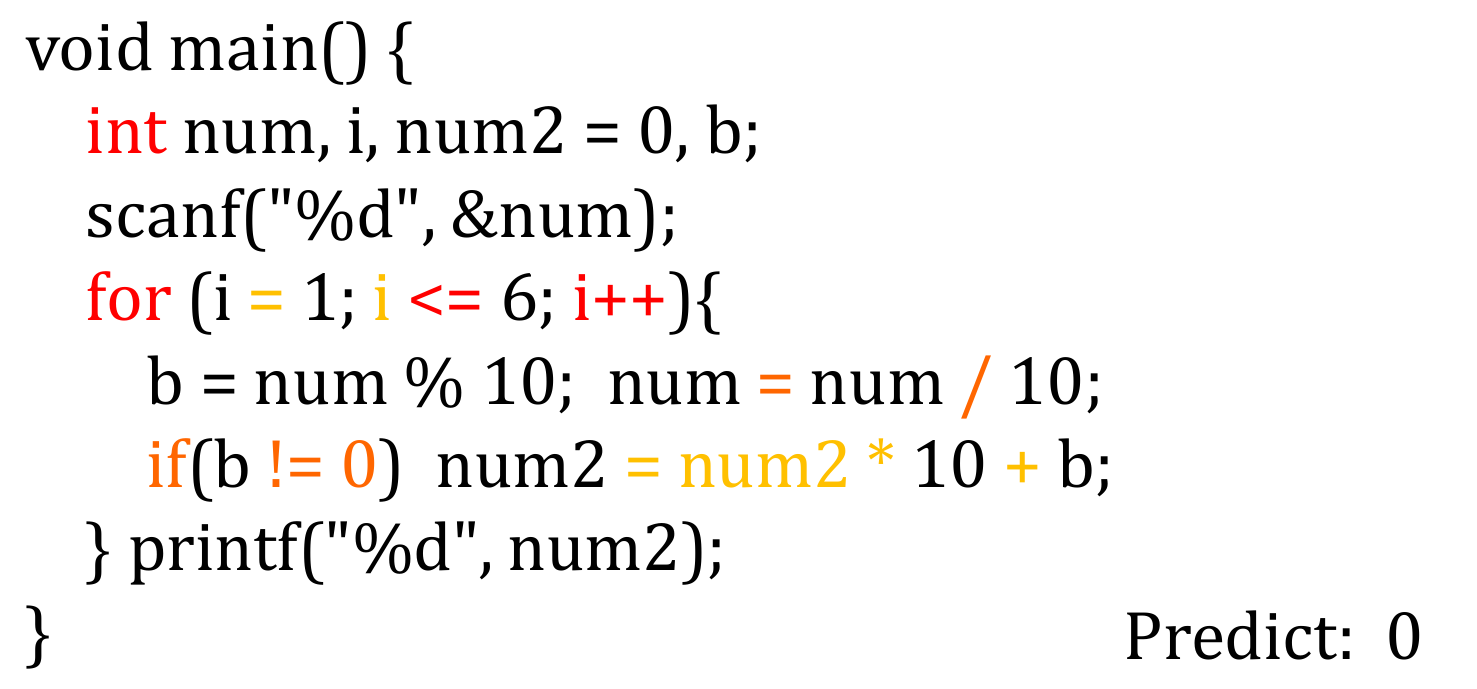}
      \label{fig:cla-32212-code2seq}
    } \\
    \subfigure[Attribution score on GGNN]{
      \includegraphics[width=0.37\columnwidth]{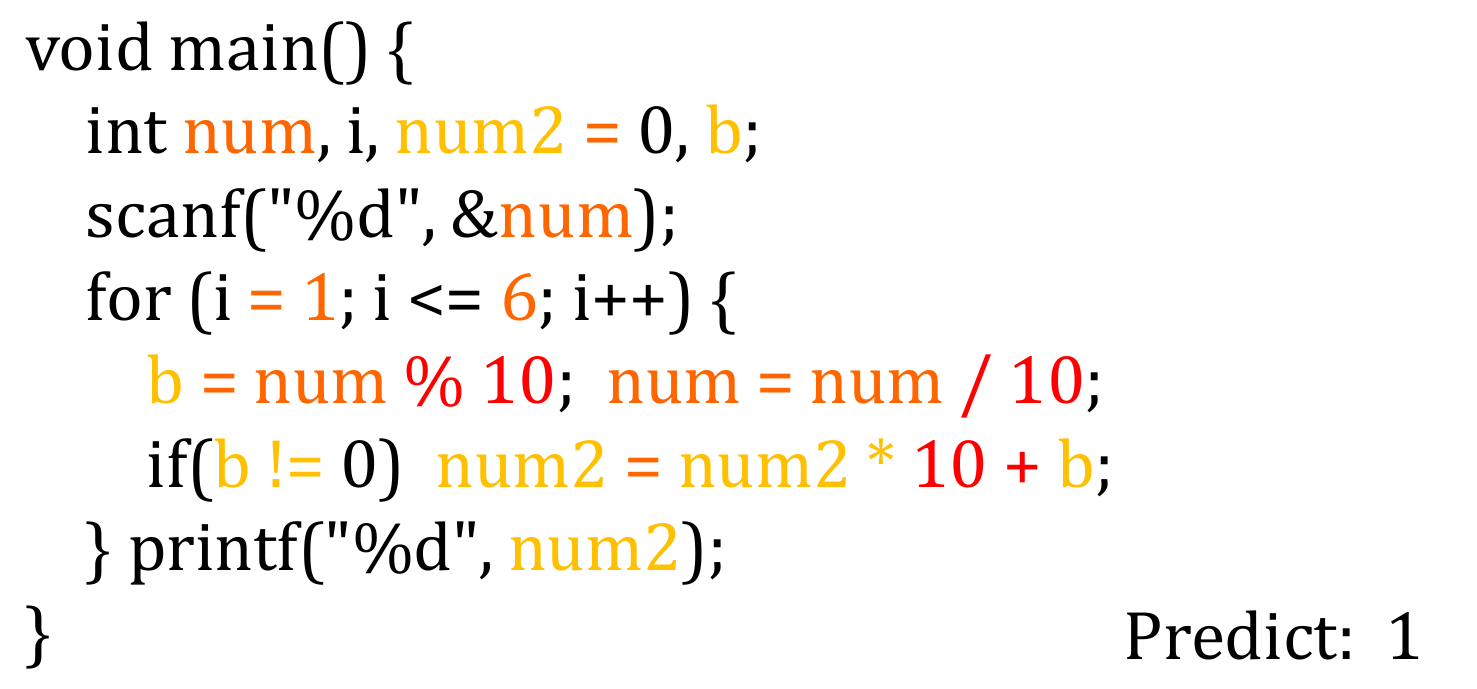}
      \label{fig:cla-32212-GGNN}
    }
    \subfigure[Attribution score on ASTNN]{
      \includegraphics[width=0.37\columnwidth]{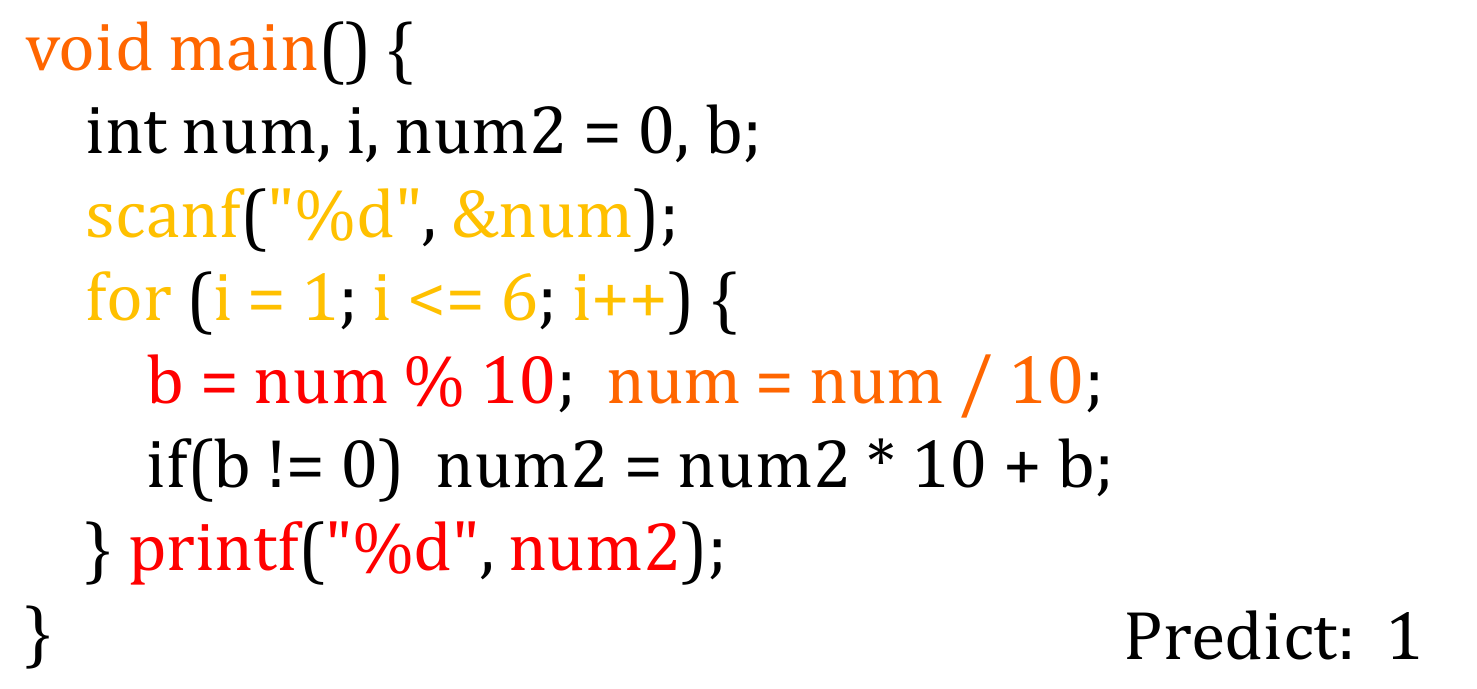}
      \label{fig:cla-32212-ASTNN}
    } 
	\caption{Attribution analysis on code classification. The function reverses the order of the digits in an integer.}
	\label{fig:cla-attr} 
\end{figure*}

\subsection{Attribution Analysis}
\subsubsection{\textbf{Code Classification}}

We first perform attribution analysis on the code classification task and explore what elements in the source code are responsible for correct or incorrect functionality prediction. The difference between the output of a program and the output of a baseline input is calculated as the sum of the difference between corresponding values in the two output vectors. We select a simple program that reverses the order of the digits in an integer for visualization. The attribution results for all the models are shown in Fig.~\ref{fig:cla-attr}, where a 1 prediction indicates a correct classification and a 0 prediction indicates a wrong classification. More results can be read in the supplementary material.

We can observe that two parts of elements are vital for the models to correctly classify this program. The first part is the tokens in the \texttt{for} loop statement (line 4), which iterates over the digits in the input integer. Among the four models that correctly predict the functionality, LSTM, TBCNN and ASTNN give high credits to the tokens in the statement. In contrast, 3 out of the 4 models that fail to predict the functionality have not captured the keyword \texttt{for} and give next-to-zero credits to the tokens in the statement. The second part are the tokens in the body of the \texttt{for} loop (line 5-6), which conduct the actual reversal process. Both ASTNN and GGNN give high credits to the tokens in the two statements. The three failed models, Transformer, code2vec and code2seq, have not captured much information of these tokens. For the latter two models, the prediction is first distributed to the AST paths, which is in turn distributed to the AST nodes. Thus if a key path is not given a high credit, all the tokens in the path might be down-weighted. On the other hand, although AutoenCODE gives high credits to the reversal statements, it fails to predict the functionality. We guess this is because it uses a greedy bottom-up approach to aggregate the token information and neglects the flow of control and data as well as the AST's structure, as explained in Section~\ref{sec:cc}.

The results indicate that to correctly predict the functionality of a program, the embedding models should capture the key features pertaining to the control statements and the statements that actually realize the functionality. Moreover, neglecting the AST's structure and the flow of control and data hinders the correct prediction, although some key features are captured.

\subsubsection{\textbf{Code Clone Detection}}

\begin{figure*}[h]
    \centering
	\subfigure[Program-1 in the clone pair]{
      \includegraphics[width=0.42\columnwidth]{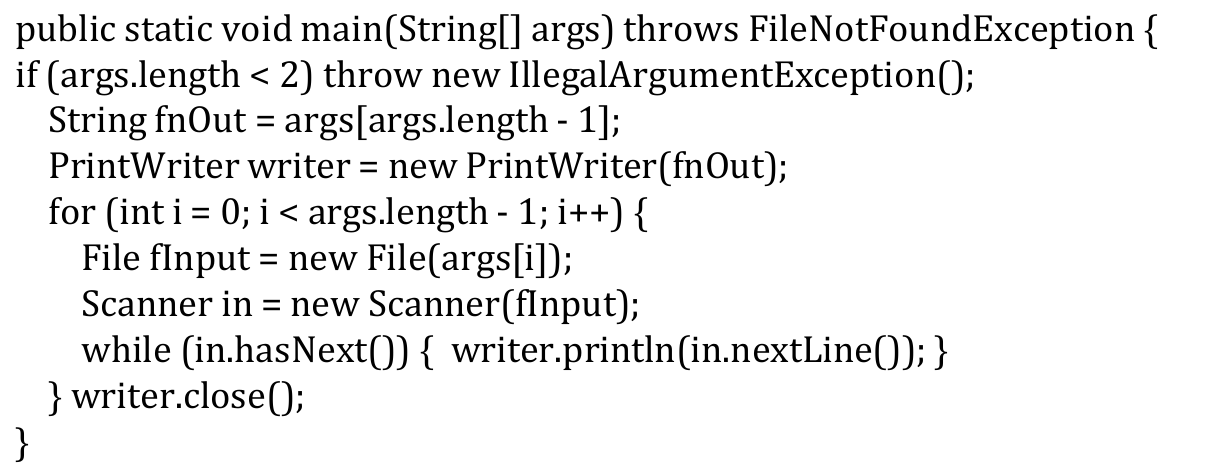}
      \label{fig:clo-2518655}
    } \\
    \subfigure[Attribution score on LSTM]{
      \includegraphics[width=0.42\columnwidth]{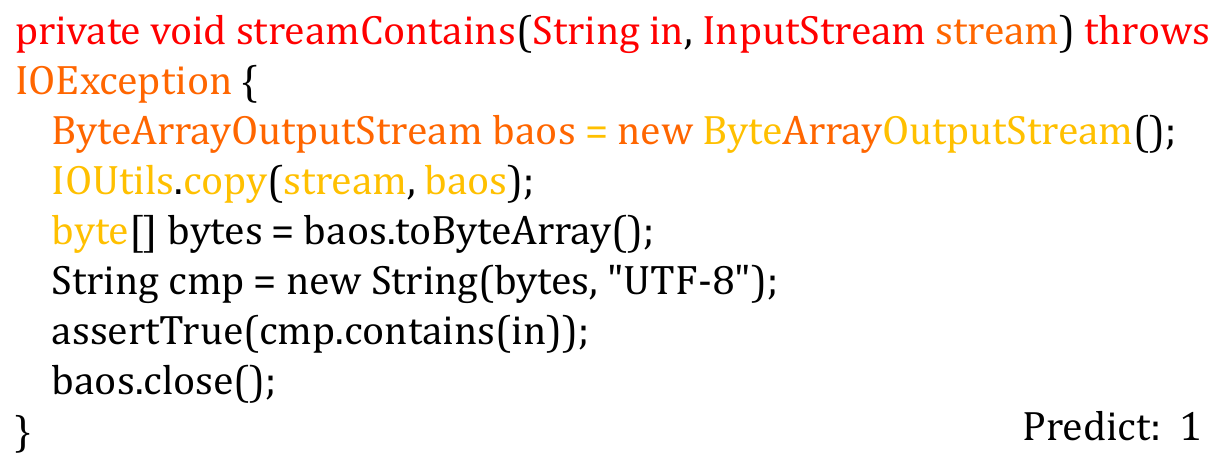}
      \label{fig:clo-11562165-LSTM}
    } 
    \subfigure[Attribution score on Transformer]{
      \includegraphics[width=0.42\columnwidth]{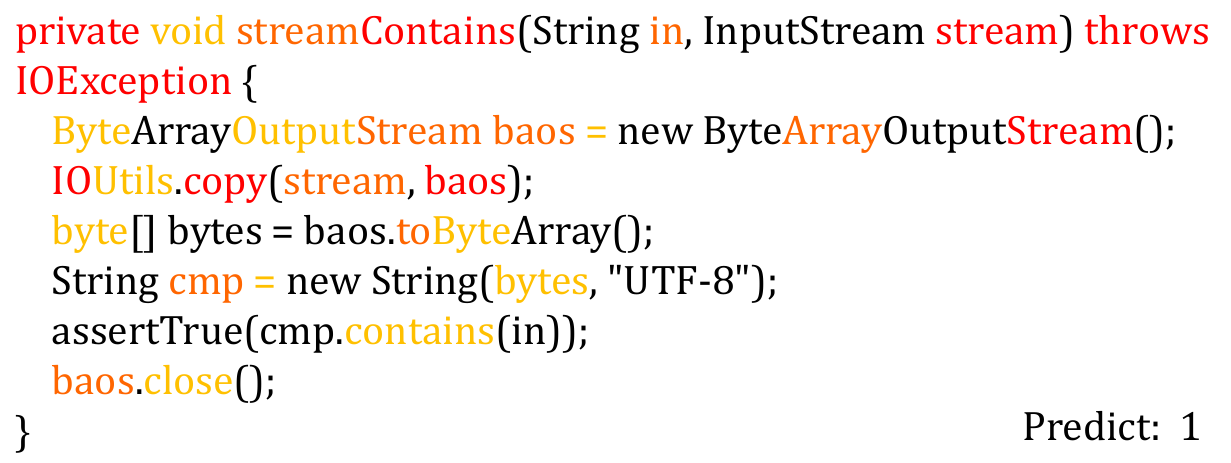}
      \label{fig:clo-11562165-Transformer}
    } \\
    \subfigure[Attribution score on TBCNN]{
      \includegraphics[width=0.42\columnwidth]{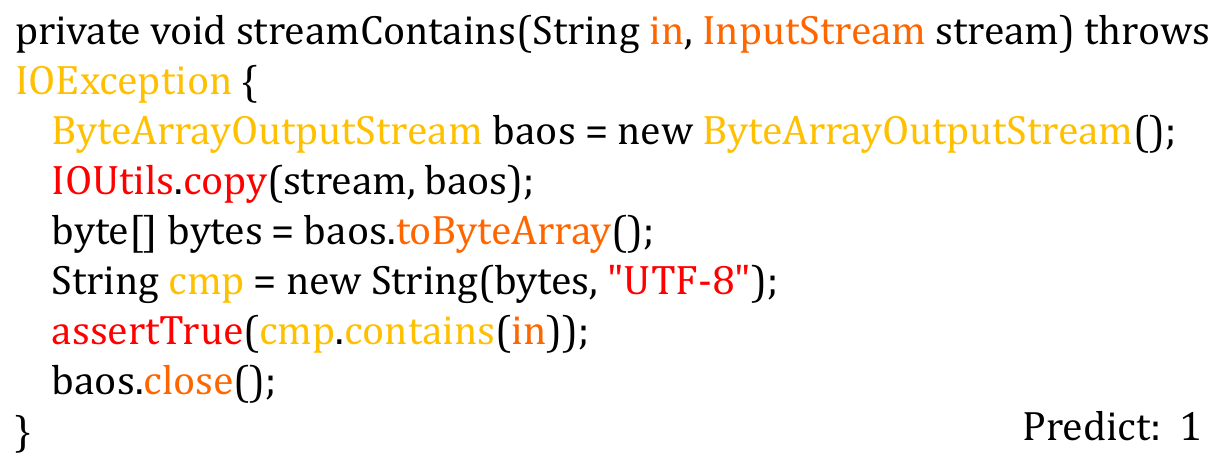}
      \label{fig:clo-11562165-TBCNN}
    } 
    \subfigure[Attribution score on AutoenCODE]{
      \includegraphics[width=0.42\columnwidth]{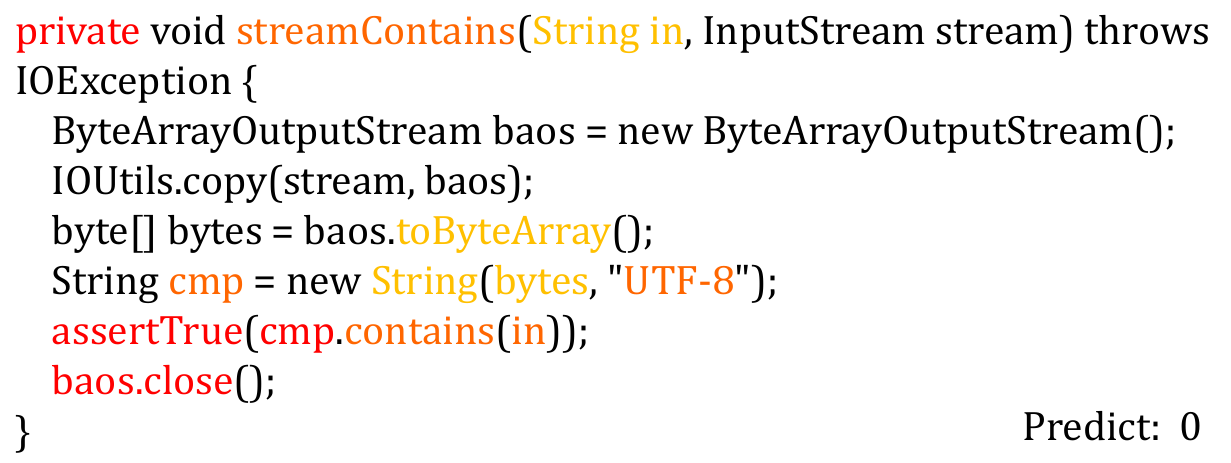}
      \label{fig:clo-11562165-autoencoder}
    }\\
    \subfigure[Attribution score on code2vec]{
      \includegraphics[width=0.42\columnwidth]{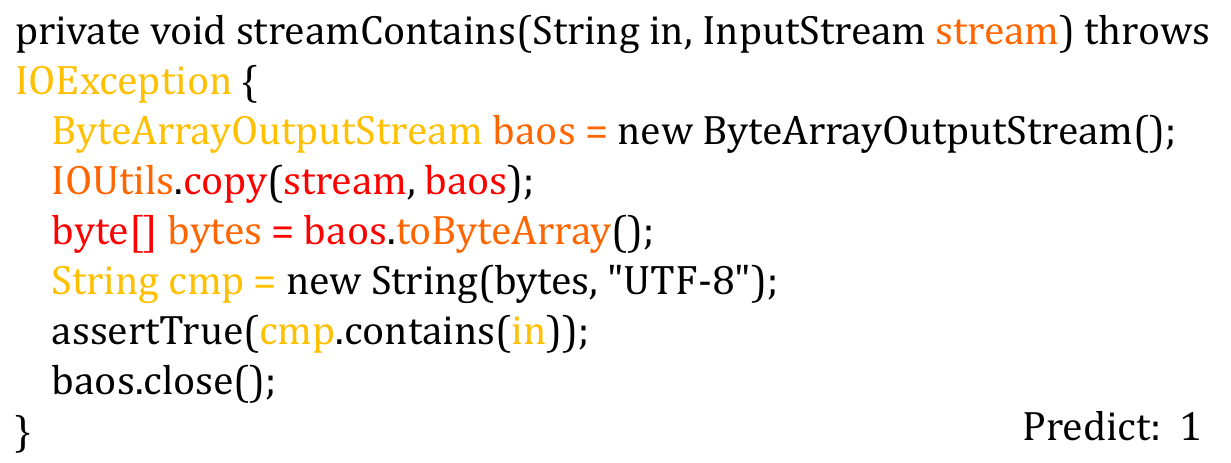}
      \label{fig:clo-11562165-code2vec}
    } 
    \subfigure[Attribution score on code2seq]{
      \includegraphics[width=0.42\columnwidth]{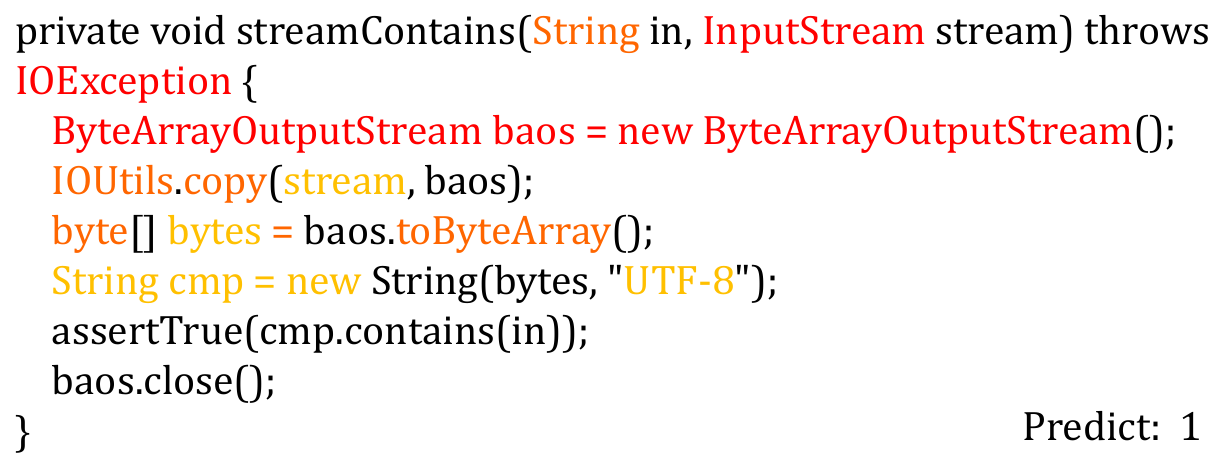}
      \label{fig:clo-11562165-code2seq}
    } \\
    \subfigure[Attribution score on GGNN]{
      \includegraphics[width=0.42\columnwidth]{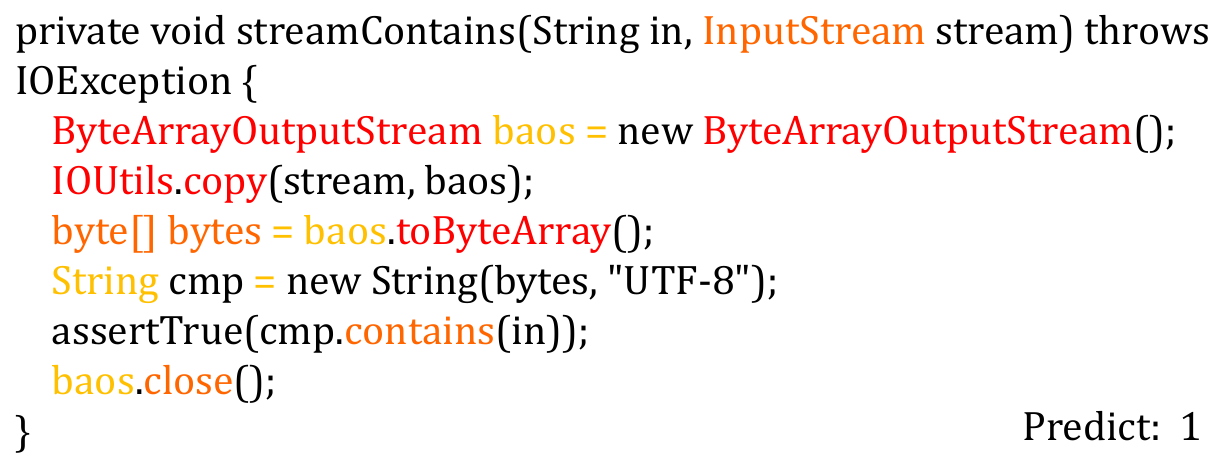}
      \label{fig:clo-11562165-GGNN}
    }
    \subfigure[Attribution score on ASTNN]{
      \includegraphics[width=0.42\columnwidth]{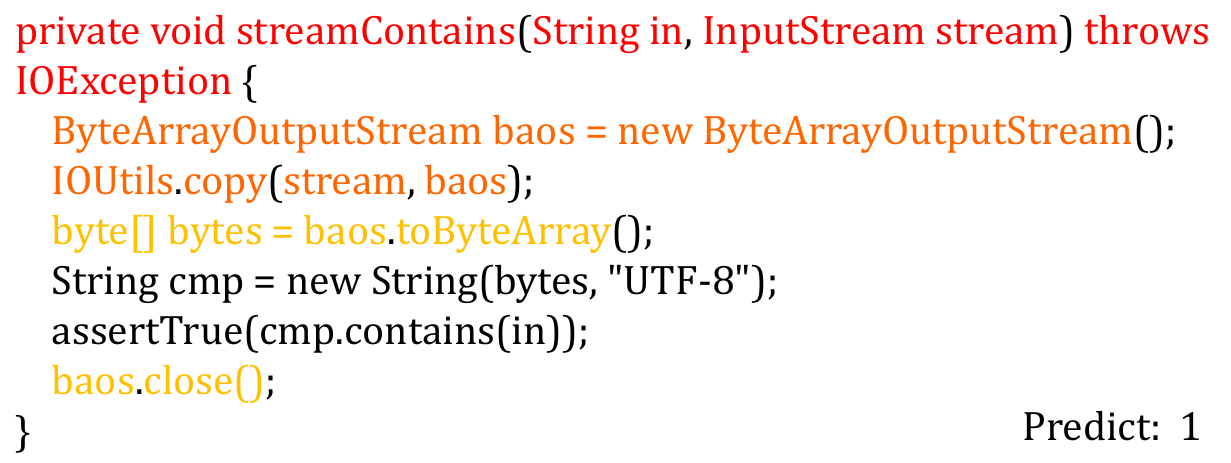}
      \label{fig:clo-11562165-ASTNN}
    }\\
	\caption{Attribution analysis on clone pairs. The functions read from input streams/files and write to output streams/files.}
	\label{fig:clo-attr} 
\end{figure*}

To visualize the prediction attribution for code clone detection, we pick a clone pair of interest and replace one program in the pair with all-zero vectors to form the baseline pair. The difference of the two outputs is calculated as the difference of the clone probability predicted by the models. Then we attribute the difference to the tokens in the program that is replaced, which could tell why the program is (or not) predicted as a clone of the other one. In the main paper, we demonstrate the prediction attribution results of a \textsc{Type-4} clone pair, where the two programs have different statements but similar functionality. The functionality is to read the content from input streams (files) and write it to output streams (files). 

The results are shown in Figure~\ref{fig:clo-attr}, where Figure~\ref{fig:clo-attr}(a) shows the fixed program in the pair and the other subfigures show the prediction attribution on the other program for the eight models. All the models except AutoenCODE successfully predict the pair is cloned. From the visualization, we can observe that the most important features to capture are the tokens regarding the input and output stream, which correspond to the tokens regarding the input and output files in Figure~\ref{fig:clo-attr}(a). AutoenCODE gives no credit to any token in line 2, which initializes an output stream, therefore it fails to predict the clone. All other models have attributed to at least one token pertaining to the input and output stream. Although LSTM attributes few credits to the tokens conducting the actual read and write, the successful capturing of the input and output stream leads to the correct prediction.


\subsubsection{\textbf{Code Search}}
\begin{figure*}[t]
    \centering
\subfigure[Attribution score on LSTM]{
      \includegraphics[width=0.37\columnwidth]{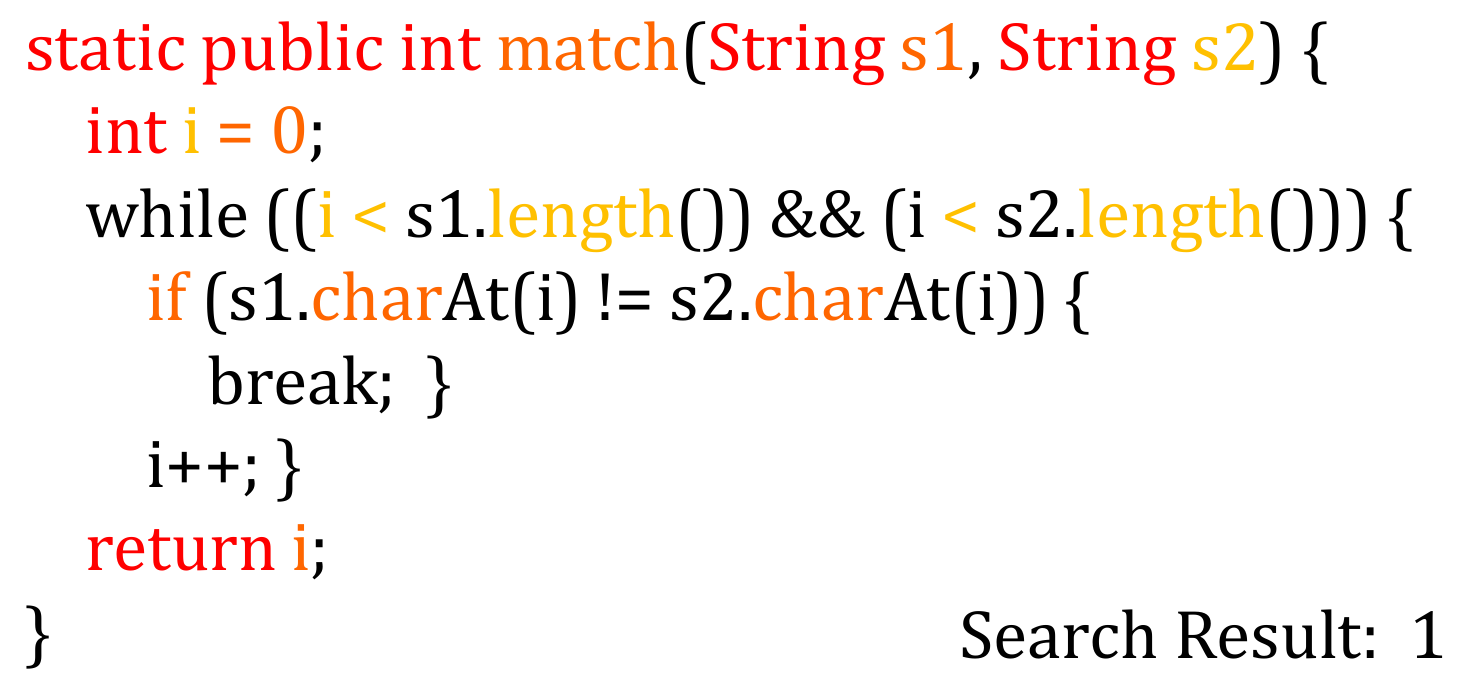}
      \label{fig:se-339813-LSTM}
    } 
    \subfigure[Attribution score on Transformer]{
      \includegraphics[width=0.37\columnwidth]{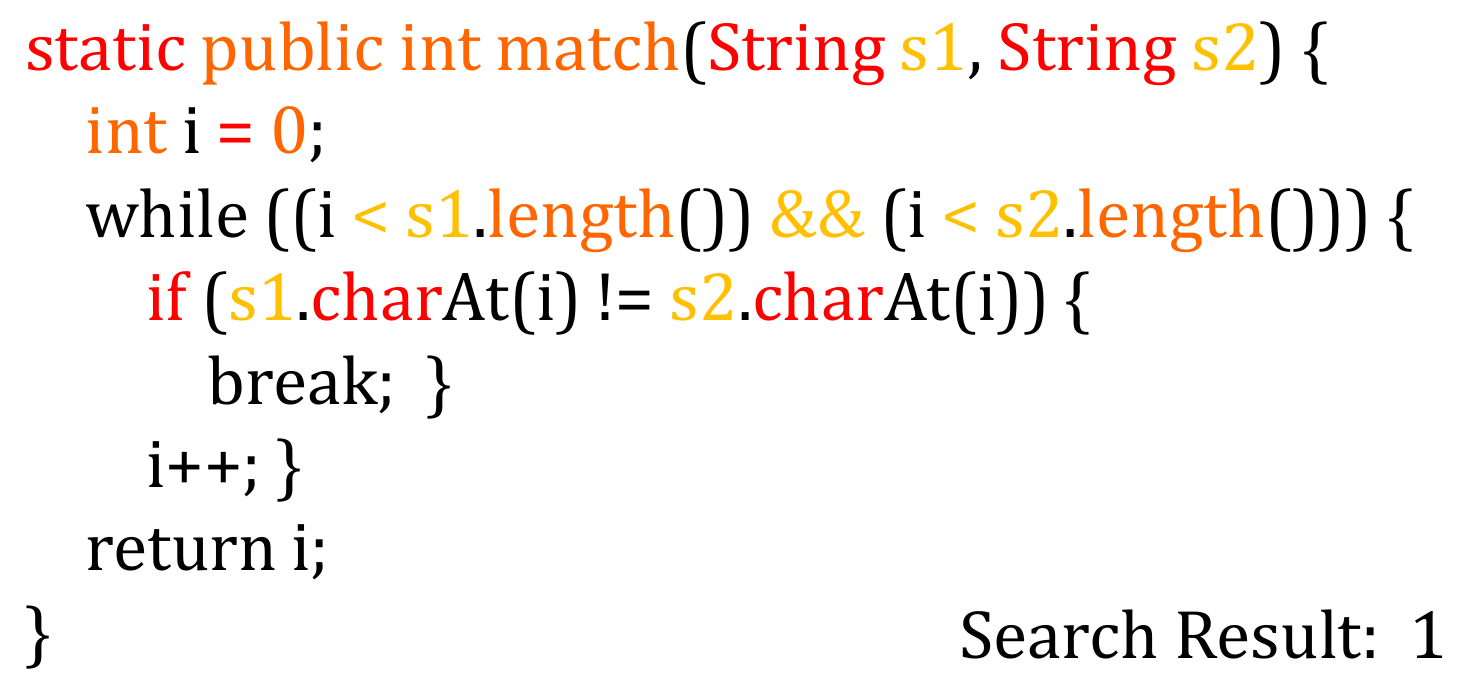}
      \label{fig:se-339813-Transformer}
    } \\
    \subfigure[Attribution score on TBCNN]{
      \includegraphics[width=0.37\columnwidth]{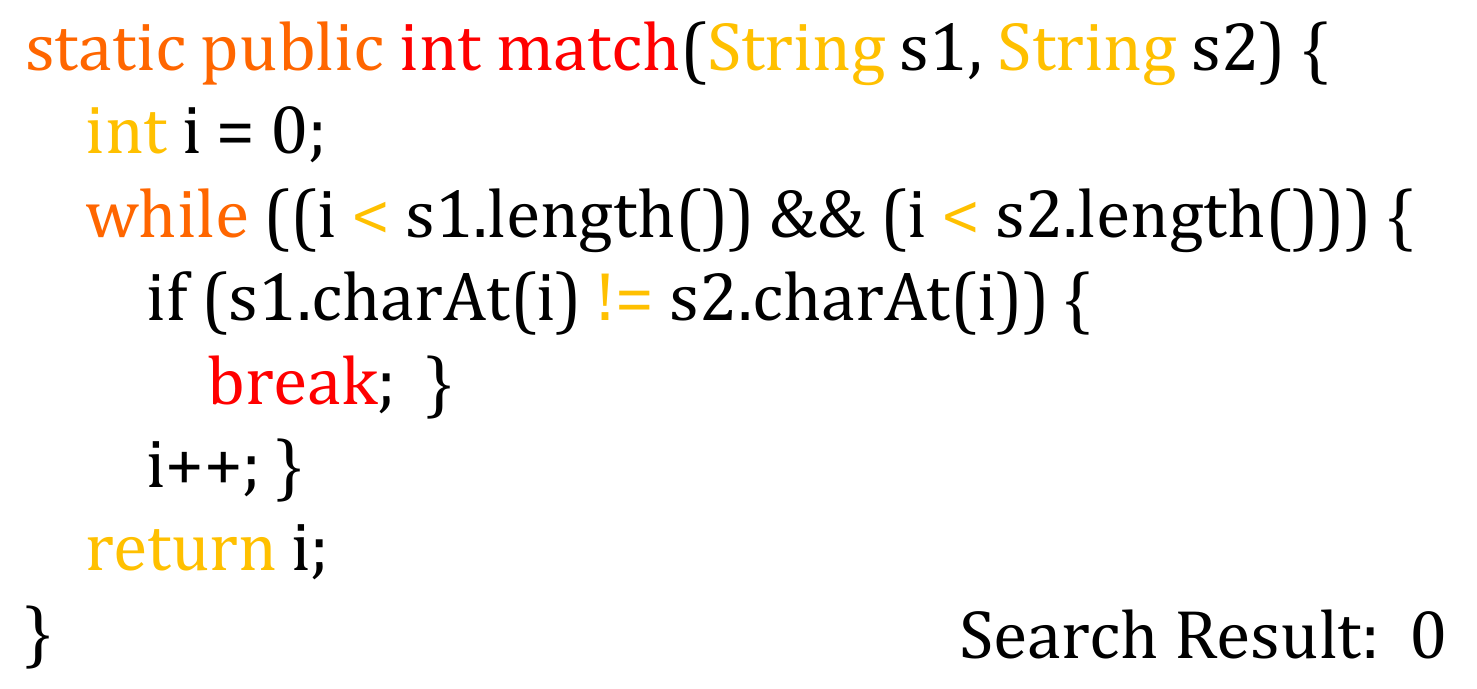}
      \label{fig:se-339813-TBCNN}
    } 
    \subfigure[Attribution score on AutoenCODE]{
      \includegraphics[width=0.37\columnwidth]{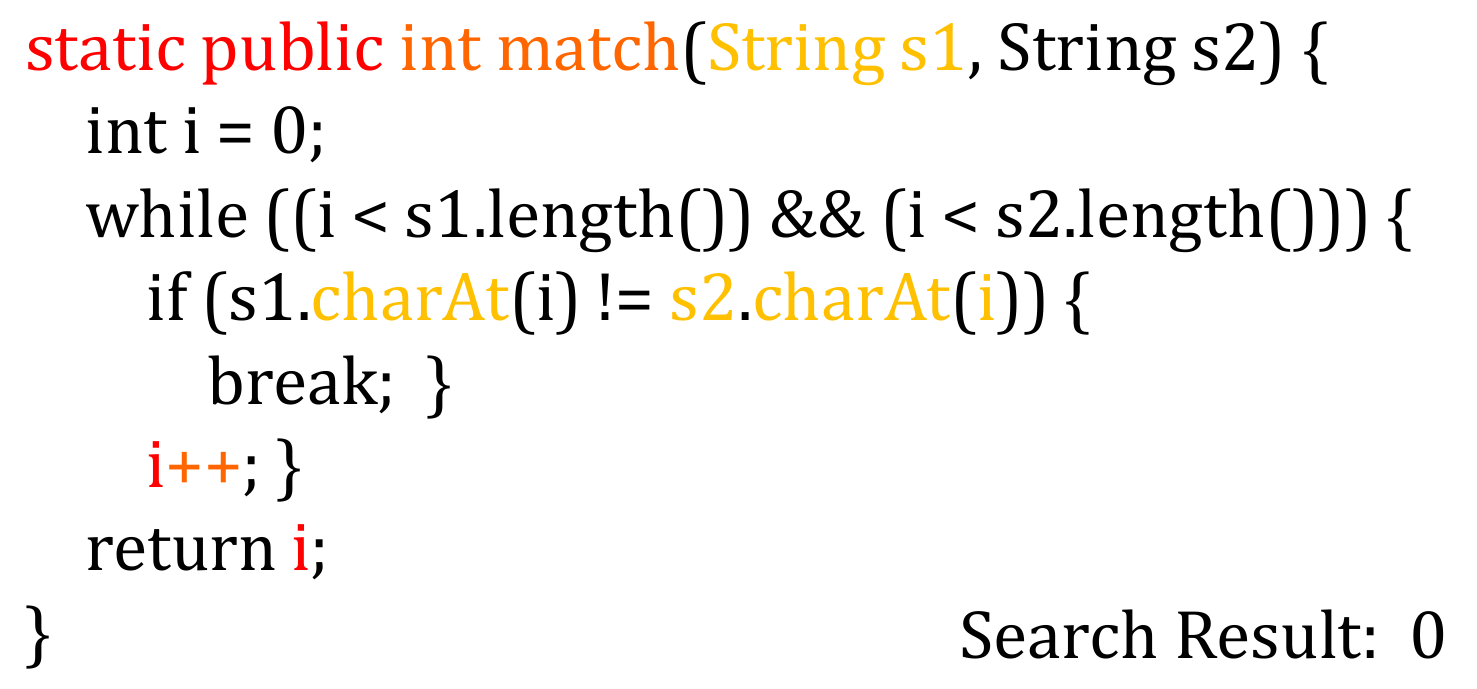}
      \label{fig:se-339813-autoencode}
    }\\
    \subfigure[Attribution score on code2vec]{
      \includegraphics[width=0.37\columnwidth]{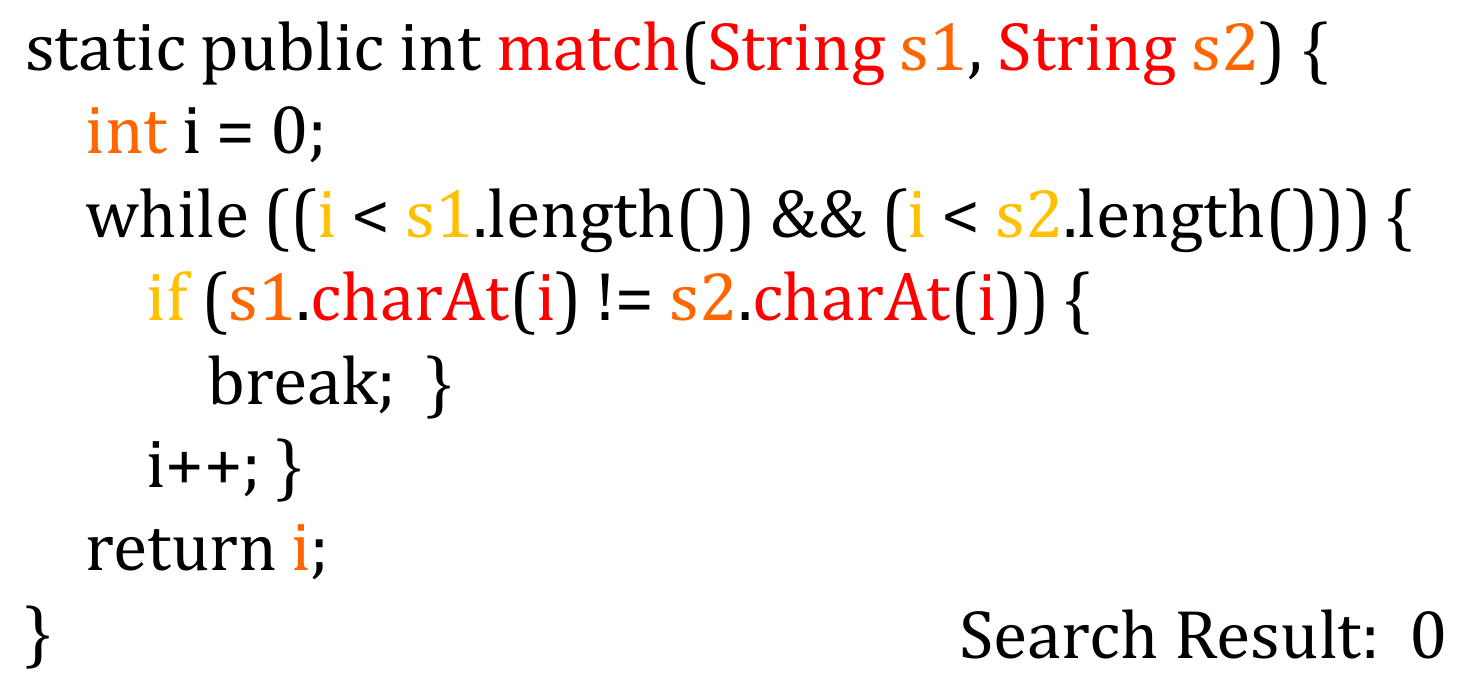}
      \label{fig:se-339813-code2vec}
    } 
    \subfigure[Attribution score on code2seq]{
      \includegraphics[width=0.37\columnwidth]{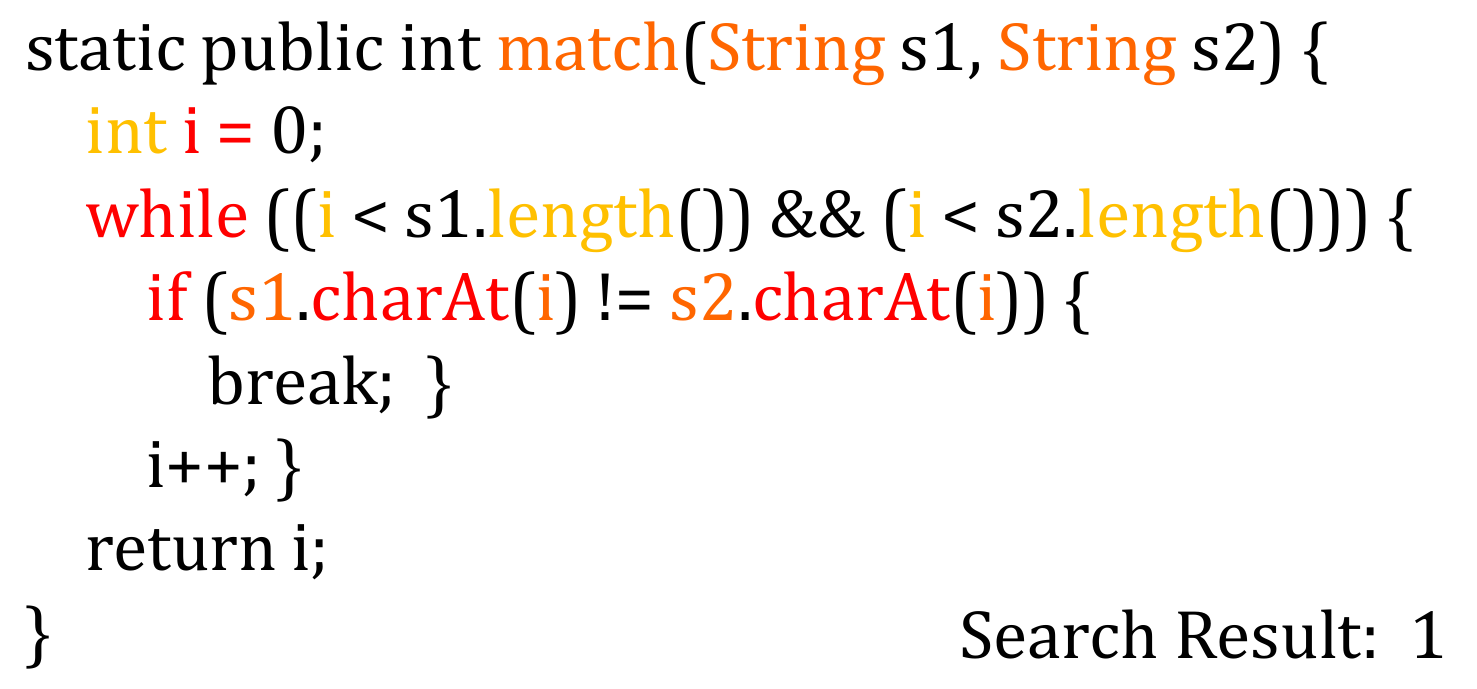}
      \label{fig:se-339813-code2seq}
    } \\
    \subfigure[Attribution score on GGNN]{
      \includegraphics[width=0.37\columnwidth]{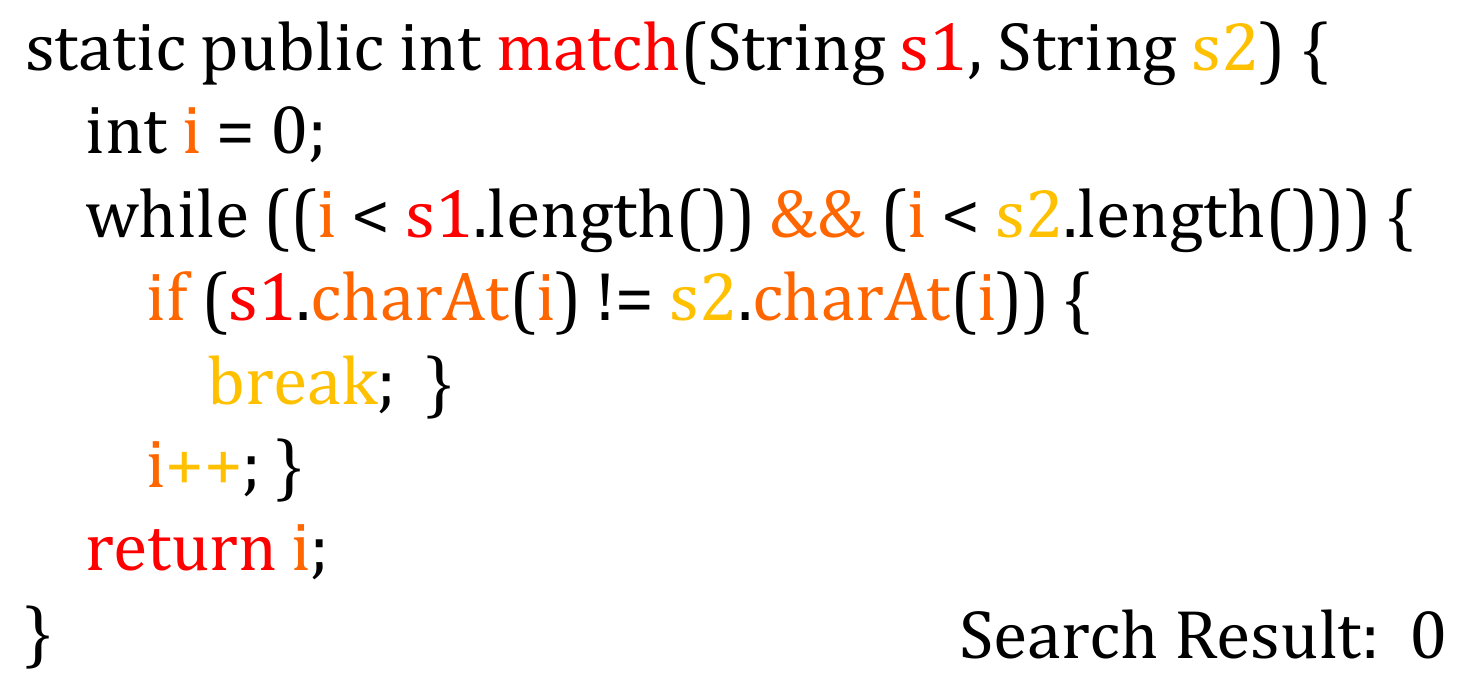}
      \label{fig:se-339813-GGNN}
    }
    \subfigure[Attribution score on ASTNN]{
      \includegraphics[width=0.37\columnwidth]{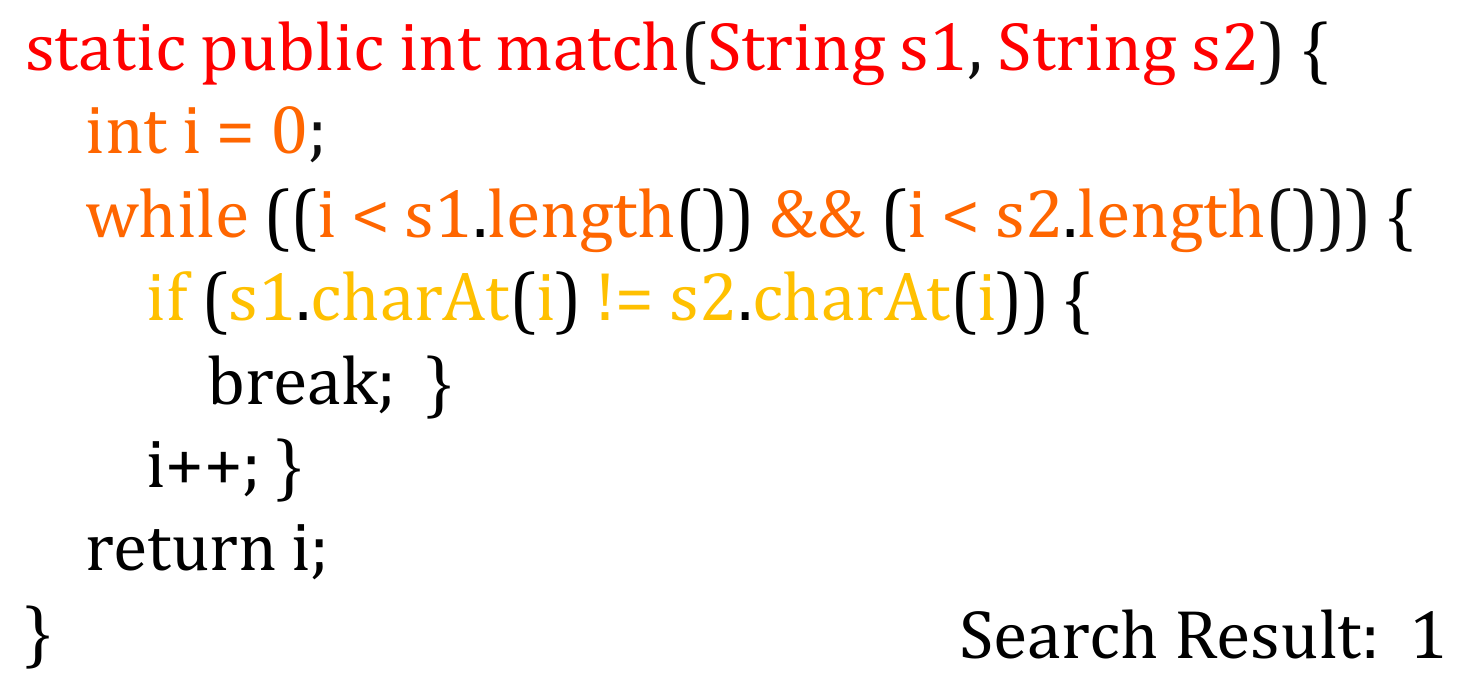}
      \label{fig:se-339813-ASTNN}
    }\\
	\caption{The example of code search. The query is ``count number of chars that match in two strings starting from front''.}
	\label{fig:se-attr} 
\end{figure*}

To visualize the prediction attribution for code search, we pick a pair of query and program and run the model to output a similarity score between them. Then we replace the program with a baseline program of all-zero vectors and obtain another similarity between the query and the baseline. The difference of the two outputs is thereby the difference of the similarity scores. We attribute the difference to the input tokens in the original program and visualize the results. The selected natural language query is ``count number of chars that match in two strings starting from front''. We may observe that the corresponding program returns the length of the longest common substring in the two strings from the beginning.

The attribution results are shown in Figure~\ref{fig:se-attr}. A search result of 1 indicates the similarity score between the program and the query is among the top-10 of the candidate program list (i.e., a successful search), and a search result of 0 indicates otherwise. LSTM, Transformer, code2seq and ASTNN make the correct prediction. We believe this is because they capture the important tokens in line 1, 3 and 4, such as ``match'', ``string'', ``length'' and ``char''. These tokens and the corresponding statements are not only relevant to the functionality of the program, but also highly correlated with the tokens in the natural language query. The other four models that fail to search the program have not attributed the similarity score to all of the important tokens. For example, all the four models fail to capture the token ``length''. For TBCNN, we notice that it gives most credits to the internal nodes, therefore most tokens in the source code (corresponding to the leaf nodes) are not highlighted. 

In short, for code search, both the program tokens correlated with the query tokens and the semantic property of the program are vital for a successful search. Therefore an embedding model that extracts fine-grained token features might be preferred.

\subsubsection{\textbf{Systemic Visualization}}
Like most explanation techniques, the integrated gradients method is tailored towards explaining black-box models using individual data points, i.e., the individual programs in this study. For better understanding the overall performance, we need to know whether the findings on a single program can be applied to most other programs. As such, we try to systemically visualize the attribution results by aggregating the attribution scores in individual programs.

\begin{figure}[h]
    \centering
    \includegraphics[width=0.8\columnwidth]{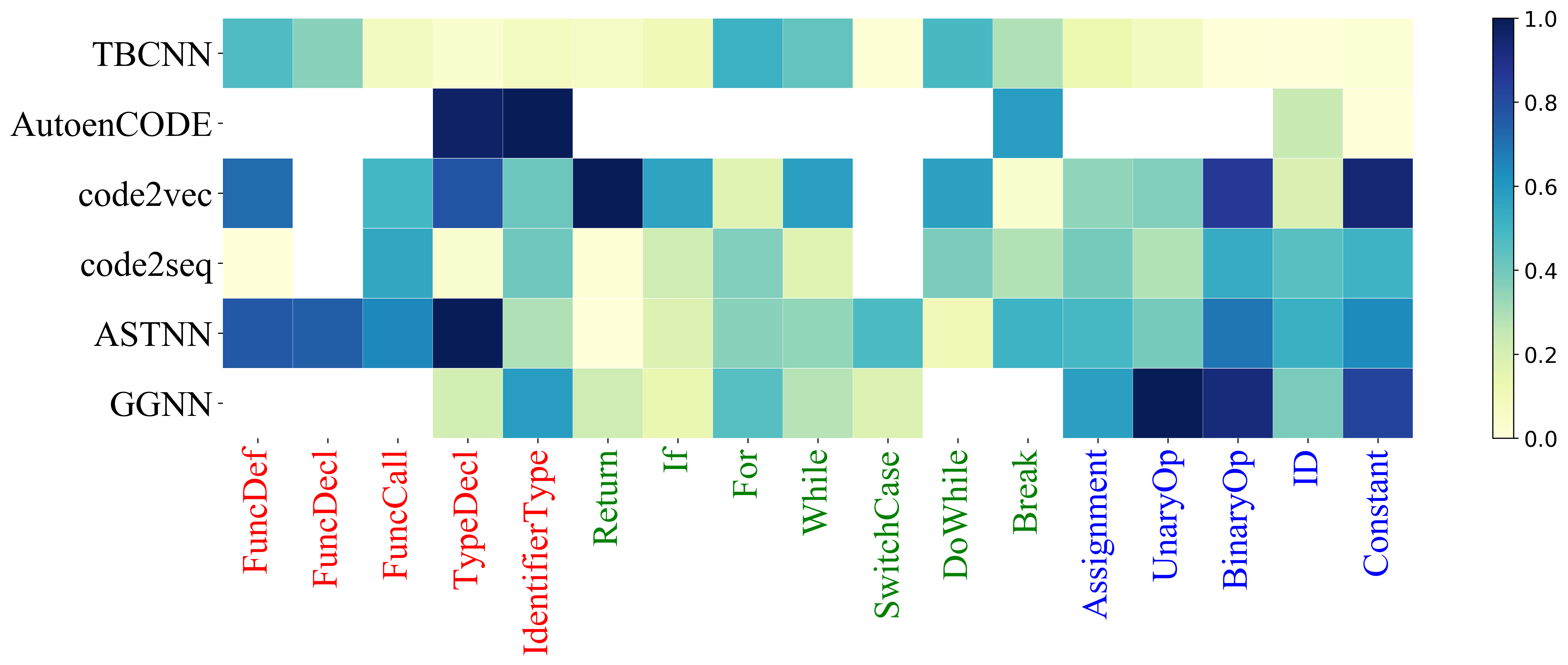}
	\caption{Overall attribution scores of token types in category-56 for code classification.}
	\label{fig:cla-type} 
\end{figure}

\begin{figure}[h]
    \centering
    \includegraphics[width=0.8\columnwidth]{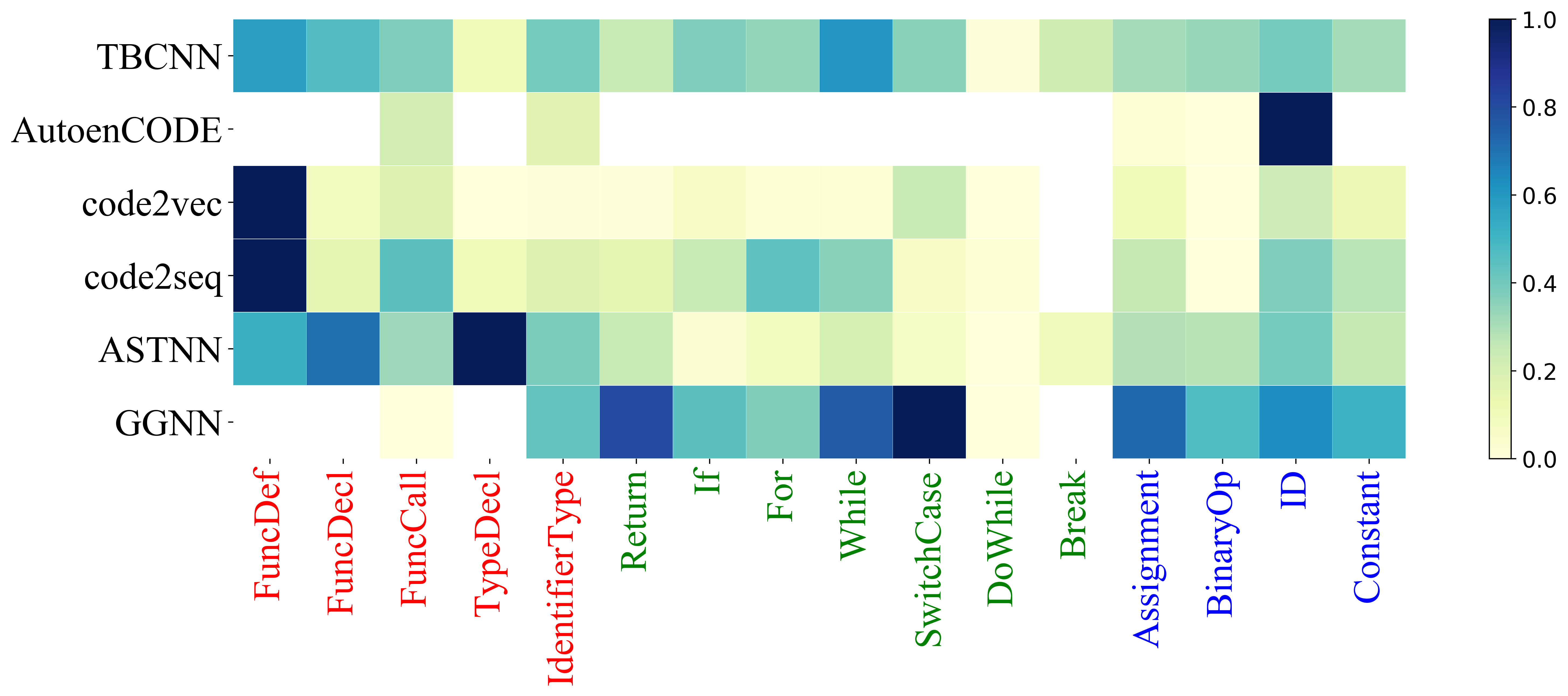}
	\caption{Overall attribution scores of token types for code clone detection.}
	\label{fig:clo-type} 
\end{figure}

\begin{figure}[h]
    \centering
    \includegraphics[width=0.8\columnwidth]{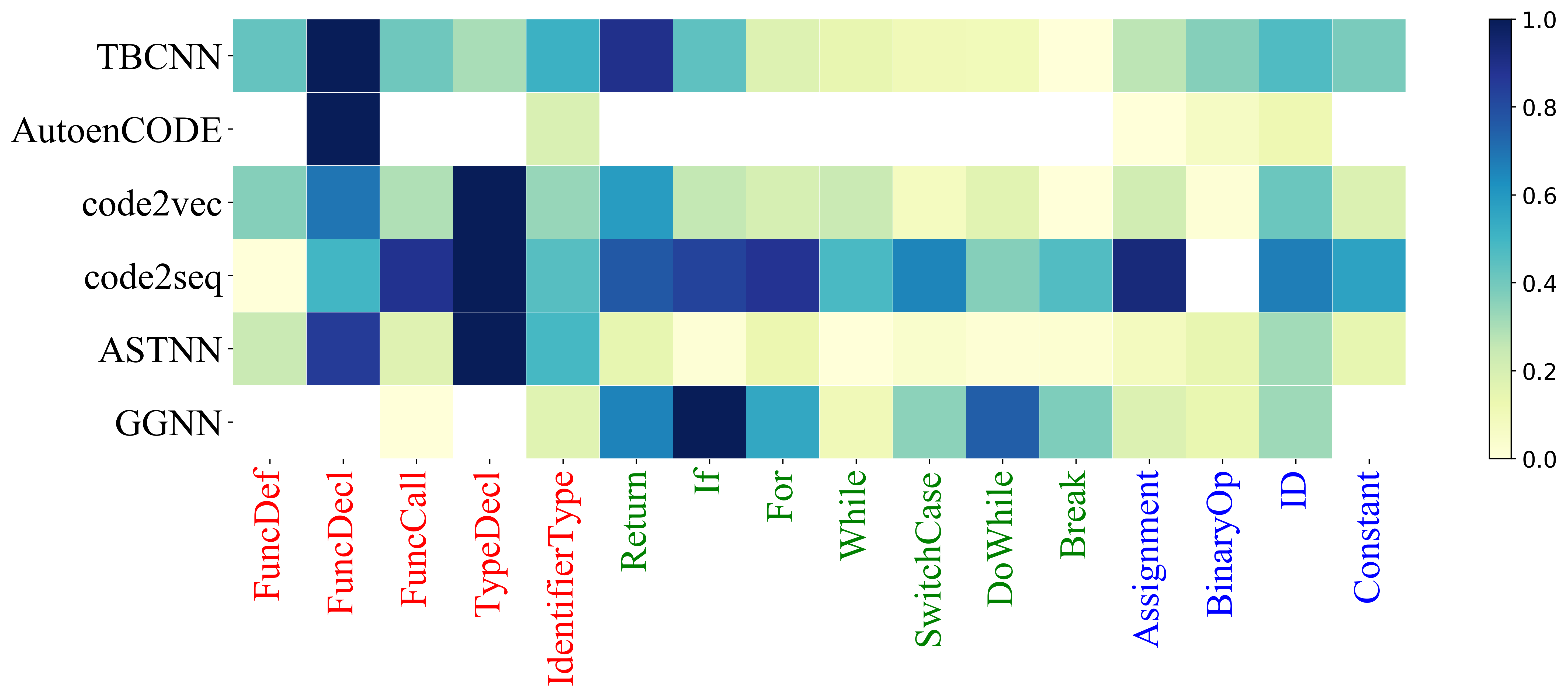}
	\caption{Overall attribution scores of token types for code search.}
	\label{fig:se-type} 
\end{figure}

It is not possible to directly aggregate the attribution scores of the elements in programs, since different programs use different data types, variable names, functions and other elements. Fortunately, each program element is encapsulated in an AST node and associated with a node type. For example, the types of \textsc{IdentifierType}, \textsc{While}, \textsc{ID} represent the type of an identifier, a \texttt{while} statement and a variable, respectively. We thus take advantage of the AST node types and visualize the average attribution scores (normalized) of the elements of each node type. The results are plotted in Figure~\ref{fig:cla-type}, \ref{fig:clo-type} and~\ref{fig:se-type} for the three tasks, where a deeper color indicates a higher score. Note that we exclude LSTM and Transformer, since they do not deal with ASTs. We roughly divide all types into three categories and pick frequent types in each category for visualization. In the figures, the types from \textsc{FuncDef} to \textsc{Assignment} are related to the elements during initialization, the types from \textsc{Return} to \textsc{Break} are related to the control statements, and the remaining types are related to the elements in the expressions. The \textsc{UnaryOp} type in C language (Figure~\ref{fig:cla-type}) is merged into the \textsc{ID} type in Java (Figure~\ref{fig:clo-type} and \ref{fig:se-type}).

For code classification, we pick all programs in category 56 for clear visualization. All programs perform the same function as the code in Figure~\ref{fig:cla-attr}. We observe in Figure~\ref{fig:cla-type} that the two best models GGNN and ASTNN give high scores to both the elements in the control statements and the elements in the expressions, which coincides with the observations in the example code of Figure~\ref{fig:cla-attr}. TBCNN, code2vec and code2seq also focus on either the control statements or the expressions, and thus have moderate performance in accordance with the results in Table~\ref{tab:cla-res}. AutoenCODE gives the least scores to these elements and hence has the worst performance among the models.

For code clone, in Figure~\ref{fig:clo-attr} we observe the important elements are the tokens pertaining to the descriptions and actual operations of functionalities. These elements often fall into the initialization and expression categories. We observe in Figure~\ref{fig:clo-type} that all models except AutoenCODE give high scores to the elements in these two categories, which is also in accordance with the performance reported in Table~\ref{tab:clo-res}.

For code search, in Figure~\ref{fig:se-attr} we observe the important elements are the source code tokens correlated with the query tokens and those relevant to the corresponding functionality. These elements are usually the function names (e.g., ``match'', ``charAt'', ``length''), data types (e.g., ``string'') and variables, which belong to the initialization and expression categories. We observe in Figure~\ref{fig:se-type} that code2seq gives the highest scores to the two categories. As such it has the best performance on the task, as reported in Table~\ref{tab:se-res}. Code2vec, ASTNN and GGNN focus less on these elements and hence perform worse than code2seq.

\section{Discussion}\label{sec:disscusion}
We discuss three key points learned from the experimental study. 

First, \textbf{no single existing AST-based model can beat the simple token-based models in all tasks.} This may indicate that the plain text of source code has revealed strong naturalness of programming language so that applying classic NLP models directly on source code could already bring pretty good performance on most tasks. Indeed, recently a programming language model \emph{CodeBert}~\cite{feng2020codebert} was trained with the plain text of source code (and natural language sentences) and showed very good performance on many tasks. The high textual similarity in the current benchmark datasets may also facilitate the training of the token-based models. Nevertheless, for a particular task, we can always derive an AST-based model that performs better than the token-based models. This can be verified by the performance of ASTNN and GGNN for code classification, TBCNN and ASTNN for code clone detection, and code2seq for code search. Therefore one may always try to explore the abundant structural information in ASTs to achieve the best performance for a specific task.

Second, \textbf{the original structure of an AST and the flow of control and data in the source code are both important features for an AST-based model to achieve good performance.} Because they capture both the information, ASTNN and GGNN have the overall best performance on the three tasks among the AST-based models. ASTNN preserves the structure of each statement subtree and feed the subtrees into the model in the order complying with the statement order in the source code. GGNN augments the original AST's structure with semantic edges that represent the control and data flow in the source code. In contrast, AutoenCODE ignores the positions of the leaf nodes in the AST and uses a greedy algorithm to aggregate the node information, which destroys the structure of the original AST. Therefore it has the overall worst performance. TBCNN, code2vec and code2seq extract only the structural information from the ASTs, and therefore have moderate overall performance.

Finally, \textbf{the prediction attribution study tells that despite the difference in the ways of manipulating the AST structure, the models with good performance all capture the key tokens to a task in the source code.} These key tokens can be the reserved words for control statements (e.g., for, if), the key variables and operators in the expressions, or the tokens that are correlated with the words in natural language. For the token-based models, all the tokens are presented in the plain text so that the models could naturally learn from the tokens. However, tokens like reserved words and function calls are often organized as the internal nodes in an AST, therefore the AST-based models may miss such information (e.g., AutoenCODE and code2vec) and result in sub-optimal performance. As such, we suggest to always fully utilize the information of the internal nodes when designing an AST-based embedding model.

\section{Threats to Validity} \label{sec:threat}
\textbf{1) Selection of datasets.}
The experimental results are related to the datasets used. A different distribution of data points or a different type of programming language of the code may bring different results for the same task. As such the conclusion drawn from the datasets experimented may not be able to generalize beyond. To mitigate the bias introduced by the datasets, we adopt widely used public datasets for all tasks, as they are recognized as the benchmark of the corresponding task. Furthermore, we evaluate on more than one dataset for a task if all the datasets are commonly used. For example, in addition to \textbf{BigCloneBench} in the main article, we also report the results on the \textbf{OJClone} dataset in the supplementary material. The similar overall results may validate the conclusions for the clone detection task.

\textbf{2) Network architecture for downstream tasks.}
The performance of a model on each task is affected by both the program embedding and the network architecture connecting the embedding and the output for the downstream task. Therefore we may not be able to completely attribute the experimental results to the program embeddings. To mitigate the influence of the network architecture, we use the structure as simple as possible for each selected task. For code classification and code clone, we only add one fully-connected layer on top of the program embedding and the difference of two embeddings, respectively. For code search, we directly calculate the cosine distance between positive and negative pairs of program embeddings, respectively. These simple strategies are also employed by the representative models for each task~\cite{mou2016convolutional,zhang2019novel,gu2018deep,sun2020pscs,husain2019codesearchnet}.

\textbf{3) Explaining with individual examples.}
Most explanation techniques including the integrated gradients method are tailored towards explaining black-box models using individual data points~\cite{serrano2019attention,wiegreffe2019attention,jacovi2020towards,baehrens2010explain,sundararajan2017axiomatic}. The findings on individual programs may not be convincing for representing the whole picture. To mitigate this problem, we take the advantage of the AST node types and systemically visualize the attribution results by aggregating the scores of input tokens of the same node type. The results are largely in accordance with the attribution findings in individual programs as well as the model performance reported in Section~\ref{sec:experiments}.

\section{Conclusions}\label{sec:conclusion}
In this experience paper, we systemically evaluate eight program embedding models on three common programming language tasks. Six of the eight models are based on abstract syntax trees, which are the focus of this paper, and the other two models are based on the plain text of source code, which are used as baseline models.

The results of performance evaluation have revealed that while the token-based models are robust to different tasks, a carefully-designed AST-based model may still achieve the best performance given a specific task. This further indicates that both the naturalness of a programming language such as data and control flow, and the structural information in the ASTs are important information for learning a good program embedding.

The explanatory study tells that whether capturing the key tokens in the source code highly influences the correctness of prediction. Given that many tokens may correspond to the internal nodes in an AST, we suggest that an AST-based model should fully make use of the internal nodes.

\bibliographystyle{ACM-Reference-Format}
\bibliography{reference}


\begin{thebibliography}{76}


\ifx \showCODEN    \undefined \def \showCODEN     #1{\unskip}     \fi
\ifx \showDOI      \undefined \def \showDOI       #1{#1}\fi
\ifx \showISBNx    \undefined \def \showISBNx     #1{\unskip}     \fi
\ifx \showISBNxiii \undefined \def \showISBNxiii  #1{\unskip}     \fi
\ifx \showISSN     \undefined \def \showISSN      #1{\unskip}     \fi
\ifx \showLCCN     \undefined \def \showLCCN      #1{\unskip}     \fi
\ifx \shownote     \undefined \def \shownote      #1{#1}          \fi
\ifx \showarticletitle \undefined \def \showarticletitle #1{#1}   \fi
\ifx \showURL      \undefined \def \showURL       {\relax}        \fi
\providecommand\bibfield[2]{#2}
\providecommand\bibinfo[2]{#2}
\providecommand\natexlab[1]{#1}
\providecommand\showeprint[2][]{arXiv:#2}

\bibitem[\protect\citeauthoryear{Ahmad, Chakraborty, Ray, and Chang}{Ahmad
  et~al\mbox{.}}{2020}]%
        {ahmad2020transformer}
\bibfield{author}{\bibinfo{person}{Wasi~Uddin Ahmad}, \bibinfo{person}{Saikat
  Chakraborty}, \bibinfo{person}{Baishakhi Ray}, {and} \bibinfo{person}{Kai-Wei
  Chang}.} \bibinfo{year}{2020}\natexlab{}.
\newblock \showarticletitle{A transformer-based approach for source code
  summarization}.
\newblock \bibinfo{journal}{\emph{arXiv preprint arXiv:2005.00653}}
  (\bibinfo{year}{2020}).
\newblock


\bibitem[\protect\citeauthoryear{Allamanis, Barr, Devanbu, and
  Sutton}{Allamanis et~al\mbox{.}}{2018a}]%
        {allamanis2018survey}
\bibfield{author}{\bibinfo{person}{Miltiadis Allamanis},
  \bibinfo{person}{Earl~T Barr}, \bibinfo{person}{Premkumar Devanbu}, {and}
  \bibinfo{person}{Charles Sutton}.} \bibinfo{year}{2018}\natexlab{a}.
\newblock \showarticletitle{A survey of machine learning for big code and
  naturalness}.
\newblock \bibinfo{journal}{\emph{ACM Computing Surveys (CSUR)}}
  \bibinfo{volume}{51}, \bibinfo{number}{4} (\bibinfo{year}{2018}),
  \bibinfo{pages}{1--37}.
\newblock


\bibitem[\protect\citeauthoryear{Allamanis, Brockschmidt, and
  Khademi}{Allamanis et~al\mbox{.}}{2018b}]%
        {allamanis2018learning}
\bibfield{author}{\bibinfo{person}{Miltiadis Allamanis}, \bibinfo{person}{Marc
  Brockschmidt}, {and} \bibinfo{person}{Mahmoud Khademi}.}
  \bibinfo{year}{2018}\natexlab{b}.
\newblock \showarticletitle{Learning to Represent Programs with Graphs}. In
  \bibinfo{booktitle}{\emph{International Conference on Learning
  Representations}}.
\newblock


\bibitem[\protect\citeauthoryear{Alon, Brody, Levy, and Yahav}{Alon
  et~al\mbox{.}}{2018}]%
        {alon2018code2seq}
\bibfield{author}{\bibinfo{person}{Uri Alon}, \bibinfo{person}{Shaked Brody},
  \bibinfo{person}{Omer Levy}, {and} \bibinfo{person}{Eran Yahav}.}
  \bibinfo{year}{2018}\natexlab{}.
\newblock \showarticletitle{code2seq: Generating Sequences from Structured
  Representations of Code}. In \bibinfo{booktitle}{\emph{International
  Conference on Learning Representations}}.
\newblock


\bibitem[\protect\citeauthoryear{Alon, Sadaka, Levy, and Yahav}{Alon
  et~al\mbox{.}}{2020}]%
        {alon2020structural}
\bibfield{author}{\bibinfo{person}{Uri Alon}, \bibinfo{person}{Roy Sadaka},
  \bibinfo{person}{Omer Levy}, {and} \bibinfo{person}{Eran Yahav}.}
  \bibinfo{year}{2020}\natexlab{}.
\newblock \showarticletitle{Structural language models of code}. In
  \bibinfo{booktitle}{\emph{International Conference on Machine Learning}}.
  PMLR, \bibinfo{pages}{245--256}.
\newblock


\bibitem[\protect\citeauthoryear{Alon, Zilberstein, Levy, and Yahav}{Alon
  et~al\mbox{.}}{2019}]%
        {alon2019code2vec}
\bibfield{author}{\bibinfo{person}{Uri Alon}, \bibinfo{person}{Meital
  Zilberstein}, \bibinfo{person}{Omer Levy}, {and} \bibinfo{person}{Eran
  Yahav}.} \bibinfo{year}{2019}\natexlab{}.
\newblock \showarticletitle{code2vec: Learning distributed representations of
  code}.
\newblock \bibinfo{journal}{\emph{Proceedings of the ACM on Programming
  Languages}} \bibinfo{volume}{3}, \bibinfo{number}{POPL}
  (\bibinfo{year}{2019}), \bibinfo{pages}{1--29}.
\newblock


\bibitem[\protect\citeauthoryear{Baehrens, Schroeter, Harmeling, Kawanabe,
  Hansen, and M{\"u}ller}{Baehrens et~al\mbox{.}}{2010}]%
        {baehrens2010explain}
\bibfield{author}{\bibinfo{person}{David Baehrens}, \bibinfo{person}{Timon
  Schroeter}, \bibinfo{person}{Stefan Harmeling}, \bibinfo{person}{Motoaki
  Kawanabe}, \bibinfo{person}{Katja Hansen}, {and}
  \bibinfo{person}{Klaus-Robert M{\"u}ller}.} \bibinfo{year}{2010}\natexlab{}.
\newblock \showarticletitle{How to explain individual classification
  decisions}.
\newblock \bibinfo{journal}{\emph{The Journal of Machine Learning Research}}
  \bibinfo{volume}{11} (\bibinfo{year}{2010}), \bibinfo{pages}{1803--1831}.
\newblock


\bibitem[\protect\citeauthoryear{Baxter, Yahin, Moura, Sant'Anna, and
  Bier}{Baxter et~al\mbox{.}}{1998}]%
        {baxter1998clone}
\bibfield{author}{\bibinfo{person}{Ira~D Baxter}, \bibinfo{person}{Andrew
  Yahin}, \bibinfo{person}{Leonardo Moura}, \bibinfo{person}{Marcelo
  Sant'Anna}, {and} \bibinfo{person}{Lorraine Bier}.}
  \bibinfo{year}{1998}\natexlab{}.
\newblock \showarticletitle{Clone detection using abstract syntax trees}. In
  \bibinfo{booktitle}{\emph{Proceedings. International Conference on Software
  Maintenance (Cat. No. 98CB36272)}}. IEEE, \bibinfo{pages}{368--377}.
\newblock


\bibitem[\protect\citeauthoryear{Ben-Nun, Jakobovits, and Hoefler}{Ben-Nun
  et~al\mbox{.}}{2018}]%
        {ben2018neural}
\bibfield{author}{\bibinfo{person}{Tal Ben-Nun},
  \bibinfo{person}{Alice~Shoshana Jakobovits}, {and} \bibinfo{person}{Torsten
  Hoefler}.} \bibinfo{year}{2018}\natexlab{}.
\newblock \showarticletitle{Neural Code Comprehension: A Learnable
  Representation of Code Semantics}.
\newblock \bibinfo{journal}{\emph{Advances in Neural Information Processing
  Systems}}  \bibinfo{volume}{31} (\bibinfo{year}{2018}),
  \bibinfo{pages}{3585--3597}.
\newblock


\bibitem[\protect\citeauthoryear{Brody, Alon, and Yahav}{Brody
  et~al\mbox{.}}{2020}]%
        {brody2020structural}
\bibfield{author}{\bibinfo{person}{Shaked Brody}, \bibinfo{person}{Uri Alon},
  {and} \bibinfo{person}{Eran Yahav}.} \bibinfo{year}{2020}\natexlab{}.
\newblock \showarticletitle{A structural model for contextual code changes}.
\newblock \bibinfo{journal}{\emph{Proceedings of the ACM on Programming
  Languages}} \bibinfo{volume}{4}, \bibinfo{number}{OOPSLA}
  (\bibinfo{year}{2020}), \bibinfo{pages}{1--28}.
\newblock


\bibitem[\protect\citeauthoryear{Bui, Jiang, and Yu}{Bui et~al\mbox{.}}{2018}]%
        {bui2018cross}
\bibfield{author}{\bibinfo{person}{Nghi~DQ Bui}, \bibinfo{person}{Lingxiao
  Jiang}, {and} \bibinfo{person}{Yijun Yu}.} \bibinfo{year}{2018}\natexlab{}.
\newblock \showarticletitle{Cross-Language Learning for Program Classification
  using Bilateral Tree-Based Convolutional Neural Networks}. In
  \bibinfo{booktitle}{\emph{The Thirty-Second AAAI Conference on Artificial
  Intelligence (AAAI-18)}}.
\newblock


\bibitem[\protect\citeauthoryear{Bui, Yu, and Jiang}{Bui et~al\mbox{.}}{2021}]%
        {bui2021infercode}
\bibfield{author}{\bibinfo{person}{Nghi~DQ Bui}, \bibinfo{person}{Yijun Yu},
  {and} \bibinfo{person}{Lingxiao Jiang}.} \bibinfo{year}{2021}\natexlab{}.
\newblock \showarticletitle{InferCode: Self-Supervised Learning of Code
  Representations by Predicting Subtrees}. In \bibinfo{booktitle}{\emph{2021
  IEEE/ACM 43rd International Conference on Software Engineering (ICSE)}}.
  IEEE, \bibinfo{pages}{1186--1197}.
\newblock


\bibitem[\protect\citeauthoryear{Cambronero, Li, Kim, Sen, and
  Chandra}{Cambronero et~al\mbox{.}}{2019}]%
        {cambronero2019deep}
\bibfield{author}{\bibinfo{person}{Jose Cambronero}, \bibinfo{person}{Hongyu
  Li}, \bibinfo{person}{Seohyun Kim}, \bibinfo{person}{Koushik Sen}, {and}
  \bibinfo{person}{Satish Chandra}.} \bibinfo{year}{2019}\natexlab{}.
\newblock \showarticletitle{When deep learning met code search}. In
  \bibinfo{booktitle}{\emph{Proceedings of the 2019 27th ACM Joint Meeting on
  European Software Engineering Conference and Symposium on the Foundations of
  Software Engineering}}. \bibinfo{pages}{964--974}.
\newblock


\bibitem[\protect\citeauthoryear{Chen, Hu, Liu, Xiao, and Jiang}{Chen
  et~al\mbox{.}}{2019}]%
        {chen2019deep}
\bibfield{author}{\bibinfo{person}{Jindong Chen}, \bibinfo{person}{Yizhou Hu},
  \bibinfo{person}{Jingping Liu}, \bibinfo{person}{Yanghua Xiao}, {and}
  \bibinfo{person}{Haiyun Jiang}.} \bibinfo{year}{2019}\natexlab{}.
\newblock \showarticletitle{Deep short text classification with knowledge
  powered attention}. In \bibinfo{booktitle}{\emph{Proceedings of the AAAI
  Conference on Artificial Intelligence}}, Vol.~\bibinfo{volume}{33}.
  \bibinfo{pages}{6252--6259}.
\newblock


\bibitem[\protect\citeauthoryear{Chen and Monperrus}{Chen and
  Monperrus}{2019}]%
        {chen2019literature}
\bibfield{author}{\bibinfo{person}{Zimin Chen} {and} \bibinfo{person}{Martin
  Monperrus}.} \bibinfo{year}{2019}\natexlab{}.
\newblock \showarticletitle{A literature study of embeddings on source code}.
\newblock \bibinfo{journal}{\emph{arXiv preprint arXiv:1904.03061}}
  (\bibinfo{year}{2019}).
\newblock


\bibitem[\protect\citeauthoryear{Chihada, Jalili, Hasheminejad, and
  Zangooei}{Chihada et~al\mbox{.}}{2015}]%
        {chihada2015source}
\bibfield{author}{\bibinfo{person}{Abdullah Chihada}, \bibinfo{person}{Saeed
  Jalili}, \bibinfo{person}{Seyed Mohammad~Hossein Hasheminejad}, {and}
  \bibinfo{person}{Mohammad~Hossein Zangooei}.}
  \bibinfo{year}{2015}\natexlab{}.
\newblock \showarticletitle{Source code and design conformance, design pattern
  detection from source code by classification approach}.
\newblock \bibinfo{journal}{\emph{Applied Soft Computing}}
  \bibinfo{volume}{26} (\bibinfo{year}{2015}), \bibinfo{pages}{357--367}.
\newblock


\bibitem[\protect\citeauthoryear{Clark, Khandelwal, Levy, and Manning}{Clark
  et~al\mbox{.}}{2019}]%
        {clark2019does}
\bibfield{author}{\bibinfo{person}{Kevin Clark}, \bibinfo{person}{Urvashi
  Khandelwal}, \bibinfo{person}{Omer Levy}, {and}
  \bibinfo{person}{Christopher~D Manning}.} \bibinfo{year}{2019}\natexlab{}.
\newblock \showarticletitle{What does BERT look at? An Analysis of BERT’s
  Attention}.
\newblock \bibinfo{journal}{\emph{ACL 2019}} (\bibinfo{year}{2019}),
  \bibinfo{pages}{276}.
\newblock


\bibitem[\protect\citeauthoryear{Dam, Tran, and Pham}{Dam
  et~al\mbox{.}}{2016}]%
        {dam2016deep}
\bibfield{author}{\bibinfo{person}{Hoa~Khanh Dam}, \bibinfo{person}{Truyen
  Tran}, {and} \bibinfo{person}{Trang Pham}.} \bibinfo{year}{2016}\natexlab{}.
\newblock \showarticletitle{A deep language model for software code}.
\newblock \bibinfo{journal}{\emph{arXiv preprint arXiv:1608.02715}}
  (\bibinfo{year}{2016}).
\newblock


\bibitem[\protect\citeauthoryear{Du, Liu, Yang, Ji, and Hu}{Du
  et~al\mbox{.}}{2019}]%
        {du2019attribution}
\bibfield{author}{\bibinfo{person}{Mengnan Du}, \bibinfo{person}{Ninghao Liu},
  \bibinfo{person}{Fan Yang}, \bibinfo{person}{Shuiwang Ji}, {and}
  \bibinfo{person}{Xia Hu}.} \bibinfo{year}{2019}\natexlab{}.
\newblock \showarticletitle{On attribution of recurrent neural network
  predictions via additive decomposition}. In \bibinfo{booktitle}{\emph{The
  World Wide Web Conference}}. \bibinfo{pages}{383--393}.
\newblock


\bibitem[\protect\citeauthoryear{Feng, Guo, Tang, Duan, Feng, Gong, Shou, Qin,
  Liu, Jiang, et~al\mbox{.}}{Feng et~al\mbox{.}}{2020}]%
        {feng2020codebert}
\bibfield{author}{\bibinfo{person}{Zhangyin Feng}, \bibinfo{person}{Daya Guo},
  \bibinfo{person}{Duyu Tang}, \bibinfo{person}{Nan Duan},
  \bibinfo{person}{Xiaocheng Feng}, \bibinfo{person}{Ming Gong},
  \bibinfo{person}{Linjun Shou}, \bibinfo{person}{Bing Qin},
  \bibinfo{person}{Ting Liu}, \bibinfo{person}{Daxin Jiang}, {et~al\mbox{.}}}
  \bibinfo{year}{2020}\natexlab{}.
\newblock \showarticletitle{Codebert: A pre-trained model for programming and
  natural languages}.
\newblock \bibinfo{journal}{\emph{arXiv preprint arXiv:2002.08155}}
  (\bibinfo{year}{2020}).
\newblock


\bibitem[\protect\citeauthoryear{Frantzeskou, MacDonell, Stamatatos, and
  Gritzalis}{Frantzeskou et~al\mbox{.}}{2008}]%
        {frantzeskou2008examining}
\bibfield{author}{\bibinfo{person}{Georgia Frantzeskou},
  \bibinfo{person}{Stephen MacDonell}, \bibinfo{person}{Efstathios Stamatatos},
  {and} \bibinfo{person}{Stefanos Gritzalis}.} \bibinfo{year}{2008}\natexlab{}.
\newblock \showarticletitle{Examining the significance of high-level
  programming features in source code author classification}.
\newblock \bibinfo{journal}{\emph{Journal of Systems and Software}}
  \bibinfo{volume}{81}, \bibinfo{number}{3} (\bibinfo{year}{2008}),
  \bibinfo{pages}{447--460}.
\newblock


\bibitem[\protect\citeauthoryear{Gu, Zhang, and Kim}{Gu et~al\mbox{.}}{2018}]%
        {gu2018deep}
\bibfield{author}{\bibinfo{person}{Xiaodong Gu}, \bibinfo{person}{Hongyu
  Zhang}, {and} \bibinfo{person}{Sunghun Kim}.}
  \bibinfo{year}{2018}\natexlab{}.
\newblock \showarticletitle{Deep code search}. In
  \bibinfo{booktitle}{\emph{2018 IEEE/ACM 40th International Conference on
  Software Engineering (ICSE)}}. IEEE, \bibinfo{pages}{933--944}.
\newblock


\bibitem[\protect\citeauthoryear{Gupta, Kanade, and Shevade}{Gupta
  et~al\mbox{.}}{2019}]%
        {gupta2019neural}
\bibfield{author}{\bibinfo{person}{Rahul Gupta}, \bibinfo{person}{Aditya
  Kanade}, {and} \bibinfo{person}{Shirish Shevade}.}
  \bibinfo{year}{2019}\natexlab{}.
\newblock \showarticletitle{Neural attribution for semantic bug-localization in
  student programs}.
\newblock \bibinfo{journal}{\emph{Network}} \bibinfo{volume}{1},
  \bibinfo{number}{P2} (\bibinfo{year}{2019}), \bibinfo{pages}{P2}.
\newblock


\bibitem[\protect\citeauthoryear{Gupta, Pal, Kanade, and Shevade}{Gupta
  et~al\mbox{.}}{2017}]%
        {gupta2017deepfix}
\bibfield{author}{\bibinfo{person}{Rahul Gupta}, \bibinfo{person}{Soham Pal},
  \bibinfo{person}{Aditya Kanade}, {and} \bibinfo{person}{Shirish Shevade}.}
  \bibinfo{year}{2017}\natexlab{}.
\newblock \showarticletitle{Deepfix: Fixing common c language errors by deep
  learning}. In \bibinfo{booktitle}{\emph{Proceedings of the aaai conference on
  artificial intelligence}}, Vol.~\bibinfo{volume}{31}.
\newblock


\bibitem[\protect\citeauthoryear{He, Tu, Wang, Wang, Lyu, and Shi}{He
  et~al\mbox{.}}{2019}]%
        {he2019towards}
\bibfield{author}{\bibinfo{person}{Shilin He}, \bibinfo{person}{Zhaopeng Tu},
  \bibinfo{person}{Xing Wang}, \bibinfo{person}{Longyue Wang},
  \bibinfo{person}{Michael Lyu}, {and} \bibinfo{person}{Shuming Shi}.}
  \bibinfo{year}{2019}\natexlab{}.
\newblock \showarticletitle{Towards Understanding Neural Machine Translation
  with Word Importance}. In \bibinfo{booktitle}{\emph{Proceedings of the 2019
  Conference on Empirical Methods in Natural Language Processing and the 9th
  International Joint Conference on Natural Language Processing
  (EMNLP-IJCNLP)}}. \bibinfo{pages}{952--961}.
\newblock


\bibitem[\protect\citeauthoryear{Huo, Li, and Zhou}{Huo et~al\mbox{.}}{2020}]%
        {huo2020control}
\bibfield{author}{\bibinfo{person}{Xuan Huo}, \bibinfo{person}{Ming Li}, {and}
  \bibinfo{person}{Zhi-Hua Zhou}.} \bibinfo{year}{2020}\natexlab{}.
\newblock \showarticletitle{Control Flow Graph Embedding Based on
  Multi-Instance Decomposition for Bug Localization}. In
  \bibinfo{booktitle}{\emph{Proceedings of the AAAI Conference on Artificial
  Intelligence}}, Vol.~\bibinfo{volume}{34}. \bibinfo{pages}{4223--4230}.
\newblock


\bibitem[\protect\citeauthoryear{Husain, Wu, Gazit, Allamanis, and
  Brockschmidt}{Husain et~al\mbox{.}}{2019}]%
        {husain2019codesearchnet}
\bibfield{author}{\bibinfo{person}{Hamel Husain}, \bibinfo{person}{Ho-Hsiang
  Wu}, \bibinfo{person}{Tiferet Gazit}, \bibinfo{person}{Miltiadis Allamanis},
  {and} \bibinfo{person}{Marc Brockschmidt}.} \bibinfo{year}{2019}\natexlab{}.
\newblock \showarticletitle{Codesearchnet challenge: Evaluating the state of
  semantic code search}.
\newblock \bibinfo{journal}{\emph{arXiv preprint arXiv:1909.09436}}
  (\bibinfo{year}{2019}).
\newblock


\bibitem[\protect\citeauthoryear{Jacovi and Goldberg}{Jacovi and
  Goldberg}{2020}]%
        {jacovi2020towards}
\bibfield{author}{\bibinfo{person}{Alon Jacovi} {and} \bibinfo{person}{Yoav
  Goldberg}.} \bibinfo{year}{2020}\natexlab{}.
\newblock \showarticletitle{Towards Faithfully Interpretable NLP Systems: How
  Should We Define and Evaluate Faithfulness?}. In
  \bibinfo{booktitle}{\emph{Proceedings of the 58th Annual Meeting of the
  Association for Computational Linguistics}}. \bibinfo{pages}{4198--4205}.
\newblock


\bibitem[\protect\citeauthoryear{Jatnika, Bijaksana, and Suryani}{Jatnika
  et~al\mbox{.}}{2019}]%
        {jatnika2019word2vec}
\bibfield{author}{\bibinfo{person}{Derry Jatnika}, \bibinfo{person}{Moch~Arif
  Bijaksana}, {and} \bibinfo{person}{Arie~Ardiyanti Suryani}.}
  \bibinfo{year}{2019}\natexlab{}.
\newblock \showarticletitle{Word2vec model analysis for semantic similarities
  in english words}.
\newblock \bibinfo{journal}{\emph{Procedia Computer Science}}
  \bibinfo{volume}{157} (\bibinfo{year}{2019}), \bibinfo{pages}{160--167}.
\newblock


\bibitem[\protect\citeauthoryear{Kamiya, Kusumoto, and Inoue}{Kamiya
  et~al\mbox{.}}{2002}]%
        {kamiya2002ccfinder}
\bibfield{author}{\bibinfo{person}{Toshihiro Kamiya}, \bibinfo{person}{Shinji
  Kusumoto}, {and} \bibinfo{person}{Katsuro Inoue}.}
  \bibinfo{year}{2002}\natexlab{}.
\newblock \showarticletitle{CCFinder: A multilinguistic token-based code clone
  detection system for large scale source code}.
\newblock \bibinfo{journal}{\emph{IEEE Transactions on Software Engineering}}
  \bibinfo{volume}{28}, \bibinfo{number}{7} (\bibinfo{year}{2002}),
  \bibinfo{pages}{654--670}.
\newblock


\bibitem[\protect\citeauthoryear{Kang, Bissyand{\'e}, and Lo}{Kang
  et~al\mbox{.}}{2019}]%
        {kang2019assessing}
\bibfield{author}{\bibinfo{person}{Hong~Jin Kang},
  \bibinfo{person}{Tegawend{\'e}~F Bissyand{\'e}}, {and} \bibinfo{person}{David
  Lo}.} \bibinfo{year}{2019}\natexlab{}.
\newblock \showarticletitle{Assessing the generalizability of code2vec token
  embeddings}. In \bibinfo{booktitle}{\emph{2019 34th IEEE/ACM International
  Conference on Automated Software Engineering (ASE)}}. IEEE,
  \bibinfo{pages}{1--12}.
\newblock


\bibitem[\protect\citeauthoryear{Kawaguchi, Garg, Matsushita, and
  Inoue}{Kawaguchi et~al\mbox{.}}{2006}]%
        {kawaguchi2006mudablue}
\bibfield{author}{\bibinfo{person}{Shinji Kawaguchi}, \bibinfo{person}{Pankaj~K
  Garg}, \bibinfo{person}{Makoto Matsushita}, {and} \bibinfo{person}{Katsuro
  Inoue}.} \bibinfo{year}{2006}\natexlab{}.
\newblock \showarticletitle{Mudablue: An automatic categorization system for
  open source repositories}.
\newblock \bibinfo{journal}{\emph{Journal of Systems and Software}}
  \bibinfo{volume}{79}, \bibinfo{number}{7} (\bibinfo{year}{2006}),
  \bibinfo{pages}{939--953}.
\newblock


\bibitem[\protect\citeauthoryear{Kenter and De~Rijke}{Kenter and
  De~Rijke}{2015}]%
        {kenter2015short}
\bibfield{author}{\bibinfo{person}{Tom Kenter} {and} \bibinfo{person}{Maarten
  De~Rijke}.} \bibinfo{year}{2015}\natexlab{}.
\newblock \showarticletitle{Short text similarity with word embeddings}. In
  \bibinfo{booktitle}{\emph{Proceedings of the 24th ACM international on
  conference on information and knowledge management}}.
  \bibinfo{pages}{1411--1420}.
\newblock


\bibitem[\protect\citeauthoryear{Kovalenko, Bogomolov, Bryksin, and
  Bacchelli}{Kovalenko et~al\mbox{.}}{2019}]%
        {kovalenko2019pathminer}
\bibfield{author}{\bibinfo{person}{Vladimir Kovalenko}, \bibinfo{person}{Egor
  Bogomolov}, \bibinfo{person}{Timofey Bryksin}, {and} \bibinfo{person}{Alberto
  Bacchelli}.} \bibinfo{year}{2019}\natexlab{}.
\newblock \showarticletitle{PathMiner: a library for mining of path-based
  representations of code}. In \bibinfo{booktitle}{\emph{Proceedings of the
  16th International Conference on Mining Software Repositories}}. IEEE Press,
  \bibinfo{pages}{13--17}.
\newblock


\bibitem[\protect\citeauthoryear{Li, Wang, Wang, Yan, Xie, and Mei}{Li
  et~al\mbox{.}}{2016}]%
        {li2016relationship}
\bibfield{author}{\bibinfo{person}{Xuan Li}, \bibinfo{person}{Zerui Wang},
  \bibinfo{person}{Qianxiang Wang}, \bibinfo{person}{Shoumeng Yan},
  \bibinfo{person}{Tao Xie}, {and} \bibinfo{person}{Hong Mei}.}
  \bibinfo{year}{2016}\natexlab{}.
\newblock \showarticletitle{Relationship-aware code search for JavaScript
  frameworks}. In \bibinfo{booktitle}{\emph{Proceedings of the 2016 24th ACM
  SIGSOFT International Symposium on Foundations of Software Engineering}}.
  \bibinfo{pages}{690--701}.
\newblock


\bibitem[\protect\citeauthoryear{Liang, Yu, Jiang, and Xie}{Liang
  et~al\mbox{.}}{2019}]%
        {liang2019seml}
\bibfield{author}{\bibinfo{person}{Hongliang Liang}, \bibinfo{person}{Yue Yu},
  \bibinfo{person}{Lin Jiang}, {and} \bibinfo{person}{Zhuosi Xie}.}
  \bibinfo{year}{2019}\natexlab{}.
\newblock \showarticletitle{Seml: A semantic lstm model for software defect
  prediction}.
\newblock \bibinfo{journal}{\emph{IEEE Access}}  \bibinfo{volume}{7}
  (\bibinfo{year}{2019}), \bibinfo{pages}{83812--83824}.
\newblock


\bibitem[\protect\citeauthoryear{Linares-V{\'a}squez, McMillan, Poshyvanyk, and
  Grechanik}{Linares-V{\'a}squez et~al\mbox{.}}{2014}]%
        {linares2014using}
\bibfield{author}{\bibinfo{person}{Mario Linares-V{\'a}squez},
  \bibinfo{person}{Collin McMillan}, \bibinfo{person}{Denys Poshyvanyk}, {and}
  \bibinfo{person}{Mark Grechanik}.} \bibinfo{year}{2014}\natexlab{}.
\newblock \showarticletitle{On using machine learning to automatically classify
  software applications into domain categories}.
\newblock \bibinfo{journal}{\emph{Empirical Software Engineering}}
  \bibinfo{volume}{19}, \bibinfo{number}{3} (\bibinfo{year}{2014}),
  \bibinfo{pages}{582--618}.
\newblock


\bibitem[\protect\citeauthoryear{Liu, Zhang, and Jin}{Liu
  et~al\mbox{.}}{2020b}]%
        {liu2020modeling}
\bibfield{author}{\bibinfo{person}{Fang Liu}, \bibinfo{person}{Lu Zhang}, {and}
  \bibinfo{person}{Zhi Jin}.} \bibinfo{year}{2020}\natexlab{b}.
\newblock \showarticletitle{Modeling programs hierarchically with
  stack-augmented LSTM}.
\newblock \bibinfo{journal}{\emph{Journal of Systems and Software}}
  \bibinfo{volume}{164} (\bibinfo{year}{2020}), \bibinfo{pages}{110547}.
\newblock


\bibitem[\protect\citeauthoryear{Liu, Kim, Murali, Chaudhuri, and Chandra}{Liu
  et~al\mbox{.}}{2019}]%
        {liu2019neural}
\bibfield{author}{\bibinfo{person}{Jason Liu}, \bibinfo{person}{Seohyun Kim},
  \bibinfo{person}{Vijayaraghavan Murali}, \bibinfo{person}{Swarat Chaudhuri},
  {and} \bibinfo{person}{Satish Chandra}.} \bibinfo{year}{2019}\natexlab{}.
\newblock \showarticletitle{Neural query expansion for code search}. In
  \bibinfo{booktitle}{\emph{Proceedings of the 3rd acm sigplan international
  workshop on machine learning and programming languages}}.
  \bibinfo{pages}{29--37}.
\newblock


\bibitem[\protect\citeauthoryear{Liu, Gao, Chen, Yiu, and Liu}{Liu
  et~al\mbox{.}}{2020a}]%
        {liu2020atom}
\bibfield{author}{\bibinfo{person}{Shangqing Liu}, \bibinfo{person}{Cuiyun
  Gao}, \bibinfo{person}{Sen Chen}, \bibinfo{person}{Nie~Lun Yiu}, {and}
  \bibinfo{person}{Yang Liu}.} \bibinfo{year}{2020}\natexlab{a}.
\newblock \showarticletitle{ATOM: Commit message generation based on abstract
  syntax tree and hybrid ranking}.
\newblock \bibinfo{journal}{\emph{IEEE Transactions on Software Engineering}}
  (\bibinfo{year}{2020}).
\newblock


\bibitem[\protect\citeauthoryear{Lu, Guo, Ren, Huang, Svyatkovskiy, Blanco,
  Clement, Drain, Jiang, Tang, et~al\mbox{.}}{Lu et~al\mbox{.}}{2021}]%
        {lu2021codexglue}
\bibfield{author}{\bibinfo{person}{Shuai Lu}, \bibinfo{person}{Daya Guo},
  \bibinfo{person}{Shuo Ren}, \bibinfo{person}{Junjie Huang},
  \bibinfo{person}{Alexey Svyatkovskiy}, \bibinfo{person}{Ambrosio Blanco},
  \bibinfo{person}{Colin Clement}, \bibinfo{person}{Dawn Drain},
  \bibinfo{person}{Daxin Jiang}, \bibinfo{person}{Duyu Tang}, {et~al\mbox{.}}}
  \bibinfo{year}{2021}\natexlab{}.
\newblock \showarticletitle{CodeXGLUE: A Machine Learning Benchmark Dataset for
  Code Understanding and Generation}.
\newblock \bibinfo{journal}{\emph{arXiv preprint arXiv:2102.04664}}
  (\bibinfo{year}{2021}).
\newblock


\bibitem[\protect\citeauthoryear{Lv, Zhang, Lou, Wang, Zhang, and Zhao}{Lv
  et~al\mbox{.}}{2015}]%
        {lv2015codehow}
\bibfield{author}{\bibinfo{person}{Fei Lv}, \bibinfo{person}{Hongyu Zhang},
  \bibinfo{person}{Jian-guang Lou}, \bibinfo{person}{Shaowei Wang},
  \bibinfo{person}{Dongmei Zhang}, {and} \bibinfo{person}{Jianjun Zhao}.}
  \bibinfo{year}{2015}\natexlab{}.
\newblock \showarticletitle{Codehow: Effective code search based on api
  understanding and extended boolean model (e)}. In
  \bibinfo{booktitle}{\emph{2015 30th IEEE/ACM International Conference on
  Automated Software Engineering (ASE)}}. IEEE, \bibinfo{pages}{260--270}.
\newblock


\bibitem[\protect\citeauthoryear{Mikolov, Chen, Corrado, and Dean}{Mikolov
  et~al\mbox{.}}{2013a}]%
        {mikolov2013efficient}
\bibfield{author}{\bibinfo{person}{Tomas Mikolov}, \bibinfo{person}{Kai Chen},
  \bibinfo{person}{Greg Corrado}, {and} \bibinfo{person}{Jeffrey Dean}.}
  \bibinfo{year}{2013}\natexlab{a}.
\newblock \showarticletitle{Efficient estimation of word representations in
  vector space}.
\newblock \bibinfo{journal}{\emph{arXiv preprint arXiv:1301.3781}}
  (\bibinfo{year}{2013}).
\newblock


\bibitem[\protect\citeauthoryear{Mikolov, Sutskever, Chen, Corrado, and
  Dean}{Mikolov et~al\mbox{.}}{2013b}]%
        {mikolov2013distributed}
\bibfield{author}{\bibinfo{person}{Tomas Mikolov}, \bibinfo{person}{Ilya
  Sutskever}, \bibinfo{person}{Kai Chen}, \bibinfo{person}{Greg~S Corrado},
  {and} \bibinfo{person}{Jeff Dean}.} \bibinfo{year}{2013}\natexlab{b}.
\newblock \showarticletitle{Distributed representations of words and phrases
  and their compositionality}. In \bibinfo{booktitle}{\emph{Advances in neural
  information processing systems}}. \bibinfo{pages}{3111--3119}.
\newblock


\bibitem[\protect\citeauthoryear{Mou, Li, Zhang, Wang, and Jin}{Mou
  et~al\mbox{.}}{2016}]%
        {mou2016convolutional}
\bibfield{author}{\bibinfo{person}{Lili Mou}, \bibinfo{person}{Ge Li},
  \bibinfo{person}{Lu Zhang}, \bibinfo{person}{Tao Wang}, {and}
  \bibinfo{person}{Zhi Jin}.} \bibinfo{year}{2016}\natexlab{}.
\newblock \showarticletitle{Convolutional neural networks over tree structures
  for programming language processing}. In
  \bibinfo{booktitle}{\emph{Proceedings of the AAAI Conference on Artificial
  Intelligence}}, Vol.~\bibinfo{volume}{30}.
\newblock


\bibitem[\protect\citeauthoryear{Murdoch, Liu, and Yu}{Murdoch
  et~al\mbox{.}}{2018}]%
        {murdoch2018beyond}
\bibfield{author}{\bibinfo{person}{W~James Murdoch}, \bibinfo{person}{Peter~J
  Liu}, {and} \bibinfo{person}{Bin Yu}.} \bibinfo{year}{2018}\natexlab{}.
\newblock \showarticletitle{Beyond Word Importance: Contextual Decomposition to
  Extract Interactions from LSTMs}. In \bibinfo{booktitle}{\emph{International
  Conference on Learning Representations}}.
\newblock


\bibitem[\protect\citeauthoryear{Rabinovich, Stern, and Klein}{Rabinovich
  et~al\mbox{.}}{2017}]%
        {rabinovich2017abstract}
\bibfield{author}{\bibinfo{person}{Maxim Rabinovich}, \bibinfo{person}{Mitchell
  Stern}, {and} \bibinfo{person}{Dan Klein}.} \bibinfo{year}{2017}\natexlab{}.
\newblock \showarticletitle{Abstract Syntax Networks for Code Generation and
  Semantic Parsing}. In \bibinfo{booktitle}{\emph{Proceedings of the 55th
  Annual Meeting of the Association for Computational Linguistics (Volume 1:
  Long Papers)}}. \bibinfo{pages}{1139--1149}.
\newblock


\bibitem[\protect\citeauthoryear{Raghothaman, Wei, and Hamadi}{Raghothaman
  et~al\mbox{.}}{2016}]%
        {raghothaman2016swim}
\bibfield{author}{\bibinfo{person}{Mukund Raghothaman}, \bibinfo{person}{Yi
  Wei}, {and} \bibinfo{person}{Youssef Hamadi}.}
  \bibinfo{year}{2016}\natexlab{}.
\newblock \showarticletitle{Swim: Synthesizing what i mean-code search and
  idiomatic snippet synthesis}. In \bibinfo{booktitle}{\emph{2016 IEEE/ACM 38th
  International Conference on Software Engineering (ICSE)}}. IEEE,
  \bibinfo{pages}{357--367}.
\newblock


\bibitem[\protect\citeauthoryear{Roy and Cordy}{Roy and Cordy}{2007}]%
        {roy2007survey}
\bibfield{author}{\bibinfo{person}{Chanchal~Kumar Roy} {and}
  \bibinfo{person}{James~R Cordy}.} \bibinfo{year}{2007}\natexlab{}.
\newblock \showarticletitle{A survey on software clone detection research}.
\newblock \bibinfo{journal}{\emph{Queen’s School of Computing TR}}
  \bibinfo{volume}{541}, \bibinfo{number}{115} (\bibinfo{year}{2007}),
  \bibinfo{pages}{64--68}.
\newblock


\bibitem[\protect\citeauthoryear{Sajnani, Saini, Svajlenko, Roy, and
  Lopes}{Sajnani et~al\mbox{.}}{2016}]%
        {sajnani2016sourcerercc}
\bibfield{author}{\bibinfo{person}{Hitesh Sajnani}, \bibinfo{person}{Vaibhav
  Saini}, \bibinfo{person}{Jeffrey Svajlenko}, \bibinfo{person}{Chanchal~K
  Roy}, {and} \bibinfo{person}{Cristina~V Lopes}.}
  \bibinfo{year}{2016}\natexlab{}.
\newblock \showarticletitle{Sourcerercc: Scaling code clone detection to
  big-code}. In \bibinfo{booktitle}{\emph{Proceedings of the 38th International
  Conference on Software Engineering}}. \bibinfo{pages}{1157--1168}.
\newblock


\bibitem[\protect\citeauthoryear{Serrano and Smith}{Serrano and Smith}{2019}]%
        {serrano2019attention}
\bibfield{author}{\bibinfo{person}{Sofia Serrano} {and} \bibinfo{person}{Noah~A
  Smith}.} \bibinfo{year}{2019}\natexlab{}.
\newblock \showarticletitle{Is Attention Interpretable?}. In
  \bibinfo{booktitle}{\emph{Proceedings of the 57th Annual Meeting of the
  Association for Computational Linguistics}}. \bibinfo{pages}{2931--2951}.
\newblock


\bibitem[\protect\citeauthoryear{Shrikumar, Greenside, and Kundaje}{Shrikumar
  et~al\mbox{.}}{2017}]%
        {shrikumar2017learning}
\bibfield{author}{\bibinfo{person}{Avanti Shrikumar}, \bibinfo{person}{Peyton
  Greenside}, {and} \bibinfo{person}{Anshul Kundaje}.}
  \bibinfo{year}{2017}\natexlab{}.
\newblock \showarticletitle{Learning important features through propagating
  activation differences}. In \bibinfo{booktitle}{\emph{International
  Conference on Machine Learning}}. PMLR, \bibinfo{pages}{3145--3153}.
\newblock


\bibitem[\protect\citeauthoryear{Simonyan, Vedaldi, and Zisserman}{Simonyan
  et~al\mbox{.}}{2014}]%
        {simonyan2014deep}
\bibfield{author}{\bibinfo{person}{Karen Simonyan}, \bibinfo{person}{Andrea
  Vedaldi}, {and} \bibinfo{person}{Andrew Zisserman}.}
  \bibinfo{year}{2014}\natexlab{}.
\newblock \showarticletitle{Deep inside convolutional networks: Visualising
  image classification models and saliency maps}.
\newblock  (\bibinfo{year}{2014}).
\newblock


\bibitem[\protect\citeauthoryear{Sivaraman, Zhang, Van~den Broeck, and
  Kim}{Sivaraman et~al\mbox{.}}{2019}]%
        {sivaraman2019active}
\bibfield{author}{\bibinfo{person}{Aishwarya Sivaraman},
  \bibinfo{person}{Tianyi Zhang}, \bibinfo{person}{Guy Van~den Broeck}, {and}
  \bibinfo{person}{Miryung Kim}.} \bibinfo{year}{2019}\natexlab{}.
\newblock \showarticletitle{Active inductive logic programming for code
  search}. In \bibinfo{booktitle}{\emph{2019 IEEE/ACM 41st International
  Conference on Software Engineering (ICSE)}}. IEEE, \bibinfo{pages}{292--303}.
\newblock


\bibitem[\protect\citeauthoryear{Sun, Liu, Yang, and Qian}{Sun
  et~al\mbox{.}}{2020}]%
        {sun2020pscs}
\bibfield{author}{\bibinfo{person}{Zhensu Sun}, \bibinfo{person}{Yan Liu},
  \bibinfo{person}{Chen Yang}, {and} \bibinfo{person}{Yu Qian}.}
  \bibinfo{year}{2020}\natexlab{}.
\newblock \showarticletitle{PSCS: A Path-based Neural Model for SemanticCode
  Search}.
\newblock \bibinfo{journal}{\emph{arXiv preprint arXiv:2008.03042}}
  (\bibinfo{year}{2020}).
\newblock


\bibitem[\protect\citeauthoryear{Sundararajan, Taly, and Yan}{Sundararajan
  et~al\mbox{.}}{2017}]%
        {sundararajan2017axiomatic}
\bibfield{author}{\bibinfo{person}{Mukund Sundararajan}, \bibinfo{person}{Ankur
  Taly}, {and} \bibinfo{person}{Qiqi Yan}.} \bibinfo{year}{2017}\natexlab{}.
\newblock \showarticletitle{Axiomatic attribution for deep networks}. In
  \bibinfo{booktitle}{\emph{International Conference on Machine Learning}}.
  PMLR, \bibinfo{pages}{3319--3328}.
\newblock


\bibitem[\protect\citeauthoryear{Svajlenko, Islam, Keivanloo, Roy, and
  Mia}{Svajlenko et~al\mbox{.}}{2014}]%
        {svajlenko2014towards}
\bibfield{author}{\bibinfo{person}{Jeffrey Svajlenko},
  \bibinfo{person}{Judith~F Islam}, \bibinfo{person}{Iman Keivanloo},
  \bibinfo{person}{Chanchal~K Roy}, {and} \bibinfo{person}{Mohammad~Mamun
  Mia}.} \bibinfo{year}{2014}\natexlab{}.
\newblock \showarticletitle{Towards a big data curated benchmark of
  inter-project code clones}. In \bibinfo{booktitle}{\emph{2014 IEEE
  International Conference on Software Maintenance and Evolution}}. IEEE,
  \bibinfo{pages}{476--480}.
\newblock


\bibitem[\protect\citeauthoryear{Svyatkovskiy, Deng, Fu, and
  Sundaresan}{Svyatkovskiy et~al\mbox{.}}{2020}]%
        {svyatkovskiy2020intellicode}
\bibfield{author}{\bibinfo{person}{Alexey Svyatkovskiy},
  \bibinfo{person}{Shao~Kun Deng}, \bibinfo{person}{Shengyu Fu}, {and}
  \bibinfo{person}{Neel Sundaresan}.} \bibinfo{year}{2020}\natexlab{}.
\newblock \showarticletitle{Intellicode compose: Code generation using
  transformer}. In \bibinfo{booktitle}{\emph{Proceedings of the 28th ACM Joint
  Meeting on European Software Engineering Conference and Symposium on the
  Foundations of Software Engineering}}. \bibinfo{pages}{1433--1443}.
\newblock


\bibitem[\protect\citeauthoryear{Tang, Sennrich, and Nivre}{Tang
  et~al\mbox{.}}{2018}]%
        {tang2018analysis}
\bibfield{author}{\bibinfo{person}{Gongbo Tang}, \bibinfo{person}{Rico
  Sennrich}, {and} \bibinfo{person}{Joakim Nivre}.}
  \bibinfo{year}{2018}\natexlab{}.
\newblock \showarticletitle{An Analysis of Attention Mechanisms: The Case of
  Word Sense Disambiguation in Neural Machine Translation}. In
  \bibinfo{booktitle}{\emph{Proceedings of the Third Conference on Machine
  Translation: Research Papers}}. \bibinfo{pages}{26--35}.
\newblock


\bibitem[\protect\citeauthoryear{Tao, Gao, Shang, Wu, Zhao, and Yan}{Tao
  et~al\mbox{.}}{2018}]%
        {tao2018get}
\bibfield{author}{\bibinfo{person}{Chongyang Tao}, \bibinfo{person}{Shen Gao},
  \bibinfo{person}{Mingyue Shang}, \bibinfo{person}{Wei Wu},
  \bibinfo{person}{Dongyan Zhao}, {and} \bibinfo{person}{Rui Yan}.}
  \bibinfo{year}{2018}\natexlab{}.
\newblock \showarticletitle{Get The Point of My Utterance! Learning Towards
  Effective Responses with Multi-Head Attention Mechanism.}. In
  \bibinfo{booktitle}{\emph{IJCAI}}. \bibinfo{pages}{4418--4424}.
\newblock


\bibitem[\protect\citeauthoryear{Vig}{Vig}{2019}]%
        {vig2019multiscale}
\bibfield{author}{\bibinfo{person}{Jesse Vig}.}
  \bibinfo{year}{2019}\natexlab{}.
\newblock \showarticletitle{A Multiscale Visualization of Attention in the
  Transformer Model}. In \bibinfo{booktitle}{\emph{Proceedings of the 57th
  Annual Meeting of the Association for Computational Linguistics: System
  Demonstrations}}. \bibinfo{pages}{37--42}.
\newblock


\bibitem[\protect\citeauthoryear{Wallace, Tuyls, Wang, Subramanian, Gardner,
  and Singh}{Wallace et~al\mbox{.}}{2019}]%
        {wallace2019allennlp}
\bibfield{author}{\bibinfo{person}{Eric Wallace}, \bibinfo{person}{Jens Tuyls},
  \bibinfo{person}{Junlin Wang}, \bibinfo{person}{Sanjay Subramanian},
  \bibinfo{person}{Matt Gardner}, {and} \bibinfo{person}{Sameer Singh}.}
  \bibinfo{year}{2019}\natexlab{}.
\newblock \showarticletitle{AllenNLP Interpret: A Framework for Explaining
  Predictions of NLP Models}. In \bibinfo{booktitle}{\emph{Proceedings of the
  2019 Conference on Empirical Methods in Natural Language Processing and the
  9th International Joint Conference on Natural Language Processing
  (EMNLP-IJCNLP): System Demonstrations}}. \bibinfo{pages}{7--12}.
\newblock


\bibitem[\protect\citeauthoryear{Wan, Zhao, Yang, Xu, Ying, Wu, and Yu}{Wan
  et~al\mbox{.}}{2018}]%
        {wan2018improving}
\bibfield{author}{\bibinfo{person}{Yao Wan}, \bibinfo{person}{Zhou Zhao},
  \bibinfo{person}{Min Yang}, \bibinfo{person}{Guandong Xu},
  \bibinfo{person}{Haochao Ying}, \bibinfo{person}{Jian Wu}, {and}
  \bibinfo{person}{Philip~S Yu}.} \bibinfo{year}{2018}\natexlab{}.
\newblock \showarticletitle{Improving automatic source code summarization via
  deep reinforcement learning}. In \bibinfo{booktitle}{\emph{Proceedings of the
  33rd ACM/IEEE International Conference on Automated Software Engineering}}.
  \bibinfo{pages}{397--407}.
\newblock


\bibitem[\protect\citeauthoryear{Wang and Su}{Wang and Su}{2020}]%
        {wang2020blended}
\bibfield{author}{\bibinfo{person}{Ke Wang} {and} \bibinfo{person}{Zhendong
  Su}.} \bibinfo{year}{2020}\natexlab{}.
\newblock \showarticletitle{Blended, precise semantic program embeddings}. In
  \bibinfo{booktitle}{\emph{Proceedings of the 41st ACM SIGPLAN Conference on
  Programming Language Design and Implementation}}. \bibinfo{pages}{121--134}.
\newblock


\bibitem[\protect\citeauthoryear{Wang, Li, Ma, Xia, and Jin}{Wang
  et~al\mbox{.}}{2020a}]%
        {wang2020detecting}
\bibfield{author}{\bibinfo{person}{Wenhan Wang}, \bibinfo{person}{Ge Li},
  \bibinfo{person}{Bo Ma}, \bibinfo{person}{Xin Xia}, {and}
  \bibinfo{person}{Zhi Jin}.} \bibinfo{year}{2020}\natexlab{a}.
\newblock \showarticletitle{Detecting code clones with graph neural network and
  flow-augmented abstract syntax tree}. In \bibinfo{booktitle}{\emph{2020 IEEE
  27th International Conference on Software Analysis, Evolution and
  Reengineering (SANER)}}. IEEE, \bibinfo{pages}{261--271}.
\newblock


\bibitem[\protect\citeauthoryear{Wang, Li, Shen, Xia, and Jin}{Wang
  et~al\mbox{.}}{2020b}]%
        {wang2020modular}
\bibfield{author}{\bibinfo{person}{Wenhan Wang}, \bibinfo{person}{Ge Li},
  \bibinfo{person}{Sijie Shen}, \bibinfo{person}{Xin Xia}, {and}
  \bibinfo{person}{Zhi Jin}.} \bibinfo{year}{2020}\natexlab{b}.
\newblock \showarticletitle{Modular tree network for source code representation
  learning}.
\newblock \bibinfo{journal}{\emph{ACM Transactions on Software Engineering and
  Methodology (TOSEM)}} \bibinfo{volume}{29}, \bibinfo{number}{4}
  (\bibinfo{year}{2020}), \bibinfo{pages}{1--23}.
\newblock


\bibitem[\protect\citeauthoryear{Wei and Li}{Wei and Li}{2017}]%
        {wei2017supervised}
\bibfield{author}{\bibinfo{person}{Huihui Wei} {and} \bibinfo{person}{Ming
  Li}.} \bibinfo{year}{2017}\natexlab{}.
\newblock \showarticletitle{Supervised Deep Features for Software Functional
  Clone Detection by Exploiting Lexical and Syntactical Information in Source
  Code.}. In \bibinfo{booktitle}{\emph{IJCAI}}. \bibinfo{pages}{3034--3040}.
\newblock


\bibitem[\protect\citeauthoryear{White, Tufano, Vendome, and Poshyvanyk}{White
  et~al\mbox{.}}{2016}]%
        {white2016deep}
\bibfield{author}{\bibinfo{person}{Martin White}, \bibinfo{person}{Michele
  Tufano}, \bibinfo{person}{Christopher Vendome}, {and} \bibinfo{person}{Denys
  Poshyvanyk}.} \bibinfo{year}{2016}\natexlab{}.
\newblock \showarticletitle{Deep learning code fragments for code clone
  detection}. In \bibinfo{booktitle}{\emph{2016 31st IEEE/ACM International
  Conference on Automated Software Engineering (ASE)}}. IEEE,
  \bibinfo{pages}{87--98}.
\newblock


\bibitem[\protect\citeauthoryear{Wiegreffe and Pinter}{Wiegreffe and
  Pinter}{2019}]%
        {wiegreffe2019attention}
\bibfield{author}{\bibinfo{person}{Sarah Wiegreffe} {and}
  \bibinfo{person}{Yuval Pinter}.} \bibinfo{year}{2019}\natexlab{}.
\newblock \showarticletitle{Attention is not not Explanation}. In
  \bibinfo{booktitle}{\emph{Proceedings of the 2019 Conference on Empirical
  Methods in Natural Language Processing and the 9th International Joint
  Conference on Natural Language Processing (EMNLP-IJCNLP)}}.
  \bibinfo{pages}{11--20}.
\newblock


\bibitem[\protect\citeauthoryear{Yan, Yu, Chen, Shen, and Jiang}{Yan
  et~al\mbox{.}}{2020}]%
        {yan2020code}
\bibfield{author}{\bibinfo{person}{Shuhan Yan}, \bibinfo{person}{Hang Yu},
  \bibinfo{person}{Yuting Chen}, \bibinfo{person}{Beijun Shen}, {and}
  \bibinfo{person}{Lingxiao Jiang}.} \bibinfo{year}{2020}\natexlab{}.
\newblock \showarticletitle{Are the code snippets what we are searching for? a
  benchmark and an empirical study on code search with natural-language
  queries}. In \bibinfo{booktitle}{\emph{2020 IEEE 27th International
  Conference on Software Analysis, Evolution and Reengineering (SANER)}}. IEEE,
  \bibinfo{pages}{344--354}.
\newblock


\bibitem[\protect\citeauthoryear{Yao, Weld, Chen, and Sun}{Yao
  et~al\mbox{.}}{2018}]%
        {yao2018staqc}
\bibfield{author}{\bibinfo{person}{Ziyu Yao}, \bibinfo{person}{Daniel~S Weld},
  \bibinfo{person}{Wei-Peng Chen}, {and} \bibinfo{person}{Huan Sun}.}
  \bibinfo{year}{2018}\natexlab{}.
\newblock \showarticletitle{Staqc: A systematically mined question-code dataset
  from stack overflow}. In \bibinfo{booktitle}{\emph{Proceedings of the 2018
  World Wide Web Conference}}. \bibinfo{pages}{1693--1703}.
\newblock


\bibitem[\protect\citeauthoryear{Ye, Bunescu, and Liu}{Ye
  et~al\mbox{.}}{2014}]%
        {ye2014learning}
\bibfield{author}{\bibinfo{person}{Xin Ye}, \bibinfo{person}{Razvan Bunescu},
  {and} \bibinfo{person}{Chang Liu}.} \bibinfo{year}{2014}\natexlab{}.
\newblock \showarticletitle{Learning to rank relevant files for bug reports
  using domain knowledge}. In \bibinfo{booktitle}{\emph{Proceedings of the 22nd
  ACM SIGSOFT International Symposium on Foundations of Software Engineering}}.
  \bibinfo{pages}{689--699}.
\newblock


\bibitem[\protect\citeauthoryear{Yin and Neubig}{Yin and Neubig}{2018}]%
        {yin2018tranx}
\bibfield{author}{\bibinfo{person}{Pengcheng Yin} {and} \bibinfo{person}{Graham
  Neubig}.} \bibinfo{year}{2018}\natexlab{}.
\newblock \showarticletitle{TRANX: A Transition-based Neural Abstract Syntax
  Parser for Semantic Parsing and Code Generation}. In
  \bibinfo{booktitle}{\emph{Proceedings of the 2018 Conference on Empirical
  Methods in Natural Language Processing: System Demonstrations}}.
  \bibinfo{pages}{7--12}.
\newblock


\bibitem[\protect\citeauthoryear{Yu, Lam, Chen, Li, Xie, and Wang}{Yu
  et~al\mbox{.}}{2019}]%
        {yu2019neural}
\bibfield{author}{\bibinfo{person}{Hao Yu}, \bibinfo{person}{Wing Lam},
  \bibinfo{person}{Long Chen}, \bibinfo{person}{Ge Li}, \bibinfo{person}{Tao
  Xie}, {and} \bibinfo{person}{Qianxiang Wang}.}
  \bibinfo{year}{2019}\natexlab{}.
\newblock \showarticletitle{Neural detection of semantic code clones via
  tree-based convolution}. In \bibinfo{booktitle}{\emph{2019 IEEE/ACM 27th
  International Conference on Program Comprehension (ICPC)}}. IEEE,
  \bibinfo{pages}{70--80}.
\newblock


\bibitem[\protect\citeauthoryear{Zhang, Li, Li, Ma, Liu, and Jin}{Zhang
  et~al\mbox{.}}{2020}]%
        {zhang2020generating}
\bibfield{author}{\bibinfo{person}{Huangzhao Zhang}, \bibinfo{person}{Zhuo Li},
  \bibinfo{person}{Ge Li}, \bibinfo{person}{Lei Ma}, \bibinfo{person}{Yang
  Liu}, {and} \bibinfo{person}{Zhi Jin}.} \bibinfo{year}{2020}\natexlab{}.
\newblock \showarticletitle{Generating adversarial examples for holding
  robustness of source code processing models}. In
  \bibinfo{booktitle}{\emph{Proceedings of the AAAI Conference on Artificial
  Intelligence}}, Vol.~\bibinfo{volume}{34}. \bibinfo{pages}{1169--1176}.
\newblock


\bibitem[\protect\citeauthoryear{Zhang, Wang, Zhang, Sun, Wang, and Liu}{Zhang
  et~al\mbox{.}}{2019}]%
        {zhang2019novel}
\bibfield{author}{\bibinfo{person}{Jian Zhang}, \bibinfo{person}{Xu Wang},
  \bibinfo{person}{Hongyu Zhang}, \bibinfo{person}{Hailong Sun},
  \bibinfo{person}{Kaixuan Wang}, {and} \bibinfo{person}{Xudong Liu}.}
  \bibinfo{year}{2019}\natexlab{}.
\newblock \showarticletitle{A novel neural source code representation based on
  abstract syntax tree}. In \bibinfo{booktitle}{\emph{2019 IEEE/ACM 41st
  International Conference on Software Engineering (ICSE)}}. IEEE,
  \bibinfo{pages}{783--794}.
\newblock


\end{thebibliography}

\appendix
\section{Attribution Analysis on two more instances}
\subsection{\textbf{Code Classification}}
\begin{figure*}[h]
    \centering
  \subfigure[Attribution score on LSTM]{
      \includegraphics[width=0.4\columnwidth]{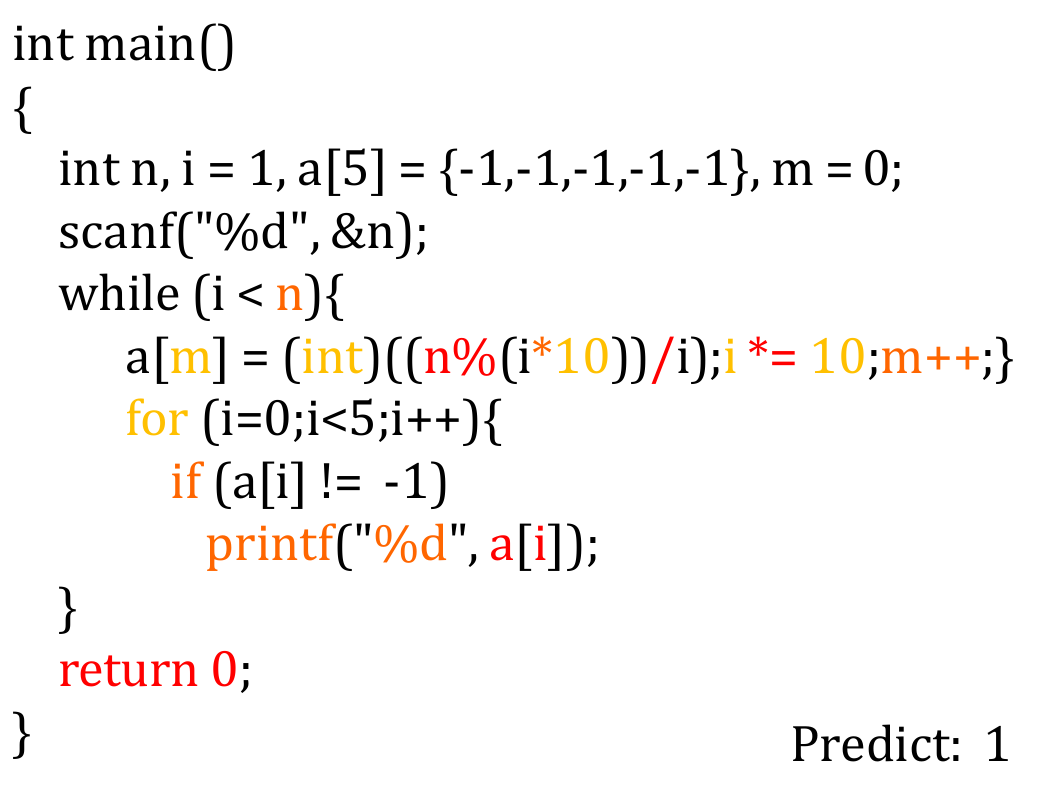}
      \label{fig:new_32232_LSTM}
    } 
    \subfigure[Attribution score on Transformer]{
      \includegraphics[width=0.4\columnwidth]{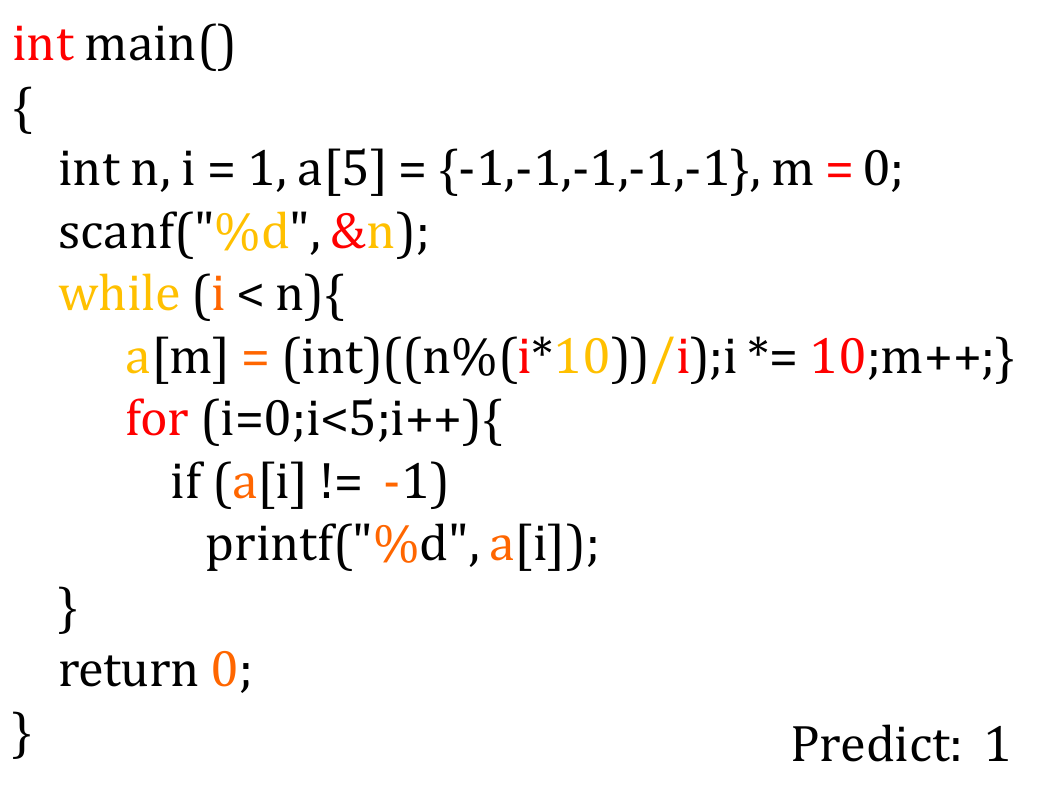}
      \label{fig:new_32232_Transformer}
    } \\
    \subfigure[Attribution score on TBCNN]{
      \includegraphics[width=0.4\columnwidth]{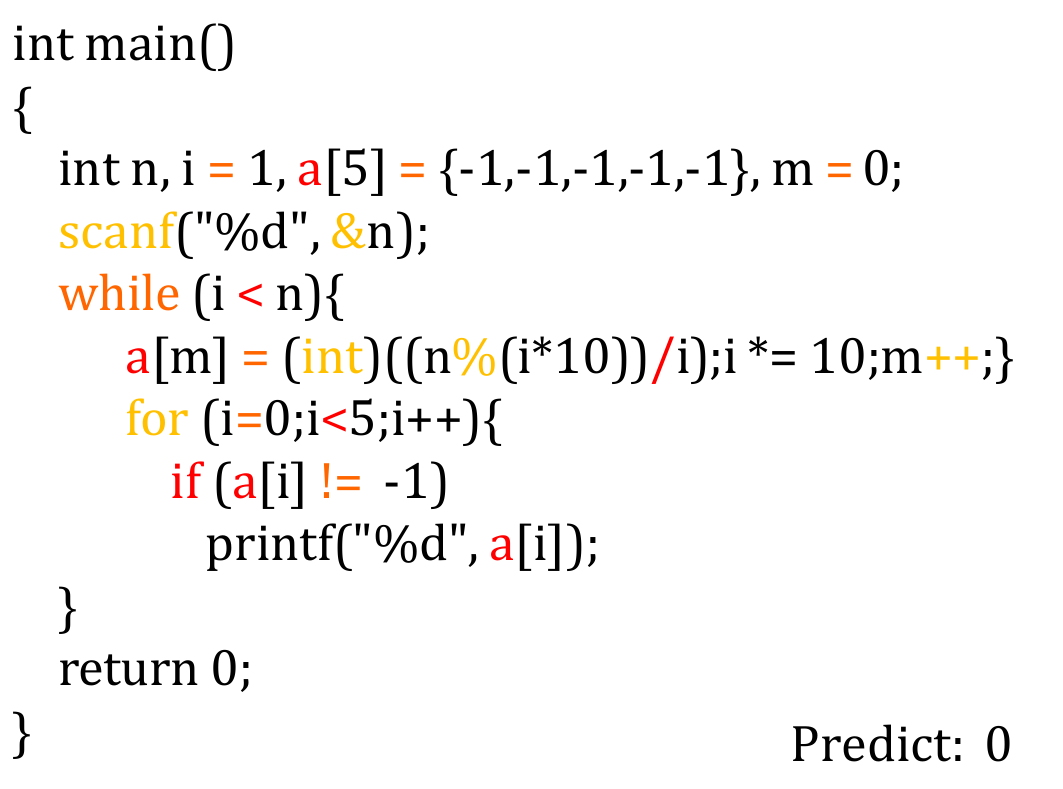}
      \label{fig:new_32232_TBCNN}
    } 
    \subfigure[Attribution score on AutoenCODE]{
      \includegraphics[width=0.4\columnwidth]{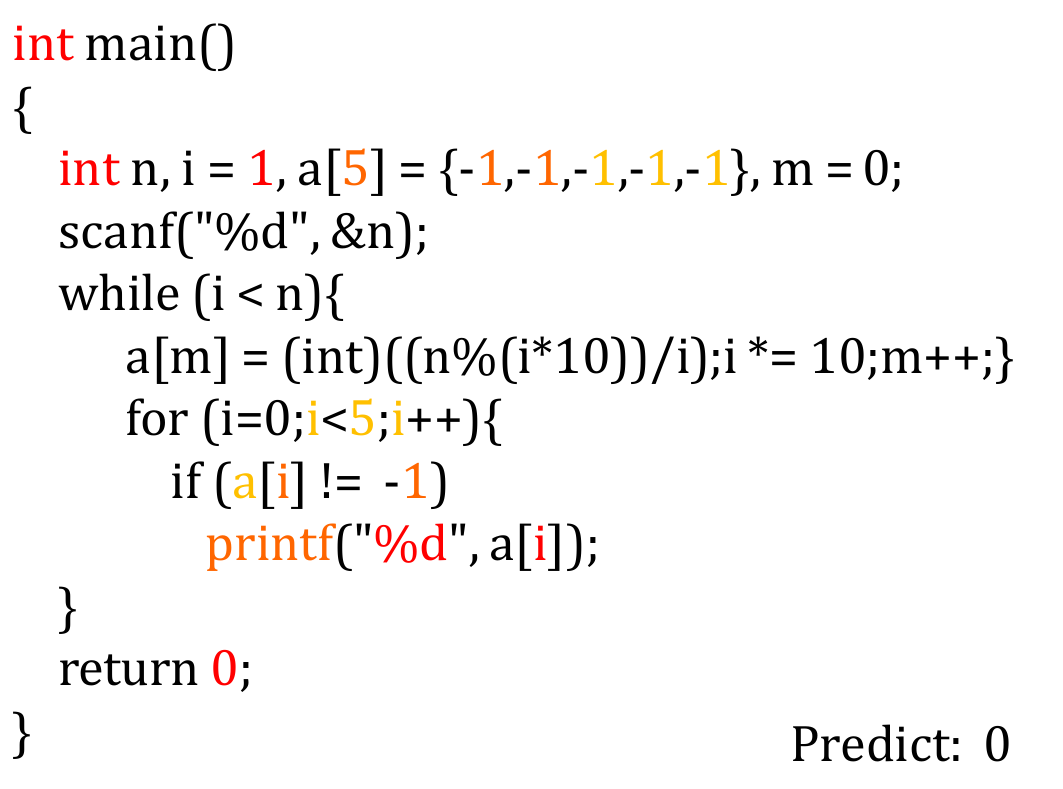}
      \label{fig:new_32232_autoencoder}
    } \\
    \subfigure[Attribution score on code2vec]{
      \includegraphics[width=0.4\columnwidth]{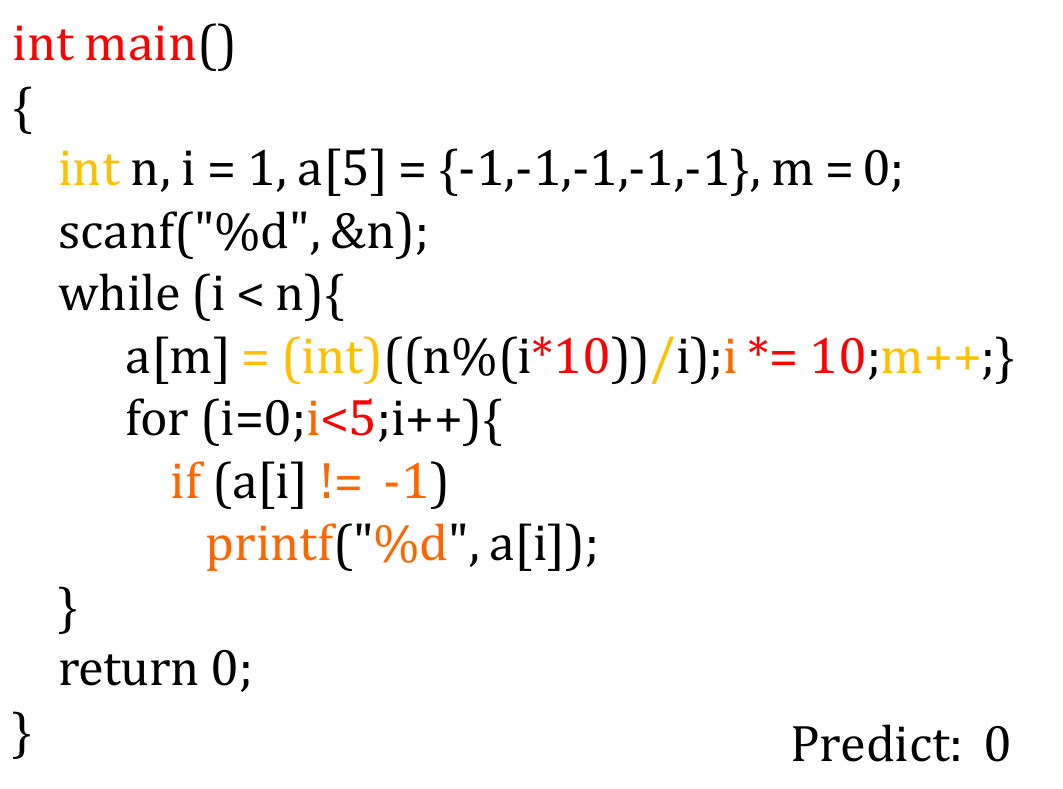}
      \label{fig:new_32232_code2vec}
    }
    \subfigure[Attribution score on code2seq]{
      \includegraphics[width=0.4\columnwidth]{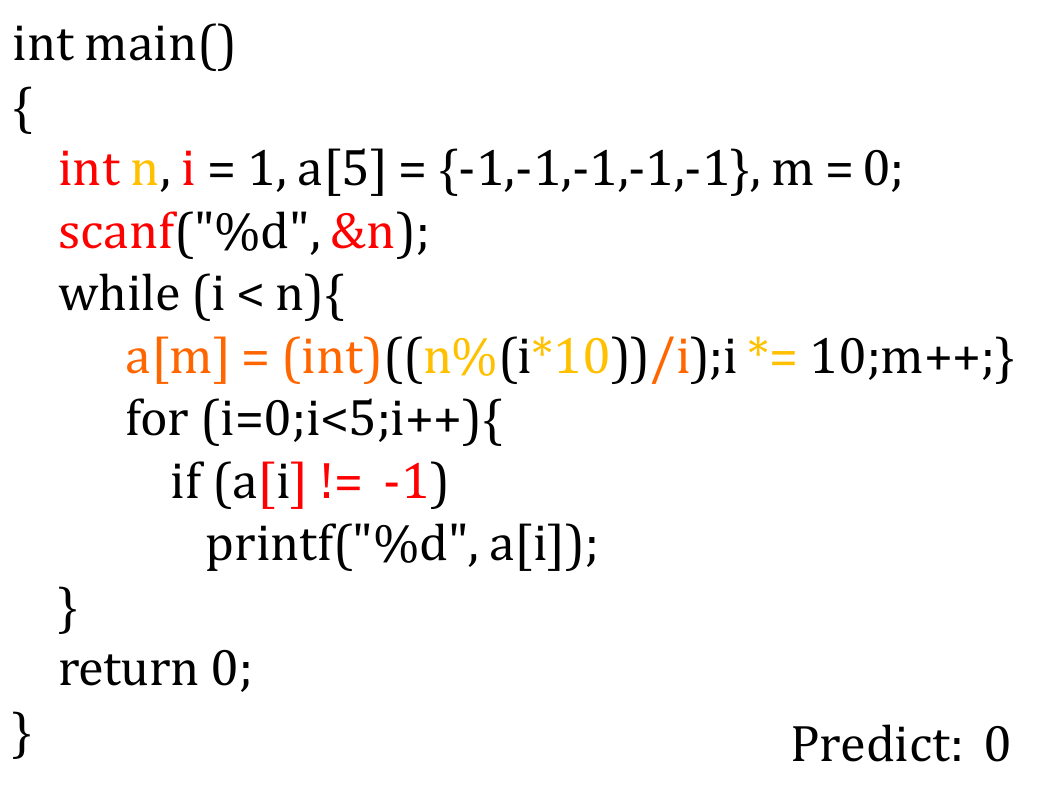}
      \label{fig:new_32232_code2seq}
    } \\
    \subfigure[Attribution score on GGNN]{
      \includegraphics[width=0.4\columnwidth]{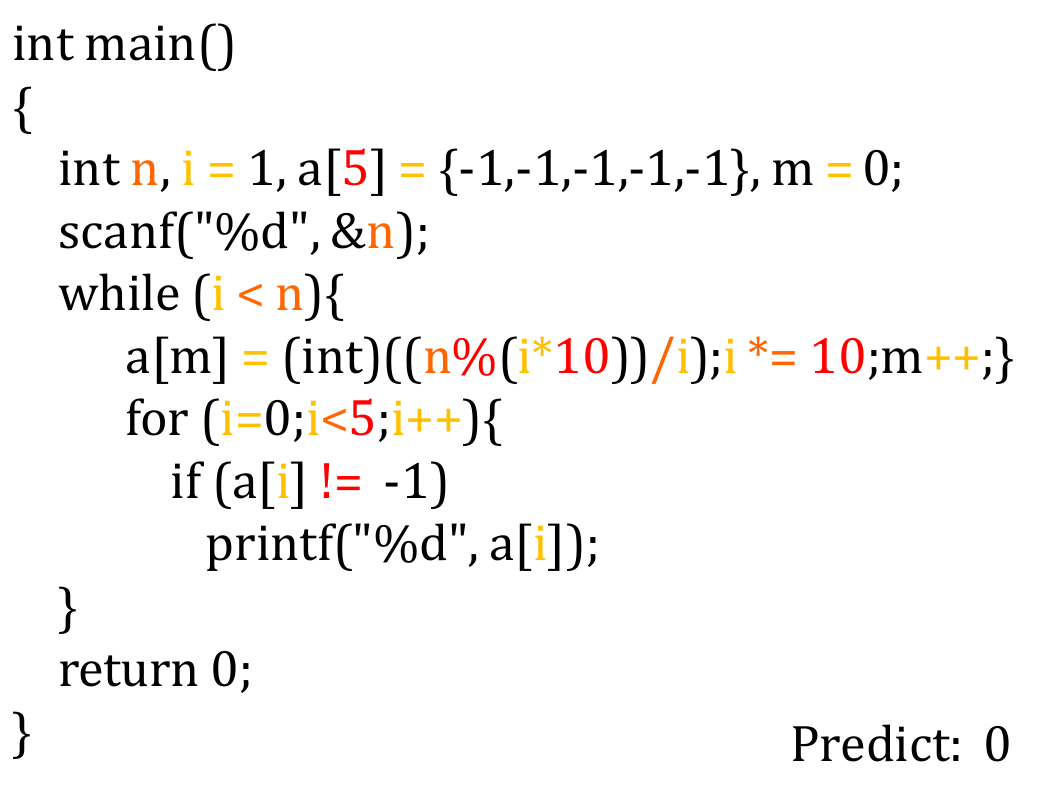}
      \label{fig:new_32232_GGNN}
    }
    \subfigure[Attribution score on ASTNN]{
      \includegraphics[width=0.4\columnwidth]{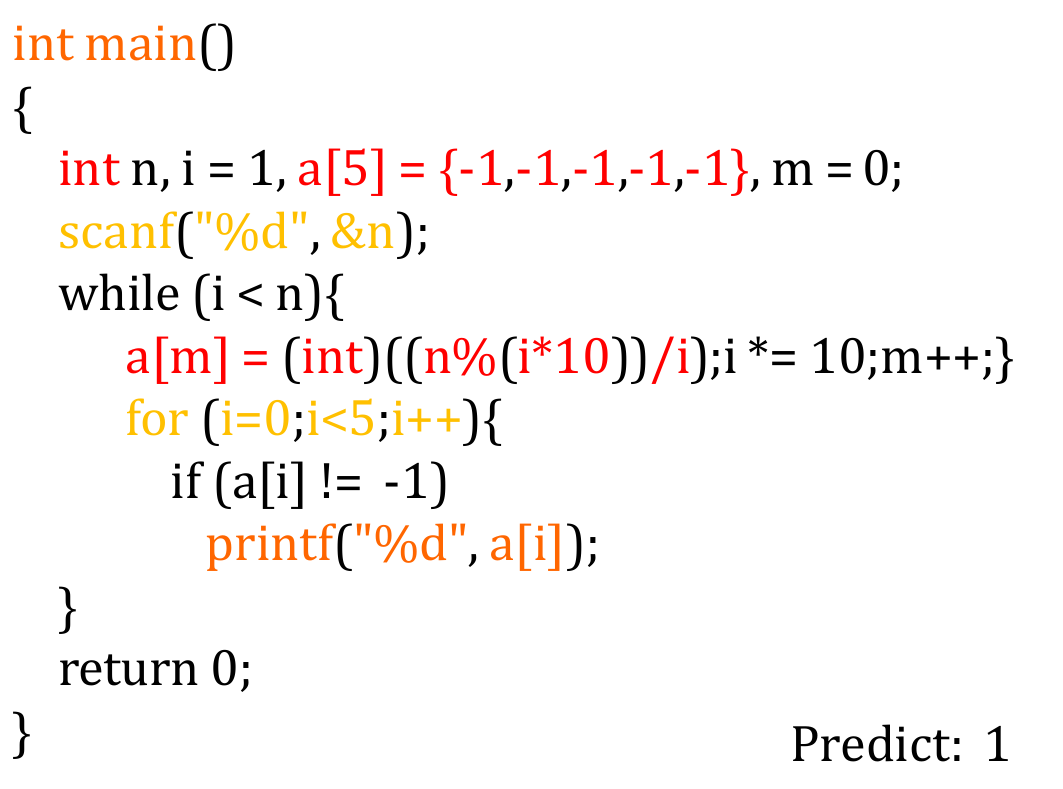}
      \label{fig:new_32232_ASTNN}
    } \\
  \caption{Attribution analysis on code classification. The functionality is to reverse a integer.}
  \label{fig:cla-attr1} 
\end{figure*}

\begin{figure*}[h]
    \centering
  \subfigure[Attribution score on LSTM]{
      \includegraphics[width=0.47\columnwidth]{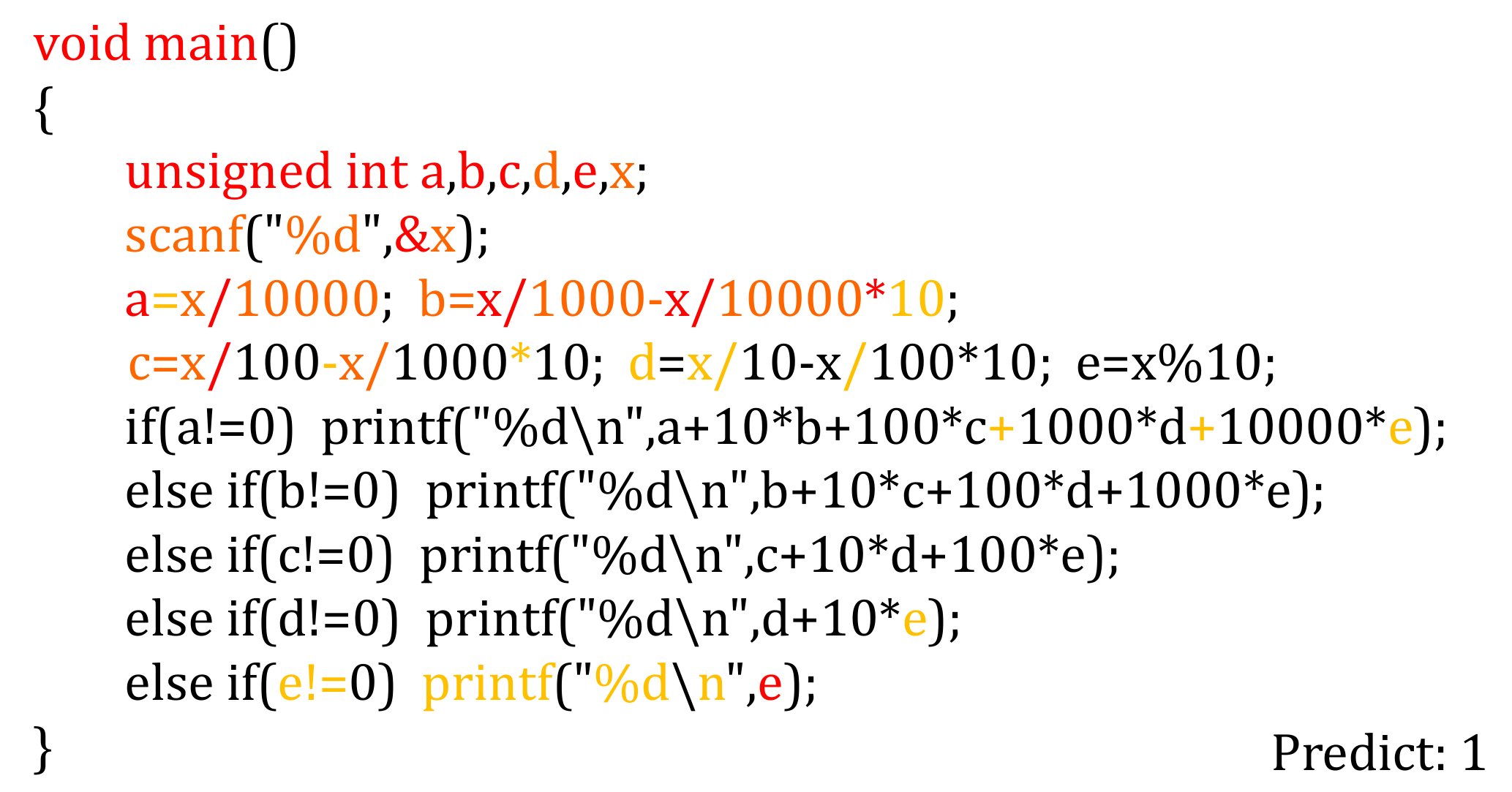}
      \label{fig:cla-32496-LSTM}
    } 
    \subfigure[Attribution score on Transformer]{
      \includegraphics[width=0.47\columnwidth]{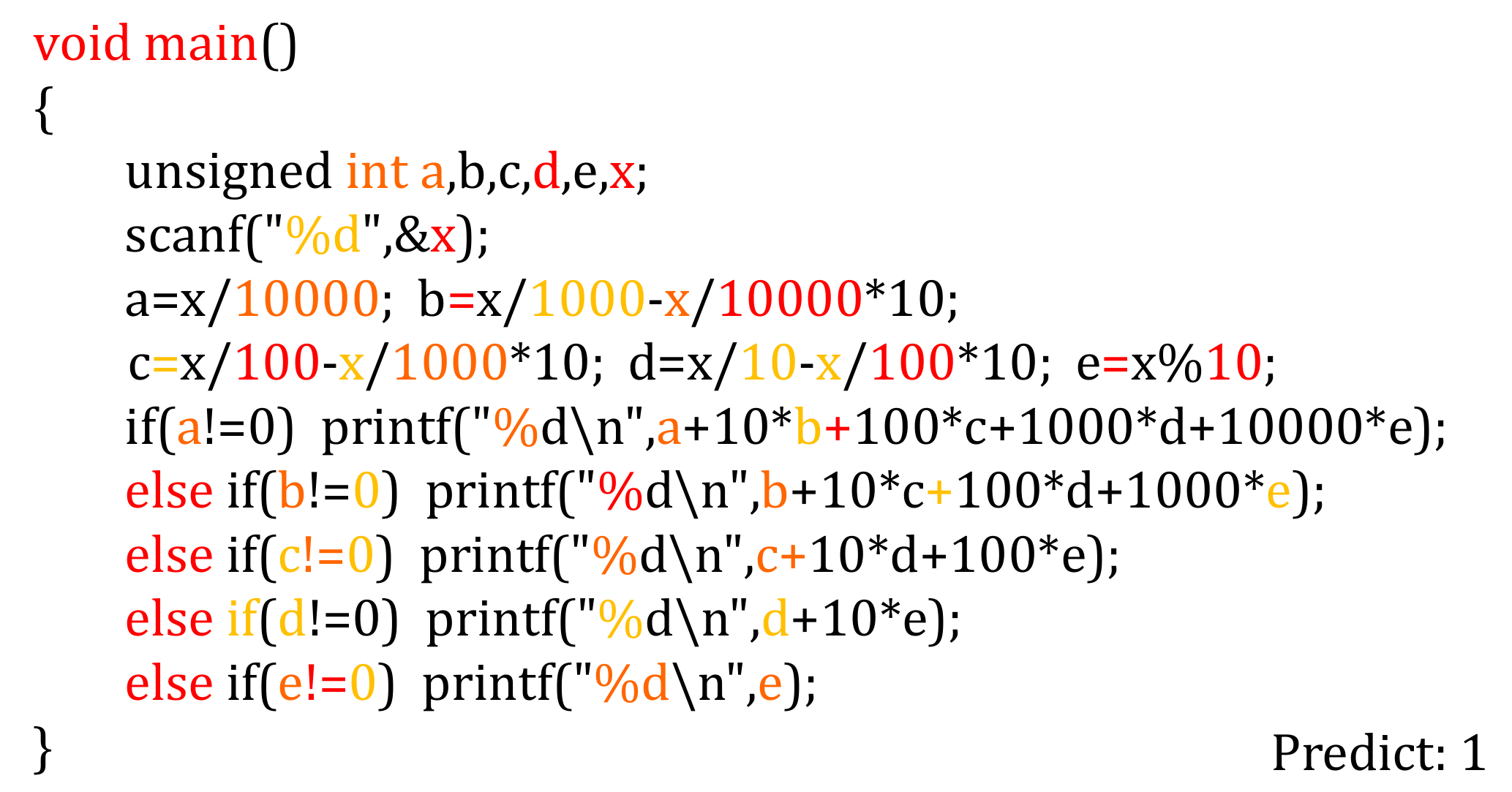}
      \label{fig:cla-32496-Transformer}
    } \\
    \subfigure[Attribution score on TBCNN]{
      \includegraphics[width=0.47\columnwidth]{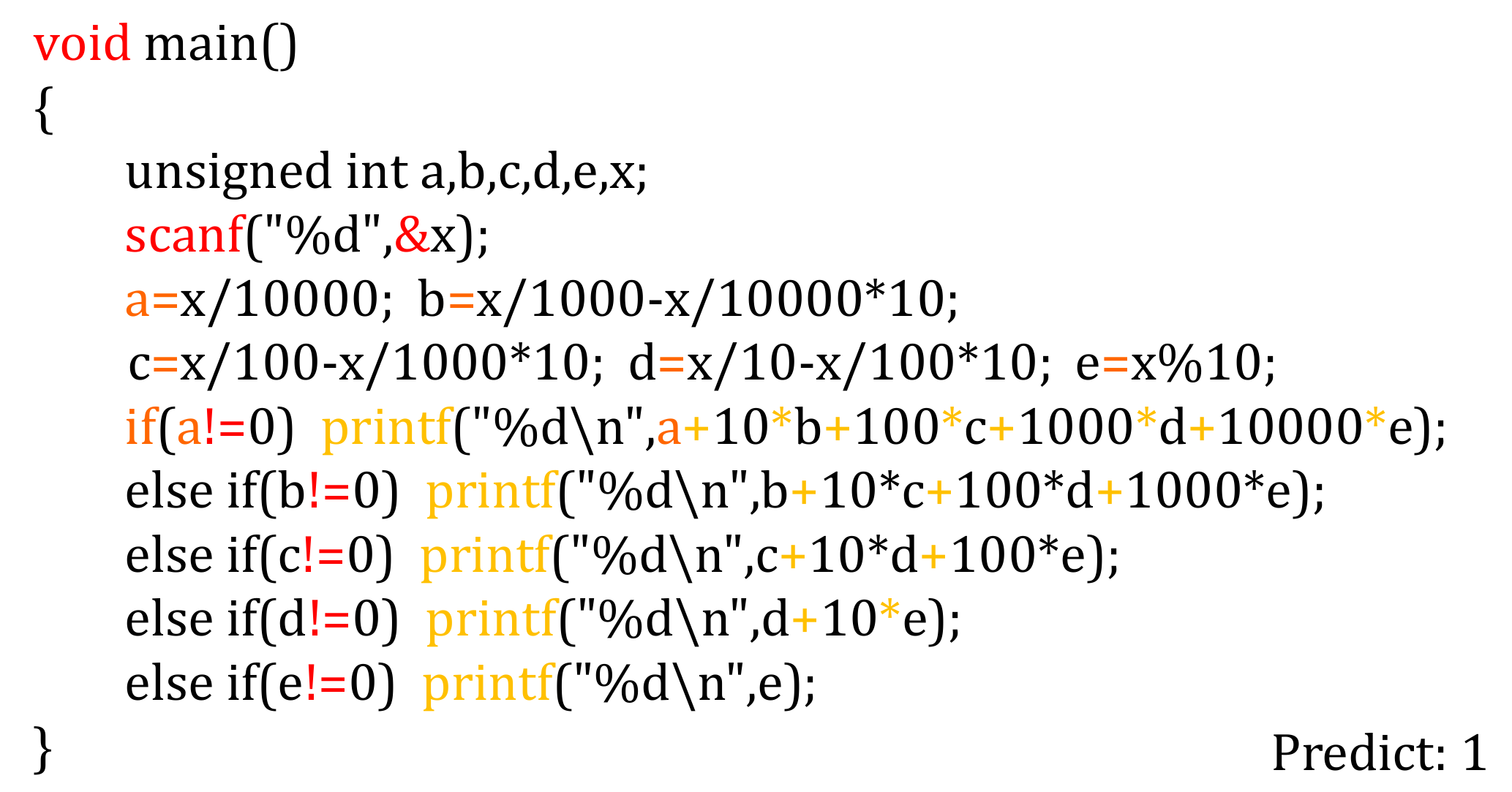}
      \label{fig:cla-32496-TBCNN}
    } 
    \subfigure[Attribution score on AutoenCODE]{
      \includegraphics[width=0.47\columnwidth]{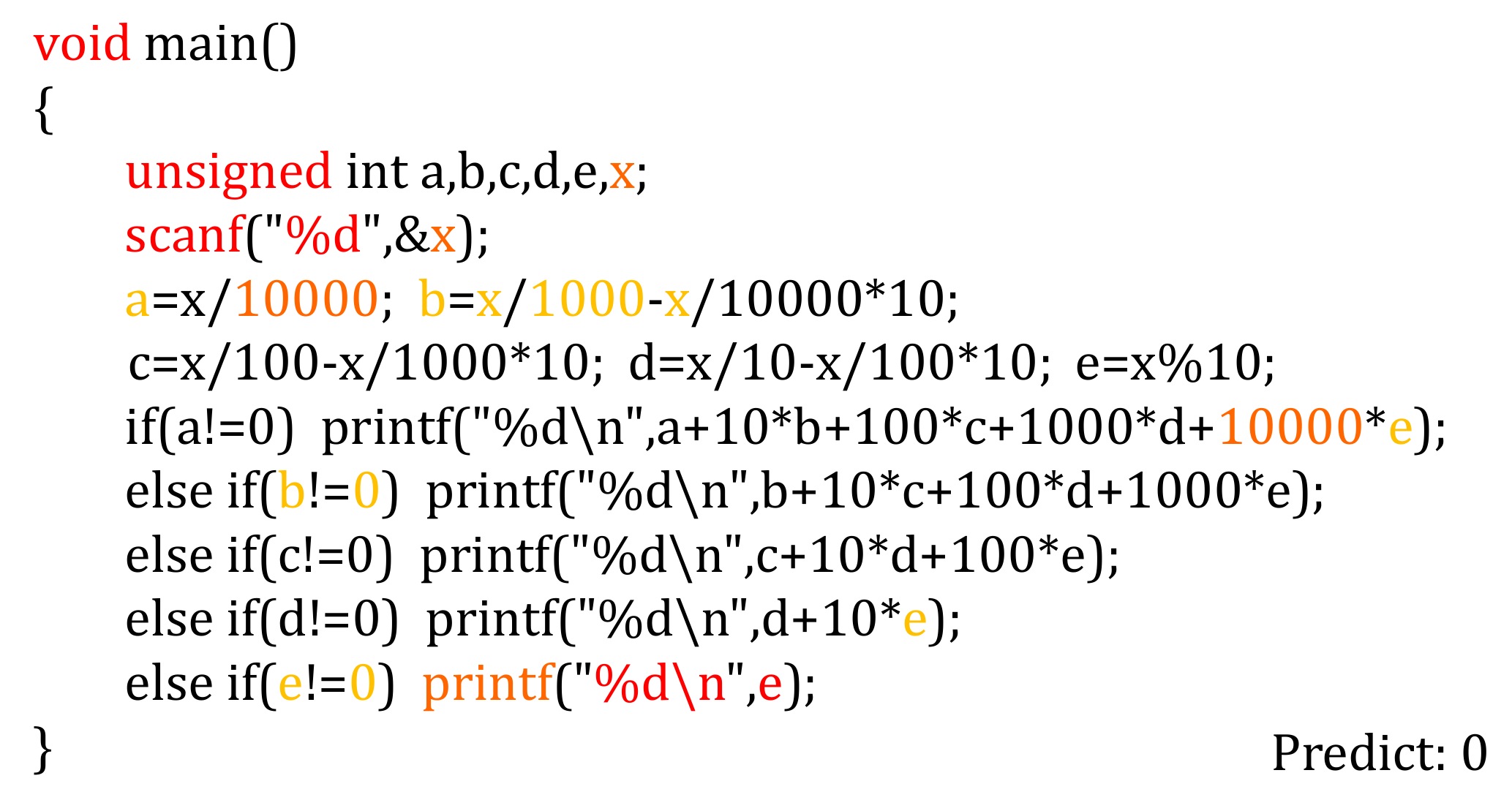}
      \label{fig:cla-32496-autoencoder}
    }  \\
    \subfigure[Attribution score on code2vec]{
      \includegraphics[width=0.47\columnwidth]{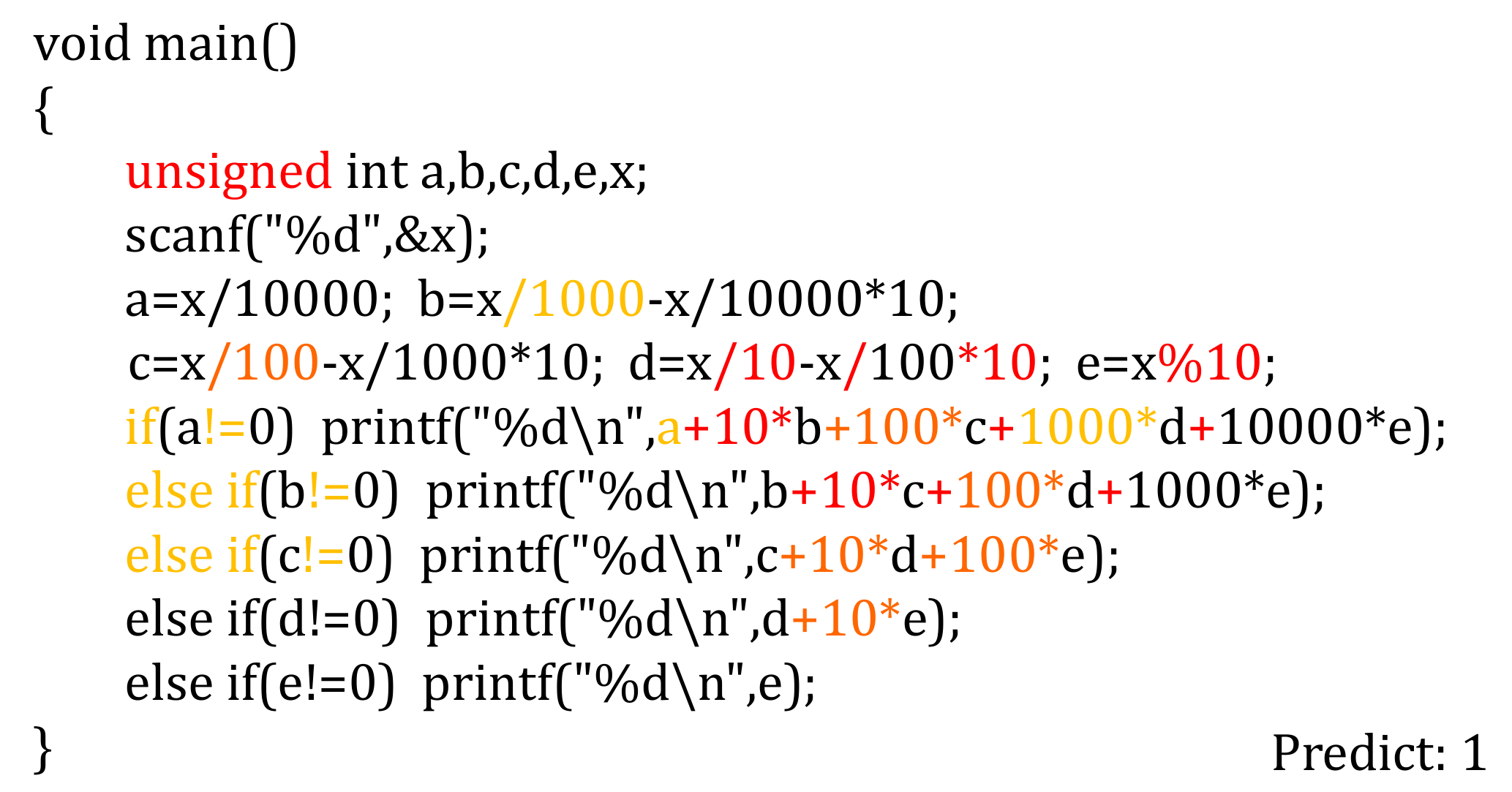}
      \label{fig:cla-32496-code2vec}
    }
    \subfigure[Attribution score on code2seq]{
      \includegraphics[width=0.47\columnwidth]{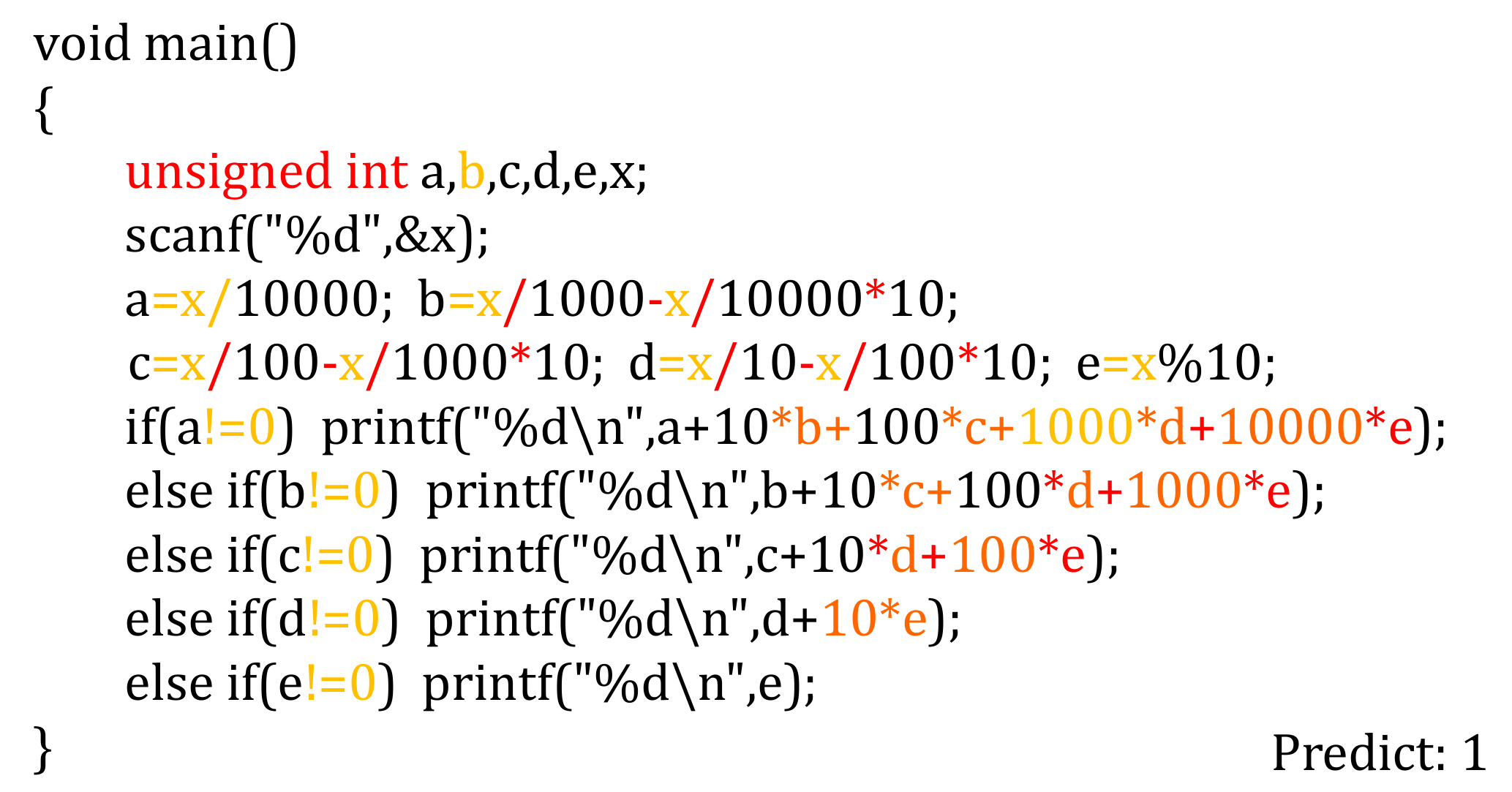}
      \label{fig:cla-32496-code2seq}
    }\\
    \subfigure[Attribution score on GGNN]{
      \includegraphics[width=0.47\columnwidth]{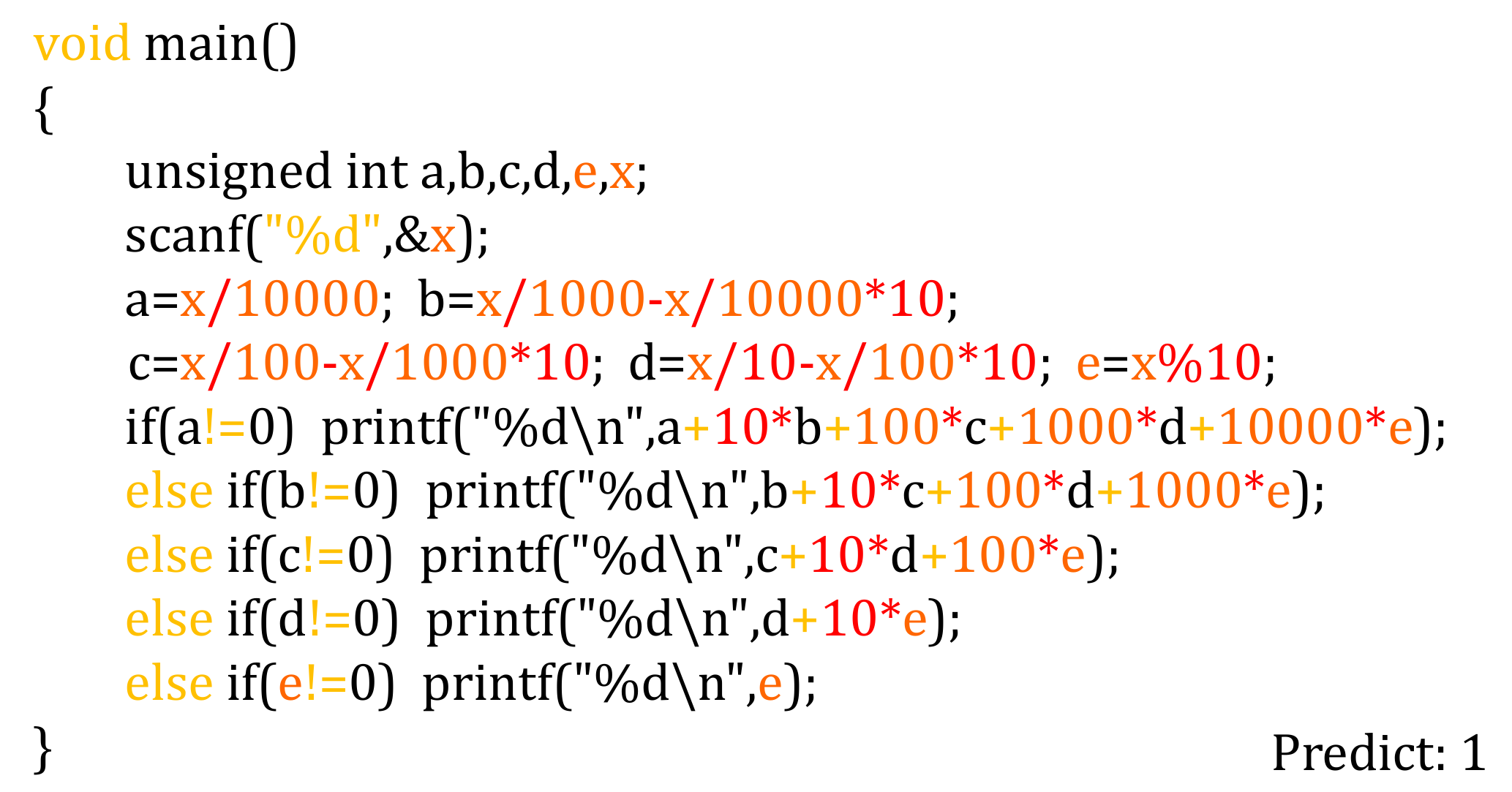}
      \label{fig:cla-32496-GGNN}
    }
    \subfigure[Attribution score on ASTNN]{
      \includegraphics[width=0.47\columnwidth]{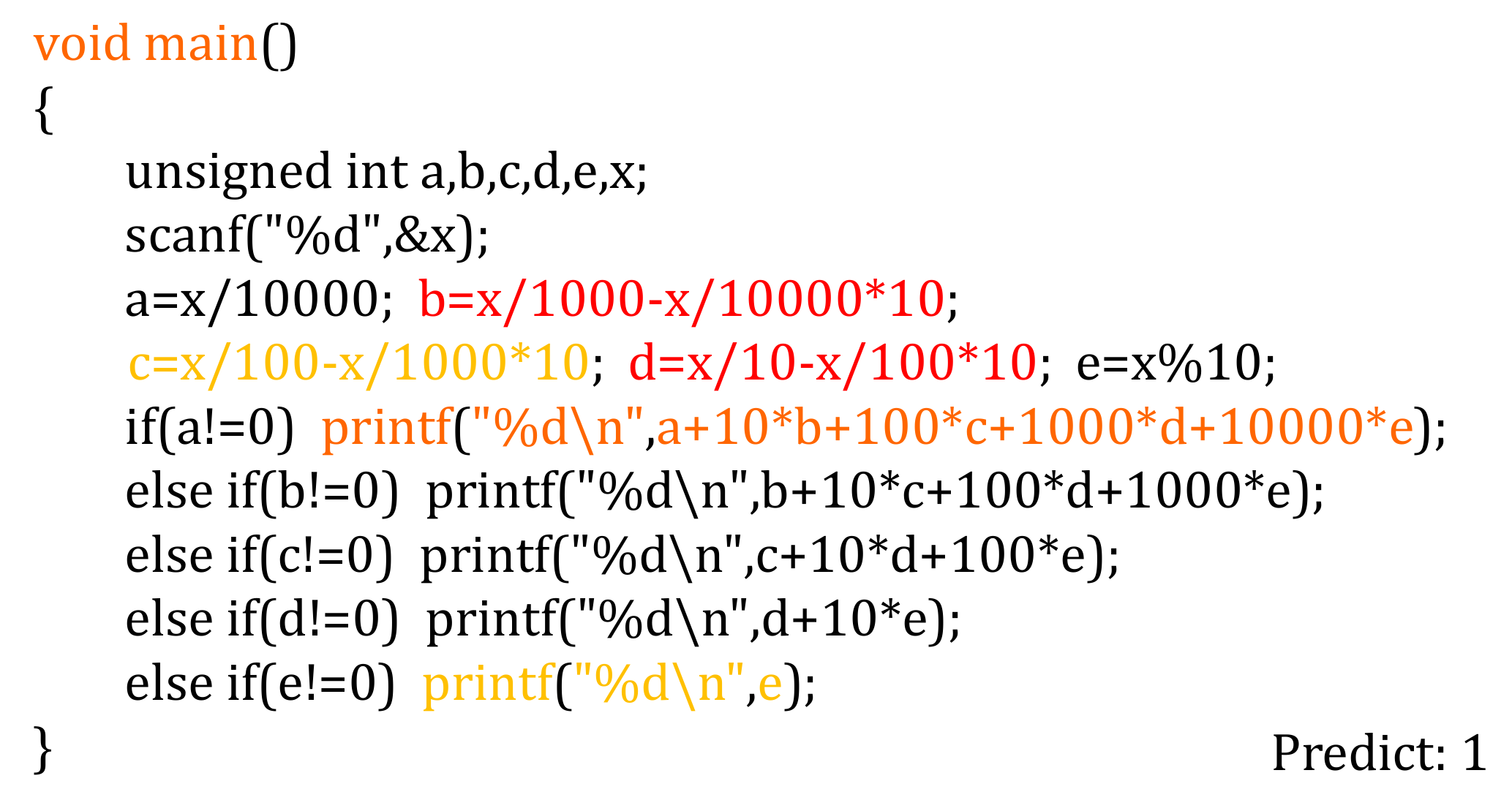}
      \label{fig:cla-32496-ASTNN}
    } \\
  \caption{Attribution analysis on code classification. The functionality is to reverse a integer.}
  \label{fig:cla-attr2} 
\end{figure*}
We select two other attribution instances visualized in Fig.~\ref{fig:cla-attr1} and Fig.~\ref{fig:cla-attr2} to support the insights in our main paper.
They preserve the same functionality of reversing the order of the digits in an integer.
In Fig.~\ref{fig:cla-attr1}, it can be found that the three models with correct prediction give high credits on the \texttt{for} loop statement whereas none of the models with incorrect prediction get a highlighted attribution score on \texttt{for} or \texttt{while}.
Another decisive part is to conduct the actual reversal process in the instance.
Among the five models with incorrect prediction, AutoenCODE perform bad on capturing both of the arithmetic operators and meaningful constants,
and the other models neglect the flow of control and disperse its attribution to more useless tokens.
In the second instance, as Fig.~\ref{fig:cla-attr2} is shown, models with correct prediction emphasize more on decisive computation consisting of operators and constants while AutoenCODE obtains higher attribution on variable declaration statements. 
With the attribution analysis on code classification task, we conclude that the embedding models should capture the key features pertaining to the control statements and the statements that actually realize the functionality. Moreover, neglecting the flow of control and data might hinder the correct prediction, although some key features are captured.

\subsubsection{\textbf{Code Clone Detection}}

\begin{figure*}[h]
    \centering
  \subfigure[Code-1 in clone pair]{
      \includegraphics[width=0.43\columnwidth]{figures/clo-attr/2518655.pdf}
      \label{fig:clo-2518655}
    } \\
    \subfigure[Attribution score on LSTM]{
      \includegraphics[width=0.45\columnwidth]{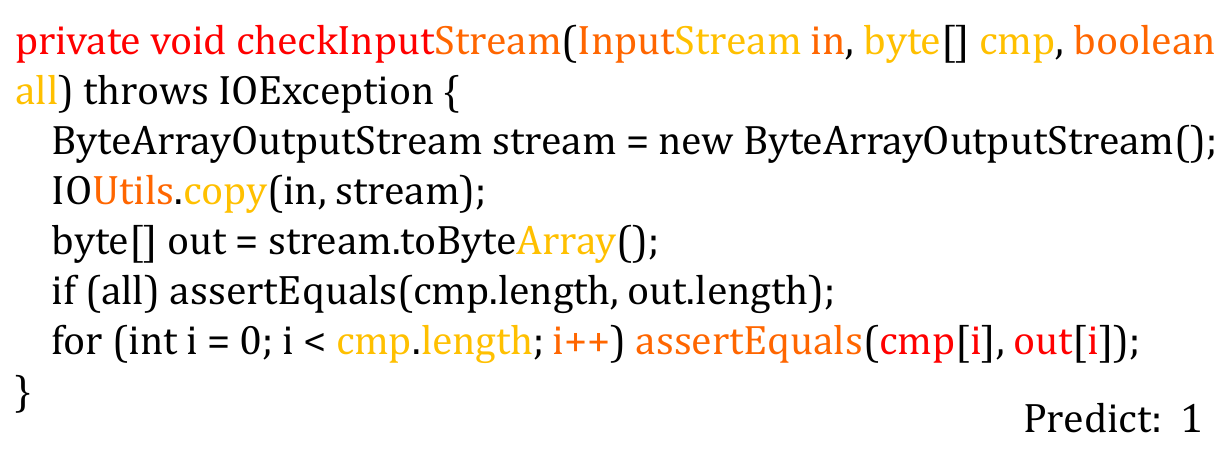}
      \label{fig:clo-11562173-LSTM}
    } 
    \subfigure[Attribution score on Transformer]{
      \includegraphics[width=0.45\columnwidth]{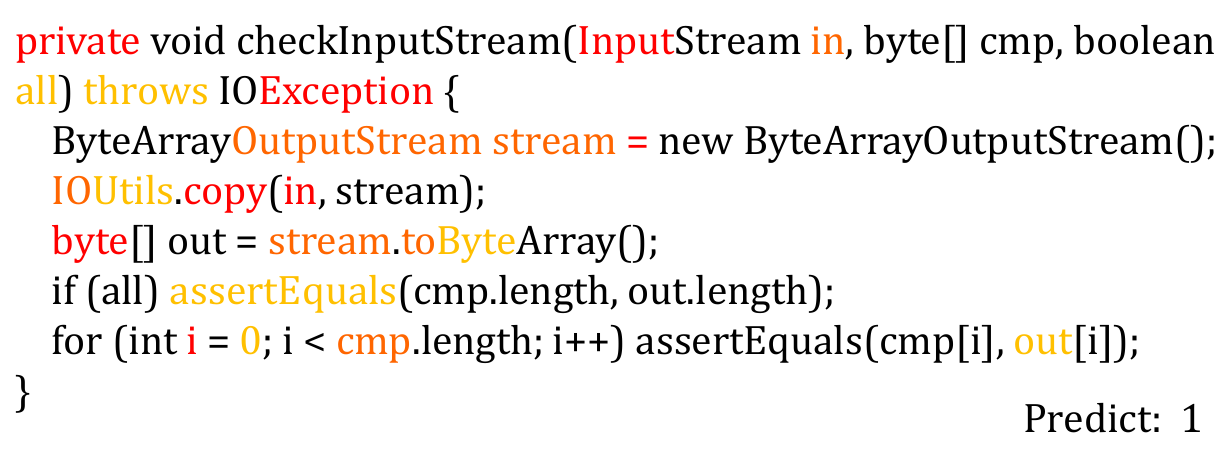}
      \label{fig:clo-11562173-Transformer}
    } \\
    \subfigure[Attribution score on TBCNN]{
      \includegraphics[width=0.45\columnwidth]{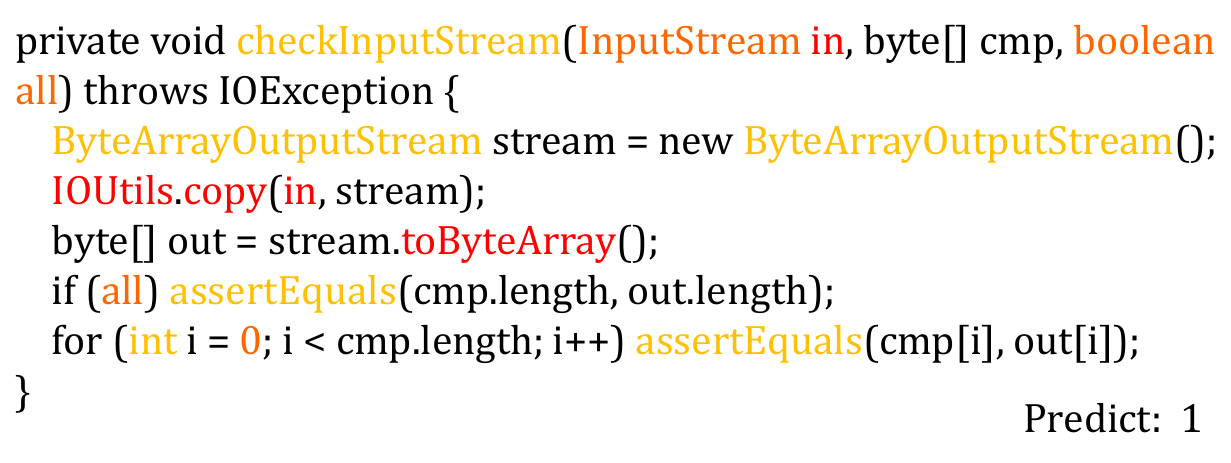}
      \label{fig:clo-11562173-TBCNN}
    } 
    \subfigure[Attribution score on AutoenCODE]{
      \includegraphics[width=0.45\columnwidth]{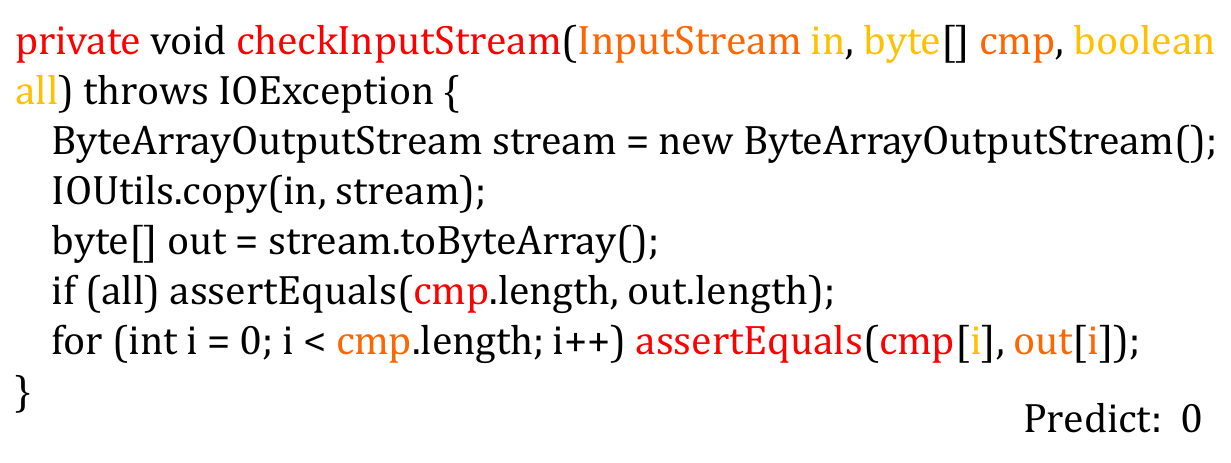}
      \label{fig:clo-11562173-autoencoder}
    }\\
    \subfigure[Attribution score on code2vec]{
      \includegraphics[width=0.45\columnwidth]{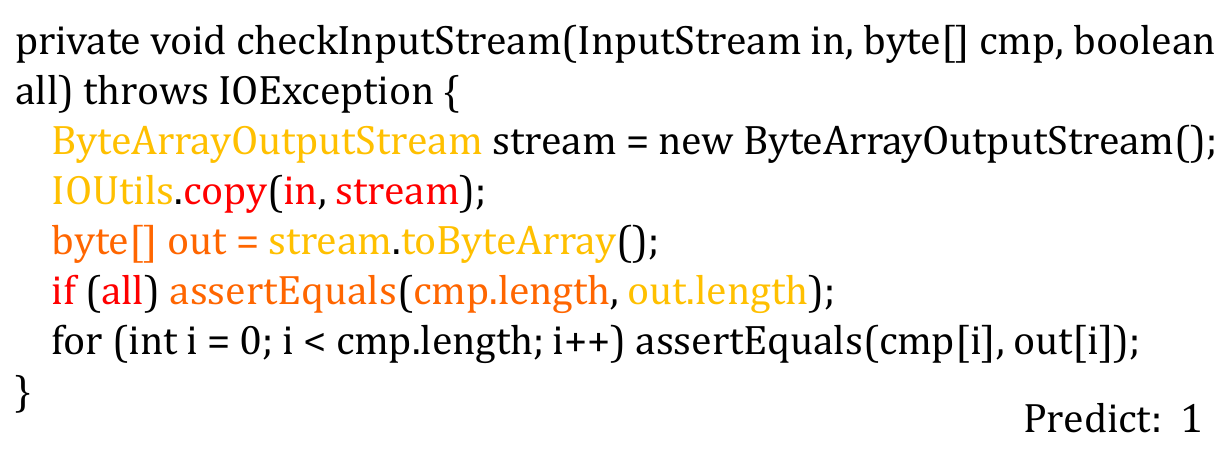}
      \label{fig:clo-11562173-code2vec}
    } 
    \subfigure[Attribution score on code2seq]{
      \includegraphics[width=0.45\columnwidth]{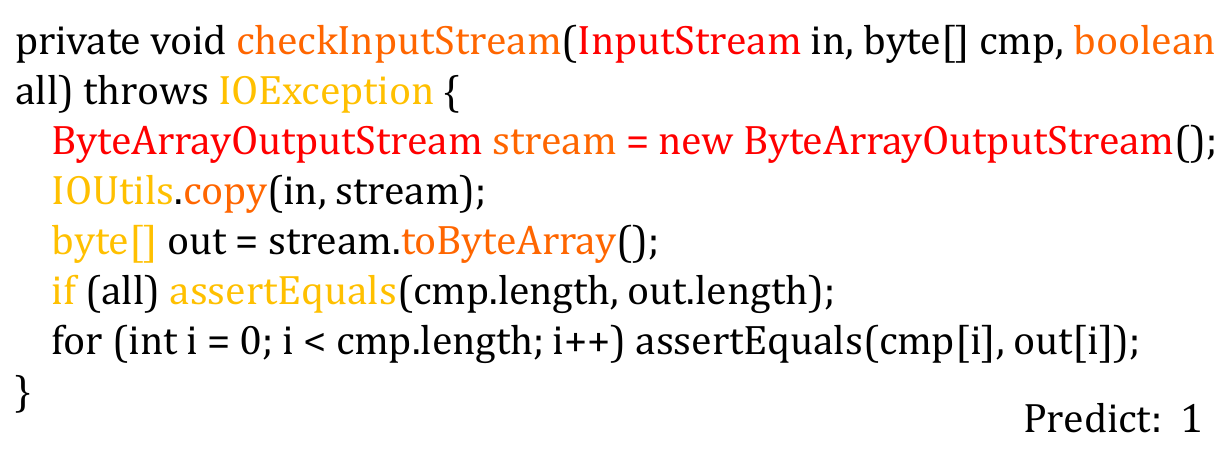}
      \label{fig:clo-11562173-code2seq}
    } \\
    \subfigure[Attribution score on GGNN]{
      \includegraphics[width=0.45\columnwidth]{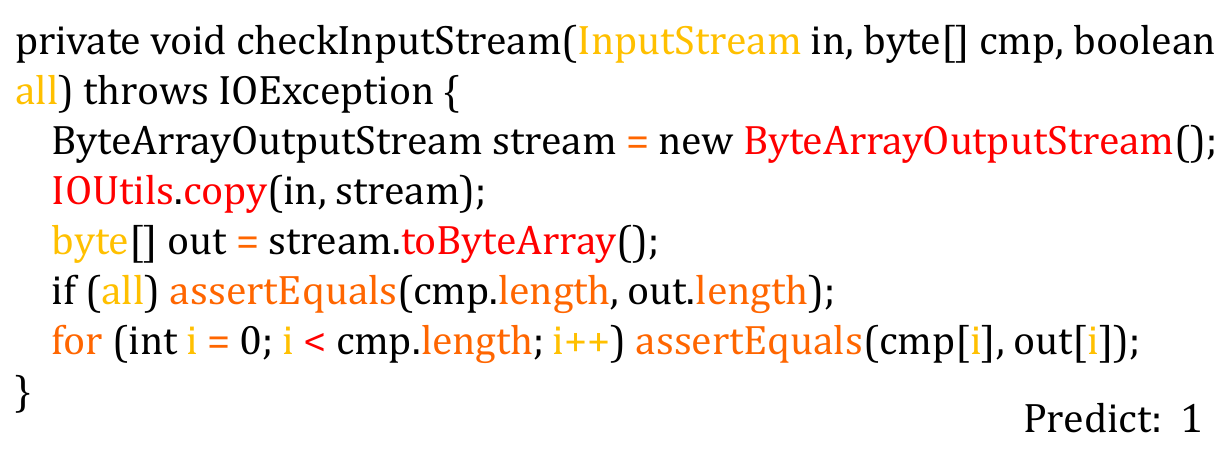}
      \label{fig:clo-11562173-GGNN}
    }
    \subfigure[Attribution score on ASTNN]{
      \includegraphics[width=0.45\columnwidth]{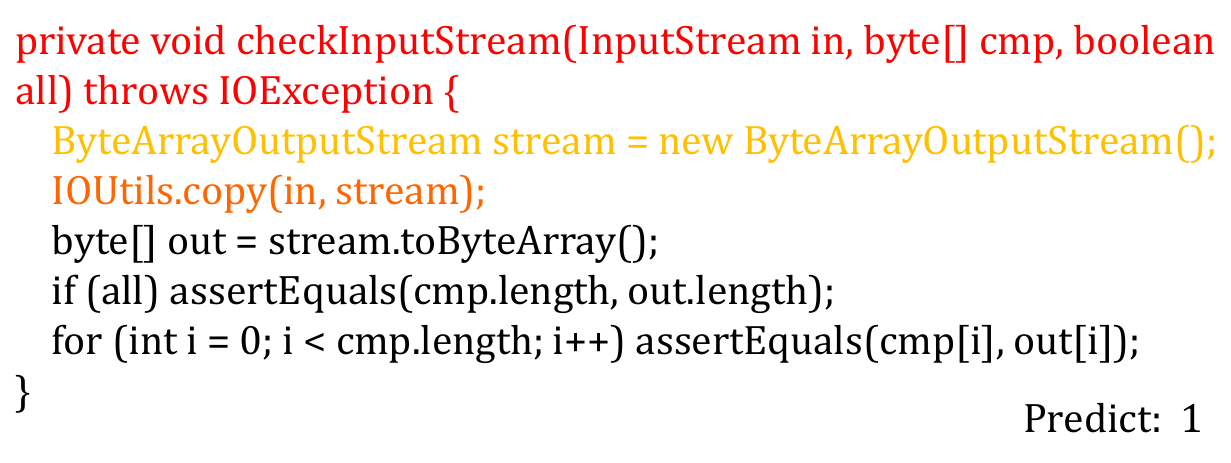}
      \label{fig:clo-11562173-ASTNN}
    }\\
  \caption{Attribution analysis on code-2 in Type-4 clone pairs.}
  \label{fig:clo-attr2} 
\end{figure*}

\begin{figure*}[h]
    \centering
  \subfigure[Code-1 in clone pair]{
      \includegraphics[width=0.48\columnwidth]{figures/clo-attr/2518655.pdf}
      \label{fig:clo-12236729}
    } \\
    \subfigure[Attribution score on LSTM]{
      \includegraphics[width=0.48\columnwidth]{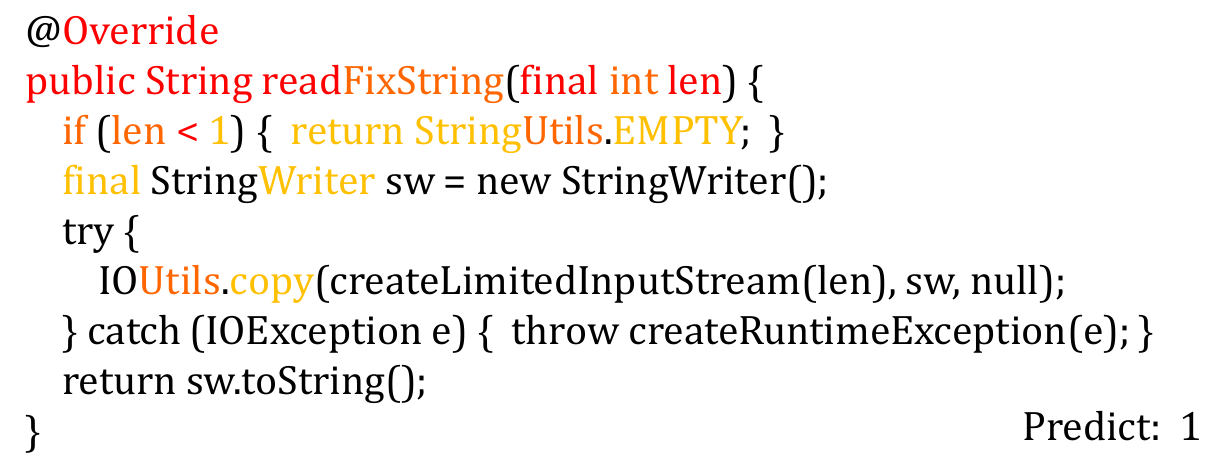}
      \label{fig:clo-22046596-LSTM}
    } 
    \subfigure[Attribution score on Transformer]{
      \includegraphics[width=0.48\columnwidth]{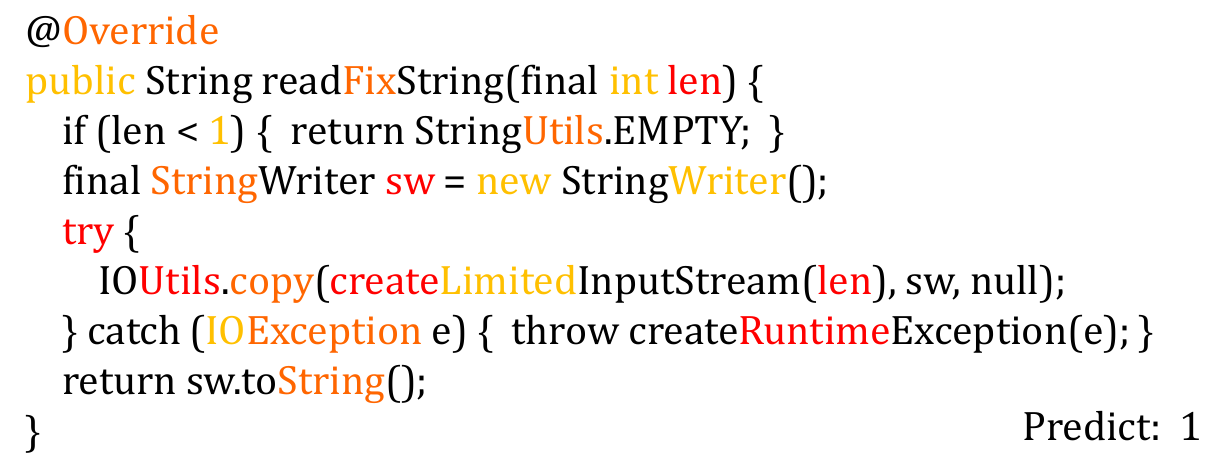}
      \label{fig:clo-22046596-Transformer}
    } \\
    \subfigure[Attribution score on TBCNN]{
      \includegraphics[width=0.48\columnwidth]{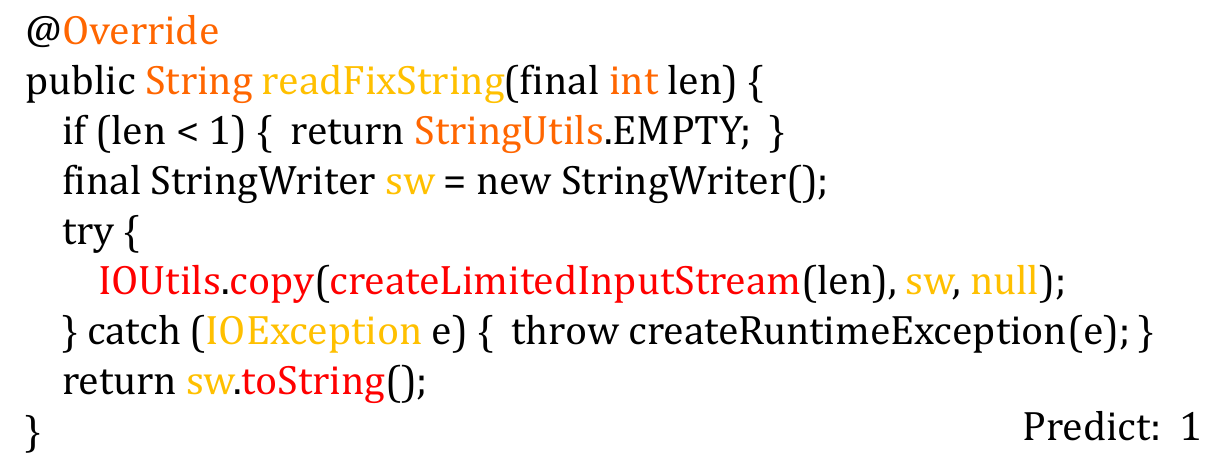}
      \label{fig:clo-22046596-TBCNN}
    } 
    \subfigure[Attribution score on AutoenCODE]{
      \includegraphics[width=0.48\columnwidth]{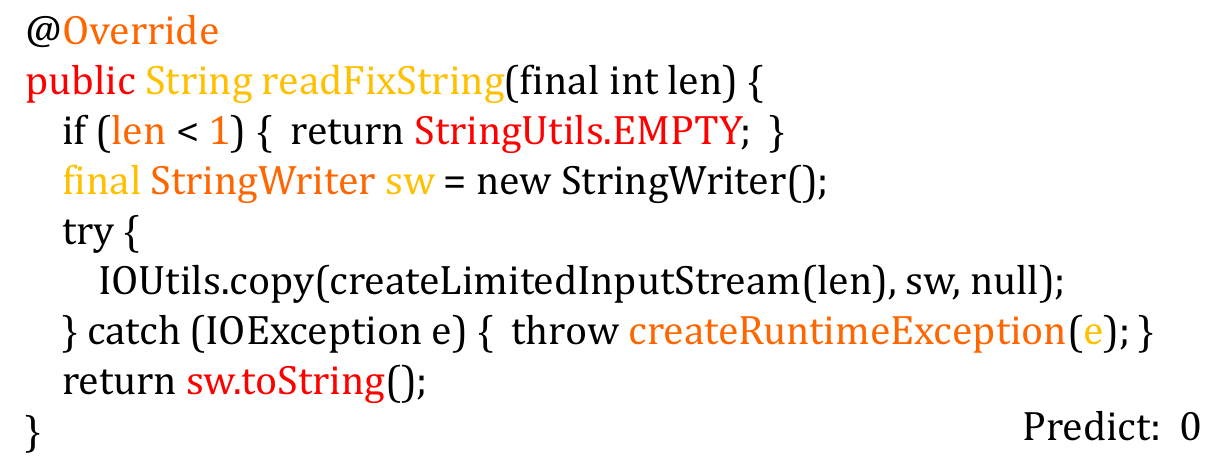}
      \label{fig:clo-22046596-autoencoder}
    }\\
    \subfigure[Attribution score on code2vec]{
      \includegraphics[width=0.48\columnwidth]{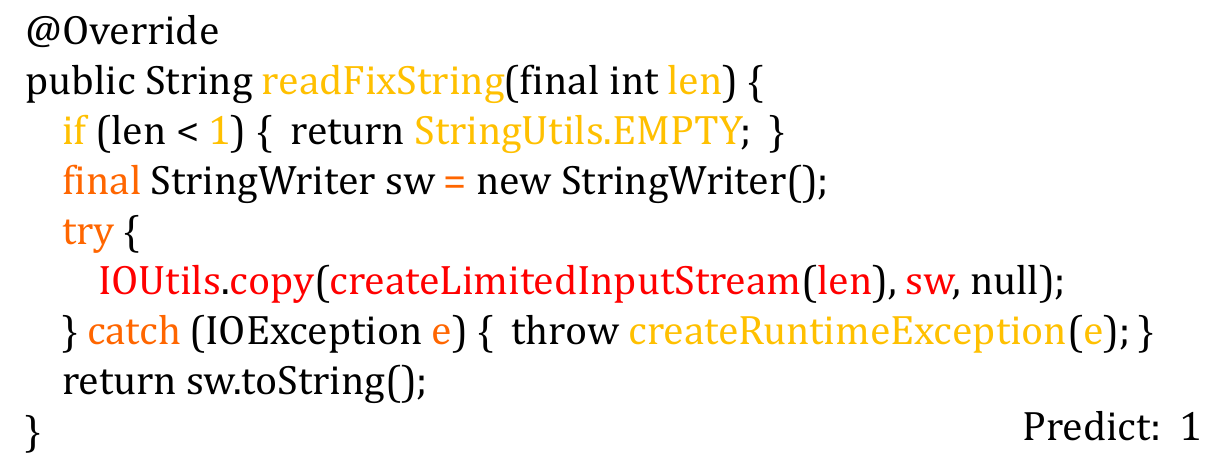}
      \label{fig:clo-22046596-code2vec}
    } 
    \subfigure[Attribution score on code2seq]{
      \includegraphics[width=0.48\columnwidth]{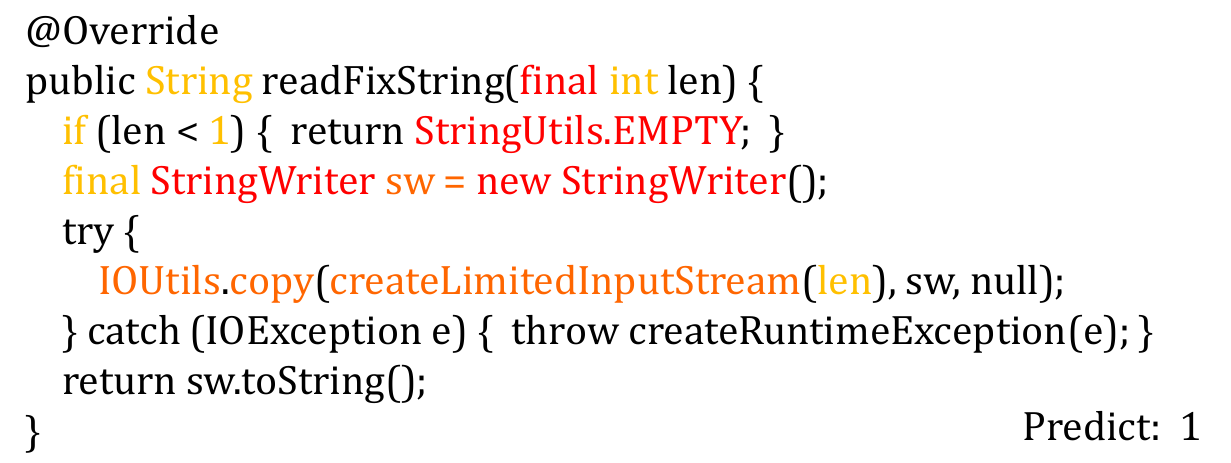}
      \label{fig:clo-22046596-code2seq}
    } \\
    \subfigure[Attribution score on GGNN]{
      \includegraphics[width=0.48\columnwidth]{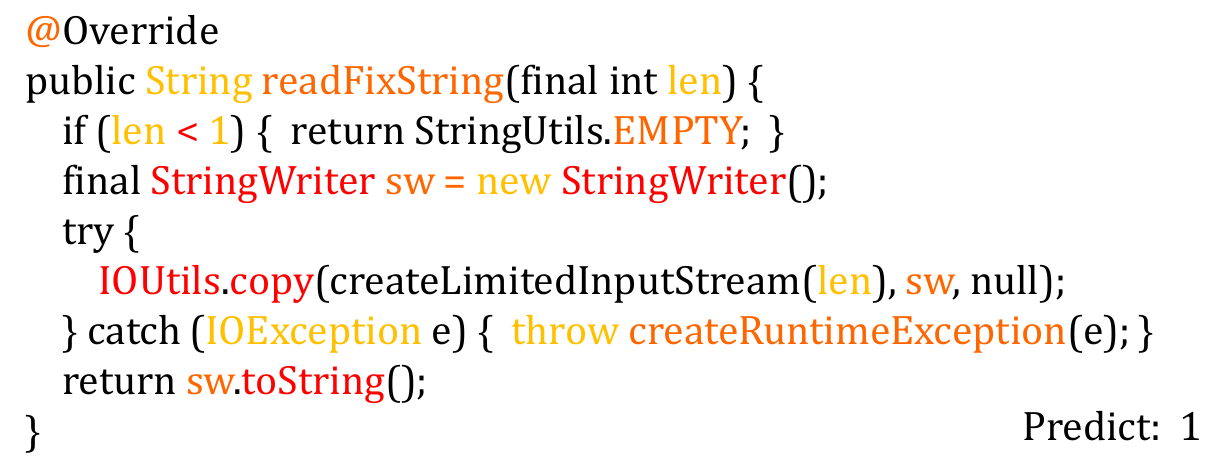}
      \label{fig:clo-22046596-GGNN}
    }
    \subfigure[Attribution score on ASTNN]{
      \includegraphics[width=0.48\columnwidth]{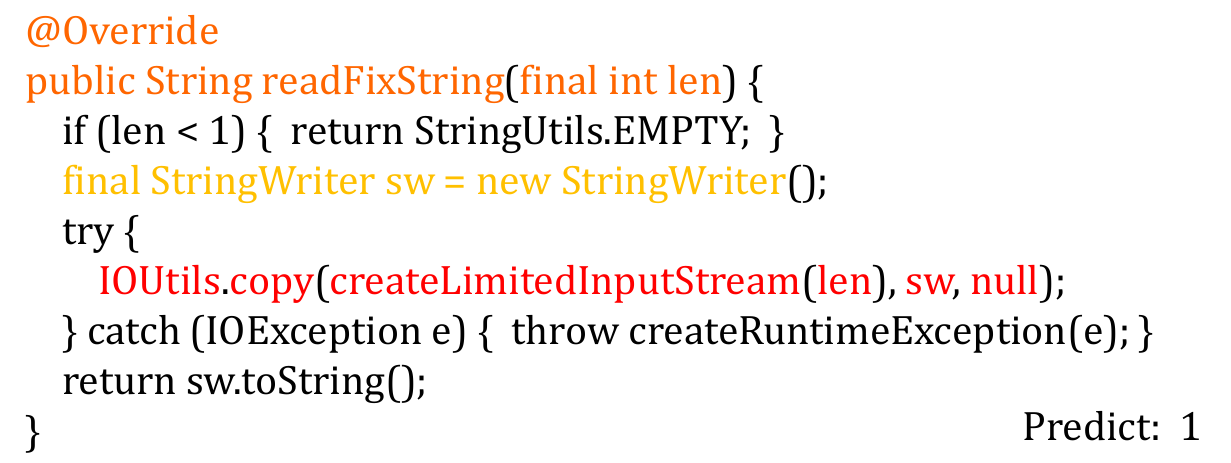}
      \label{fig:clo-22046596-ASTNN}
    }
  \caption{Attribution analysis on code-2 in Type-4 clone pairs.}
  \label{fig:clo-attr1} 
\end{figure*}

We analyze the attribution results of two other \textsc{Type-4} examples in Fig.~\ref{fig:clo-attr2} and Fig.~\ref{fig:clo-attr1}  most of the disparity of these models appear in type-4 cloned pairs.
Same as our paper, the functionality of the first snippet in the program pair is to read the content from input streams (files) and write it to output streams (files), which can serve as a copy procedure.
The key point of our comparison is whether the model can capture the tokens regarding the input and output stream and the copy semantics.
We observe from the Fig.~\ref{fig:clo-attr2} that all models except AutoenCODE have attributed the prediction to at least one token pertaining to the copy semantics and the input and output stream.
Similarly, in Fig.~\ref{fig:clo-attr1}, although vert little tokens are related to the input/output streams, most of the models can capture tokens in the statement of line 6, which conducts the process of copy.
Only AutoenCODE gives no credit to any meaningful token in these two instances, therefore it fails to predict the clone.

\subsubsection{\textbf{Code Search}}
\begin{figure*}[h]
    \centering
    \subfigure[Attribution score on LSTM]{
      \includegraphics[width=0.48\columnwidth]{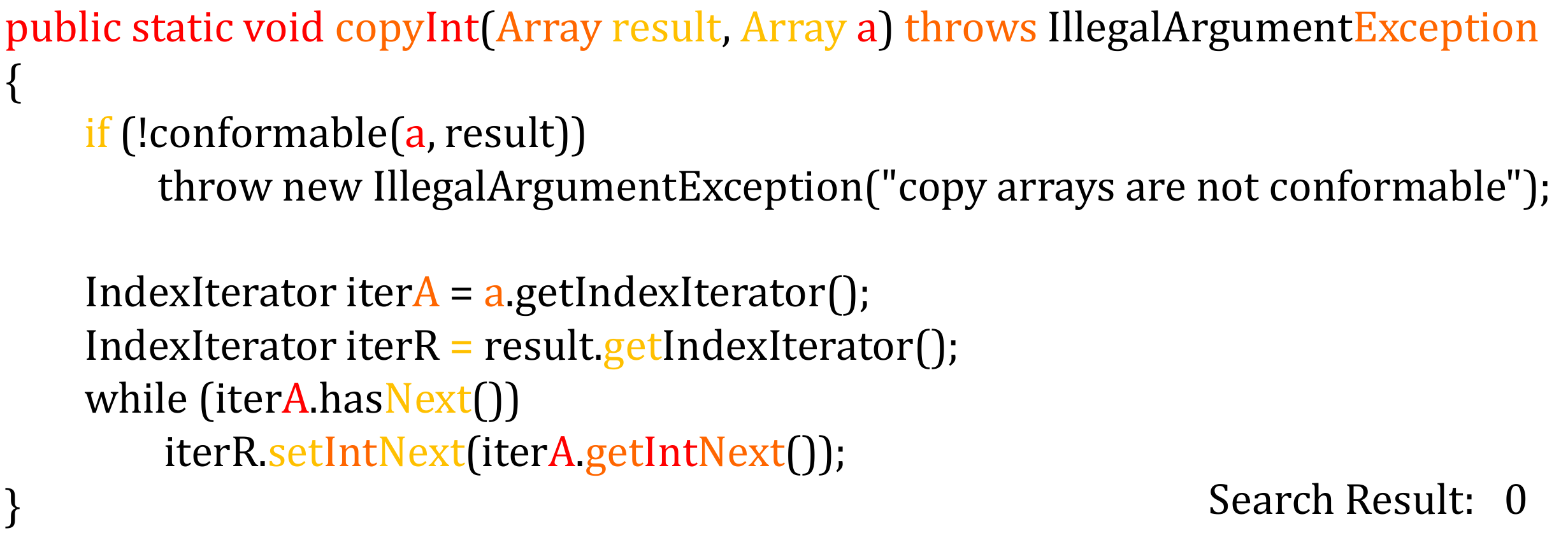}
      \label{fig:se-339161-LSTM}
    } 
    \subfigure[Attribution score on Transformer]{
      \includegraphics[width=0.48\columnwidth]{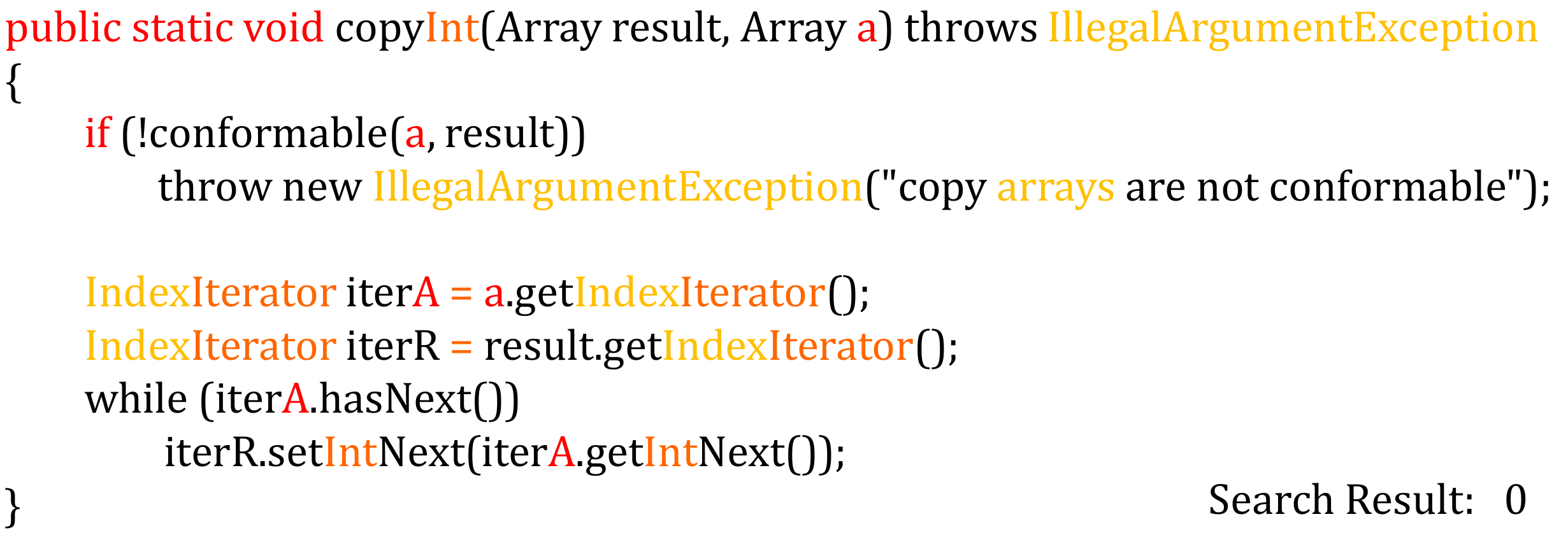}
      \label{fig:se-339161-Transformer}
    } \\
    \subfigure[Attribution score on TBCNN]{
      \includegraphics[width=0.48\columnwidth]{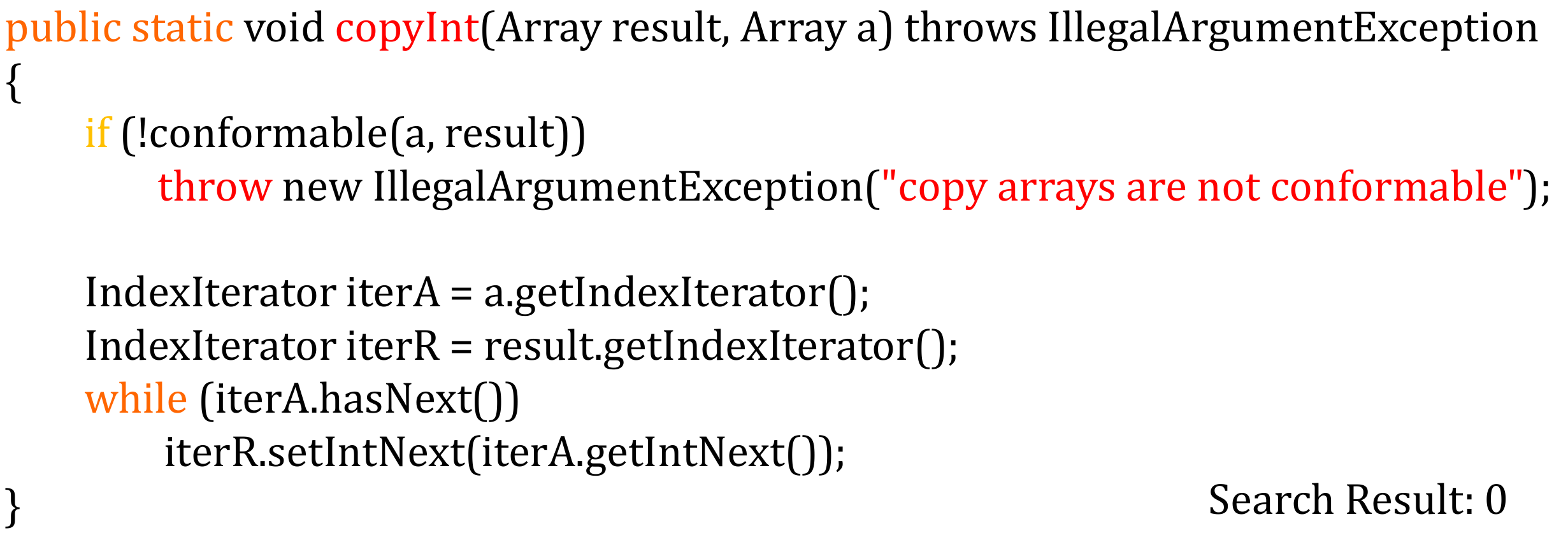}
      \label{fig:se-339161-TBCNN}
    } 
    \subfigure[Attribution score on AutoenCODE]{
      \includegraphics[width=0.48\columnwidth]{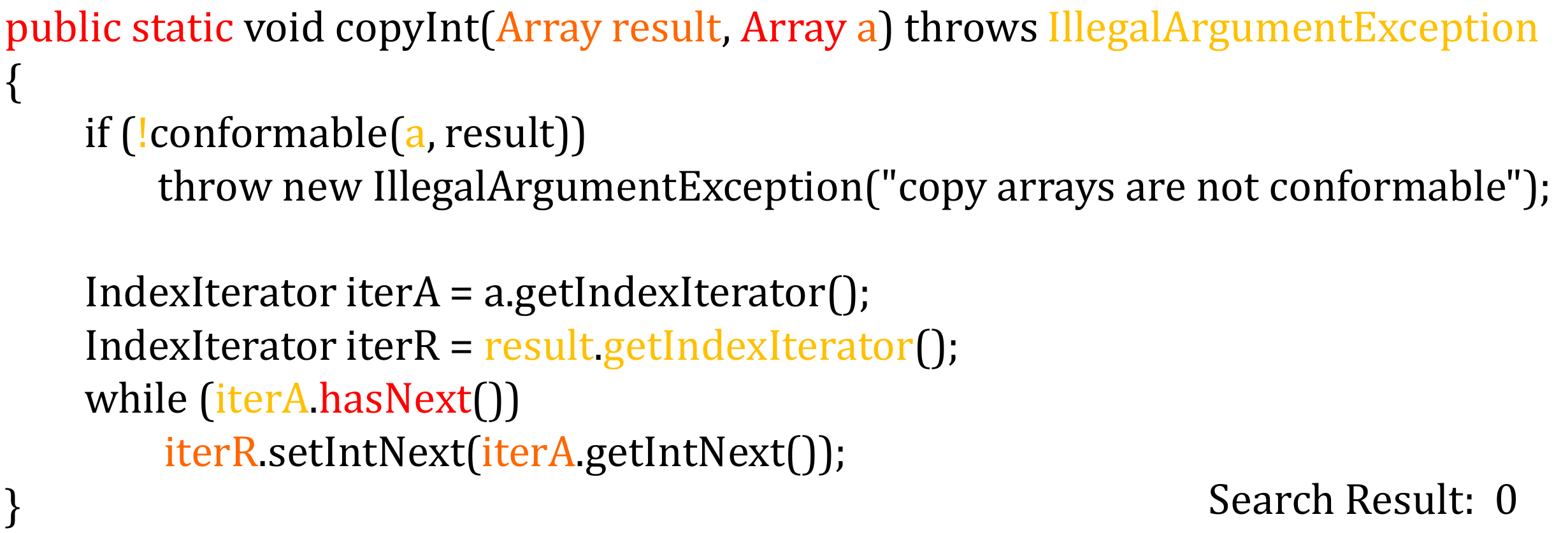}
      \label{fig:se-339161-autoencode}
    }\\
    \subfigure[Attribution score on code2vec]{
      \includegraphics[width=0.48\columnwidth]{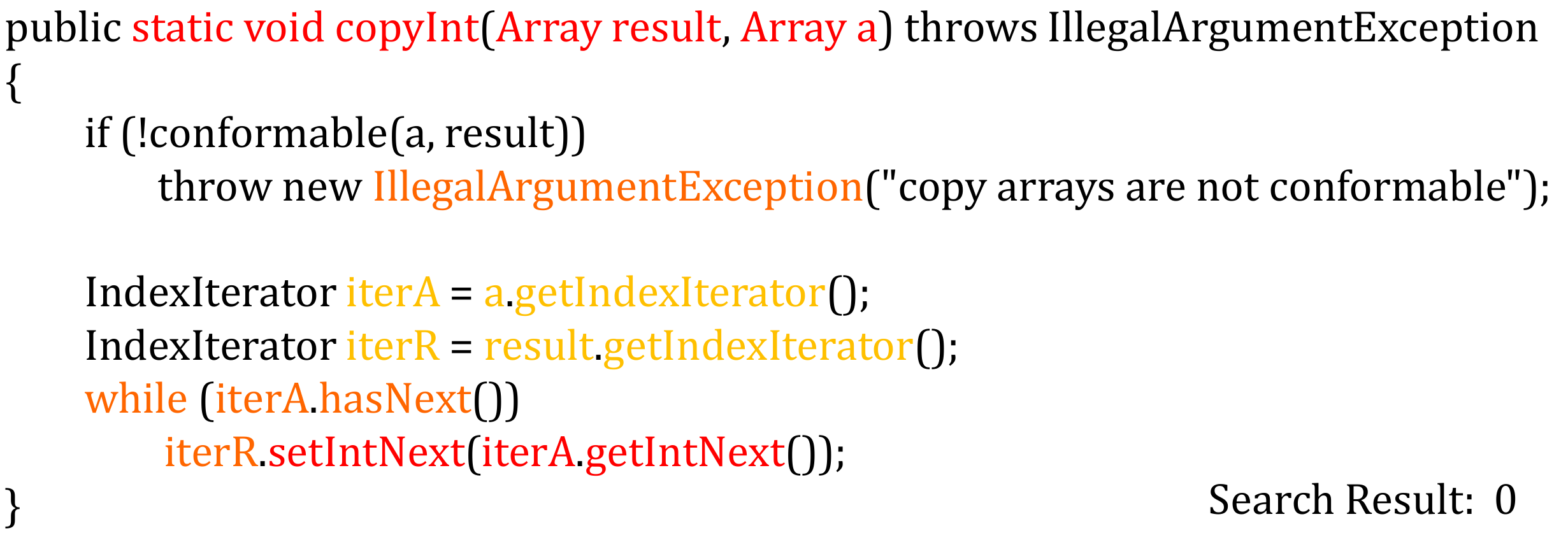}
      \label{fig:se-339161-code2vec}
    } 
    \subfigure[Attribution score on code2seq]{
      \includegraphics[width=0.48\columnwidth]{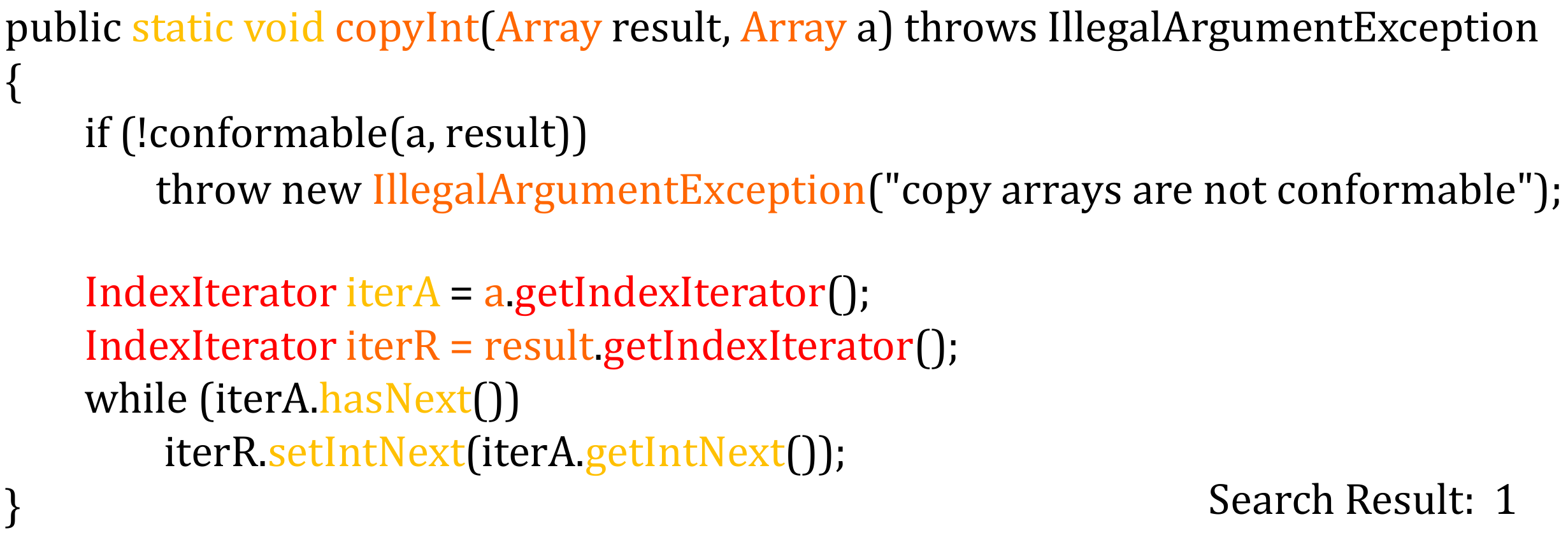}
      \label{fig:se-339161-code2seq}
    } \\
    \subfigure[Attribution score on GGNN]{
      \includegraphics[width=0.48\columnwidth]{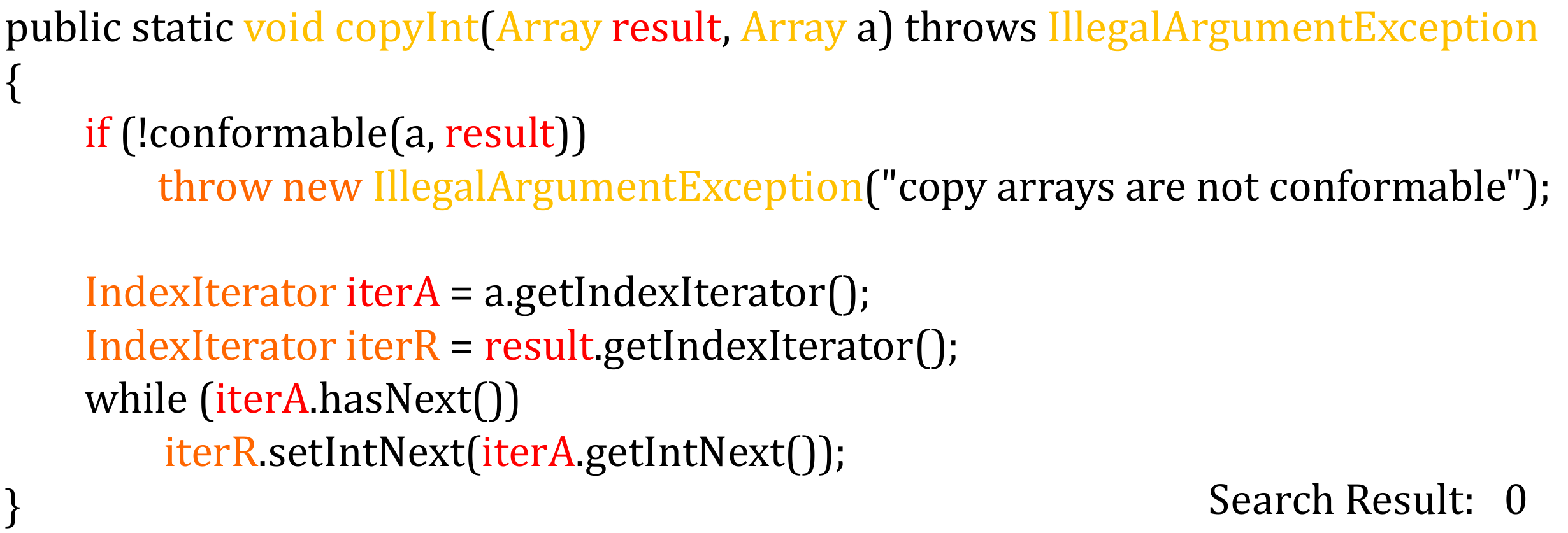}
      \label{fig:se-339161-GGNN}
    }
    \subfigure[Attribution score on ASTNN]{
      \includegraphics[width=0.48\columnwidth]{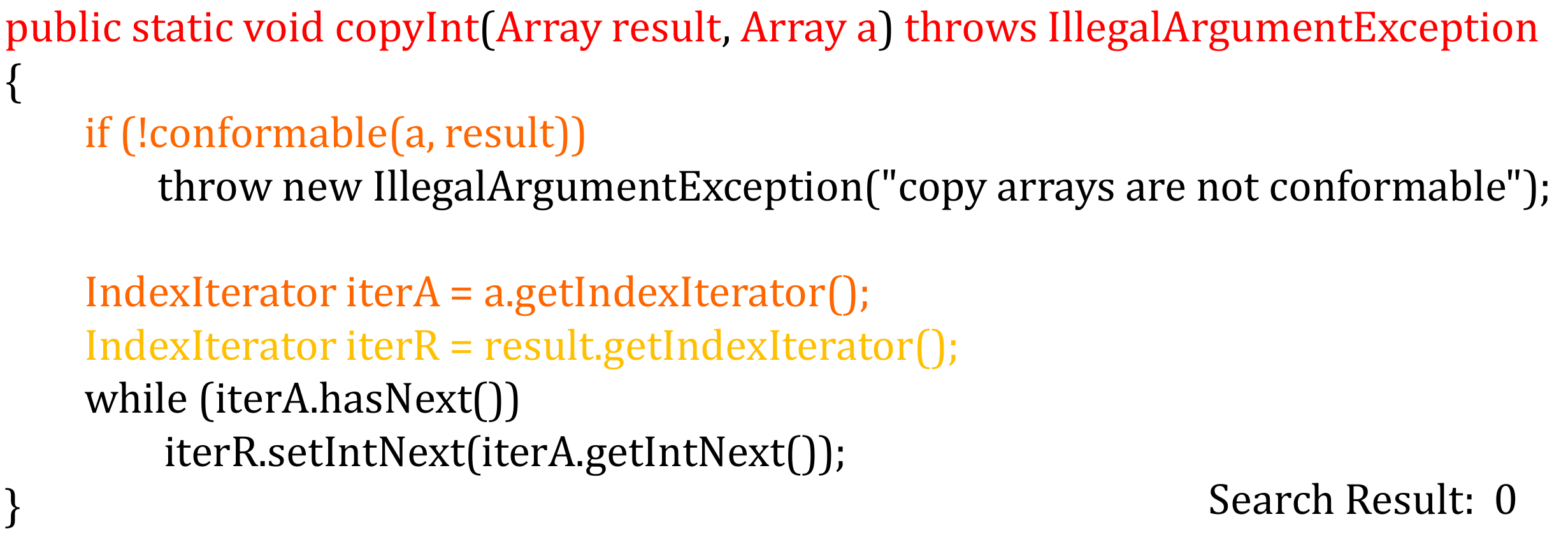}
      \label{fig:se-339161-ASTNN}
    }\\
  \caption{The example of code search. The query is "copy array a to array result as integers, the values from the arrays a are converted to integer if needed and then converted to the type of result if needed".}
  \label{fig:se-attr1} 
\end{figure*}

\begin{figure*}[h]
    \centering
    \subfigure[Attribution score on LSTM]{
      \includegraphics[width=0.48\columnwidth]{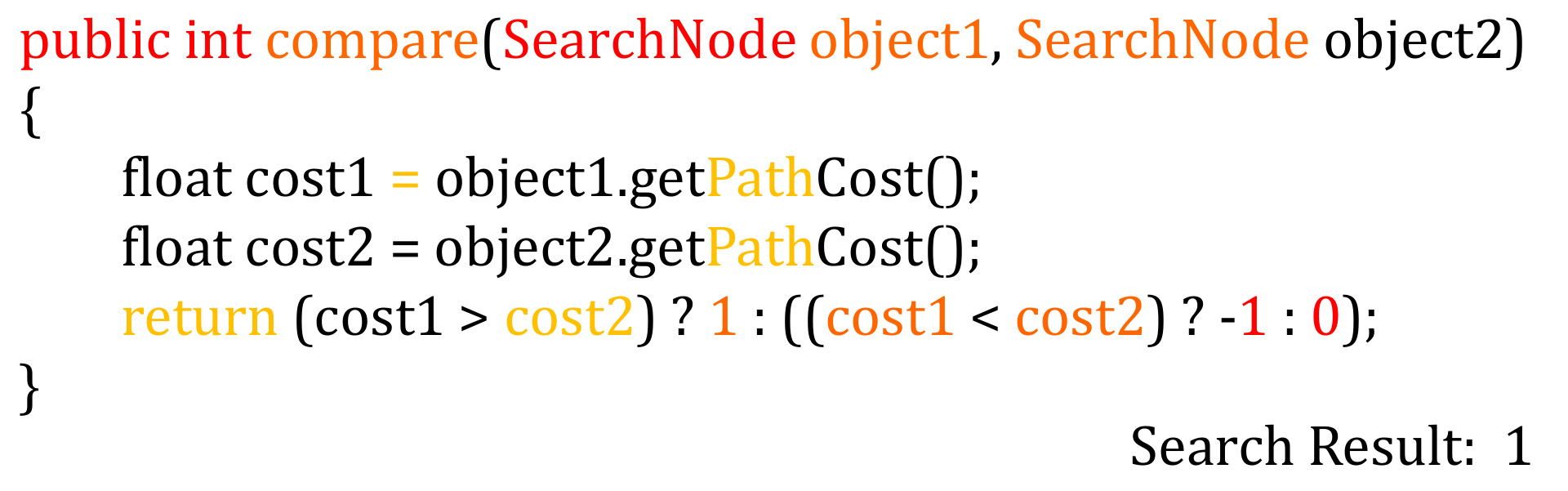}
      \label{fig:se-348281-LSTM}
    } 
    \subfigure[Attribution score on Transformer]{
      \includegraphics[width=0.48\columnwidth]{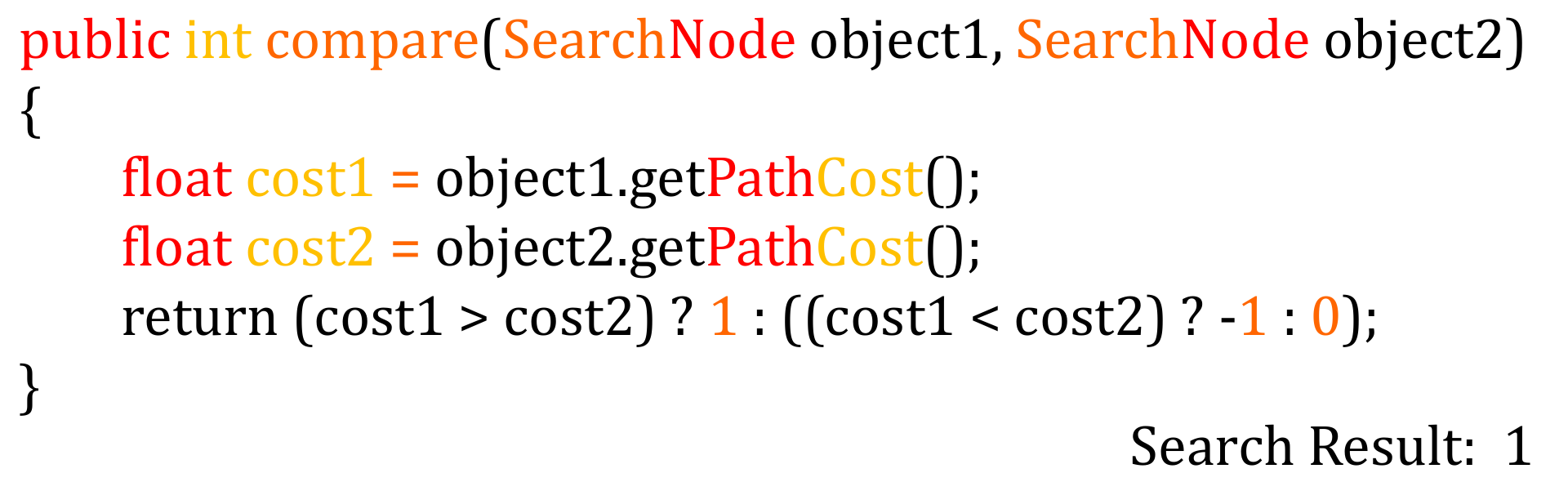}
      \label{fig:se-348281-Transformer}
    } \\
    \subfigure[Attribution score on TBCNN]{
      \includegraphics[width=0.48\columnwidth]{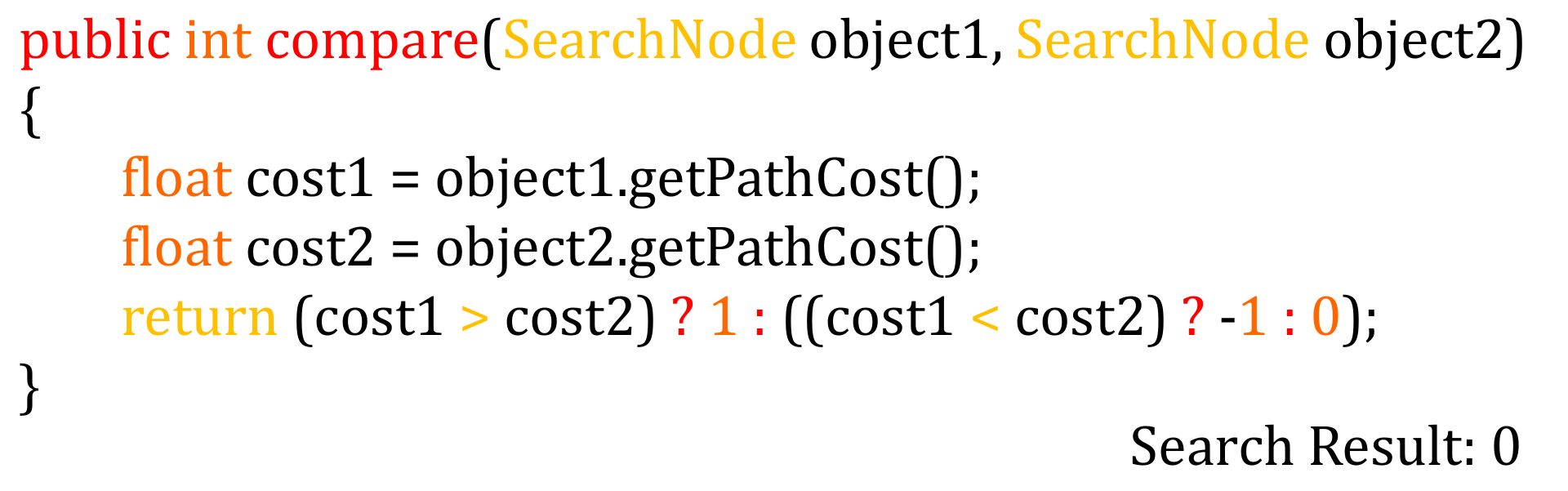}
      \label{fig:se-348281-TBCNN}
    } 
    \subfigure[Attribution score on AutoenCODE]{
      \includegraphics[width=0.48\columnwidth]{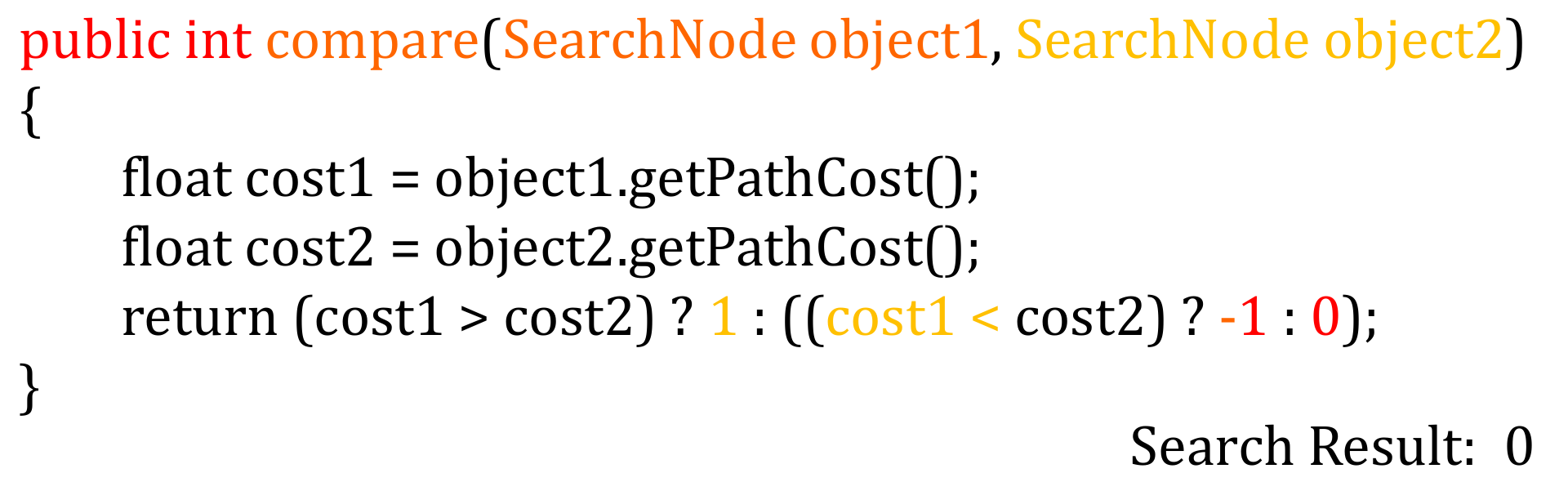}
      \label{fig:se-348281-autoencoder}
    } \\
    \subfigure[Attribution score on code2vec]{
      \includegraphics[width=0.48\columnwidth]{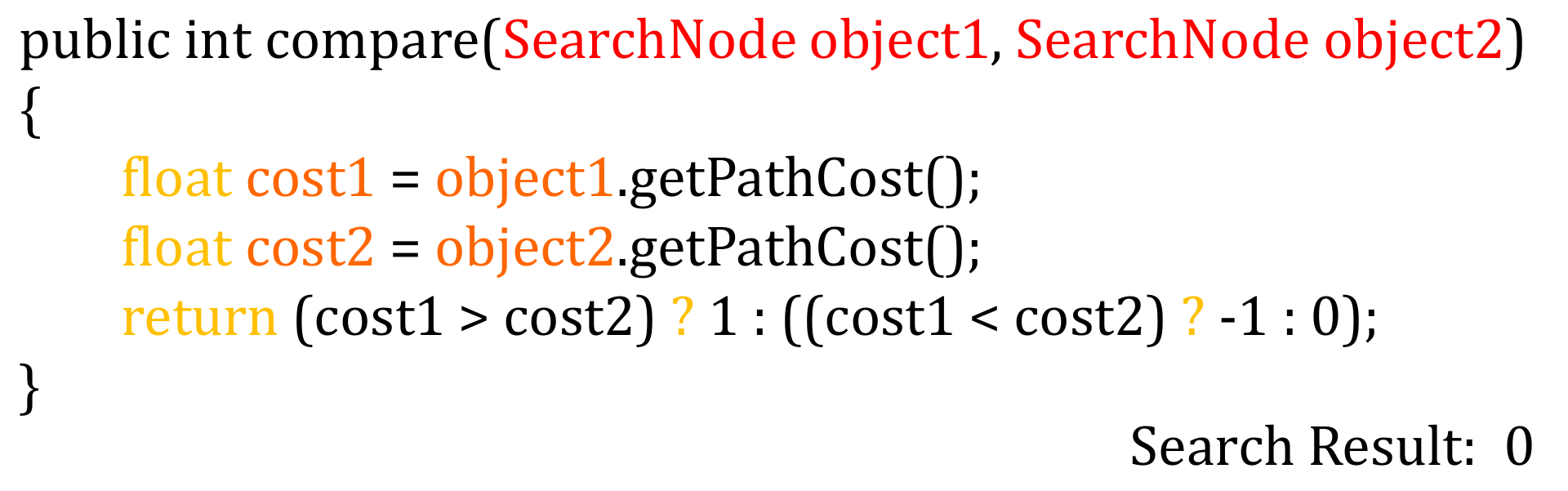}
      \label{fig:se-348281-code2vec}
    } 
    \subfigure[Attribution score on code2seq]{
      \includegraphics[width=0.48\columnwidth]{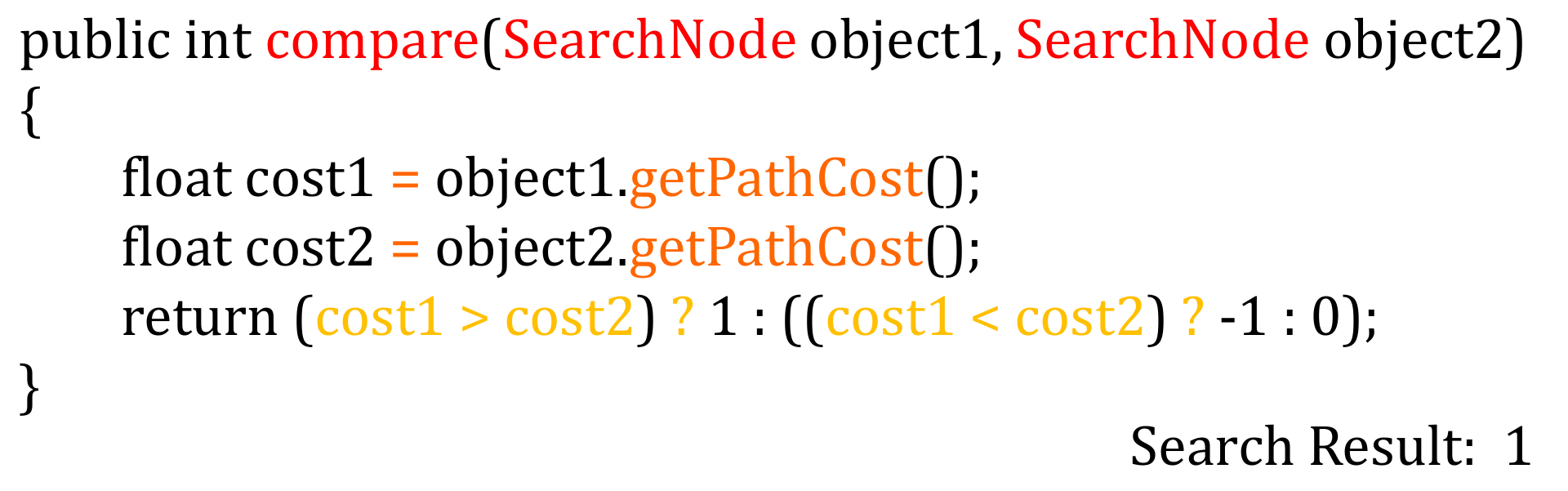}
      \label{fig:se-348281-code2seq}
    } \\
    \subfigure[Attribution score on GGNN]{
      \includegraphics[width=0.48\columnwidth]{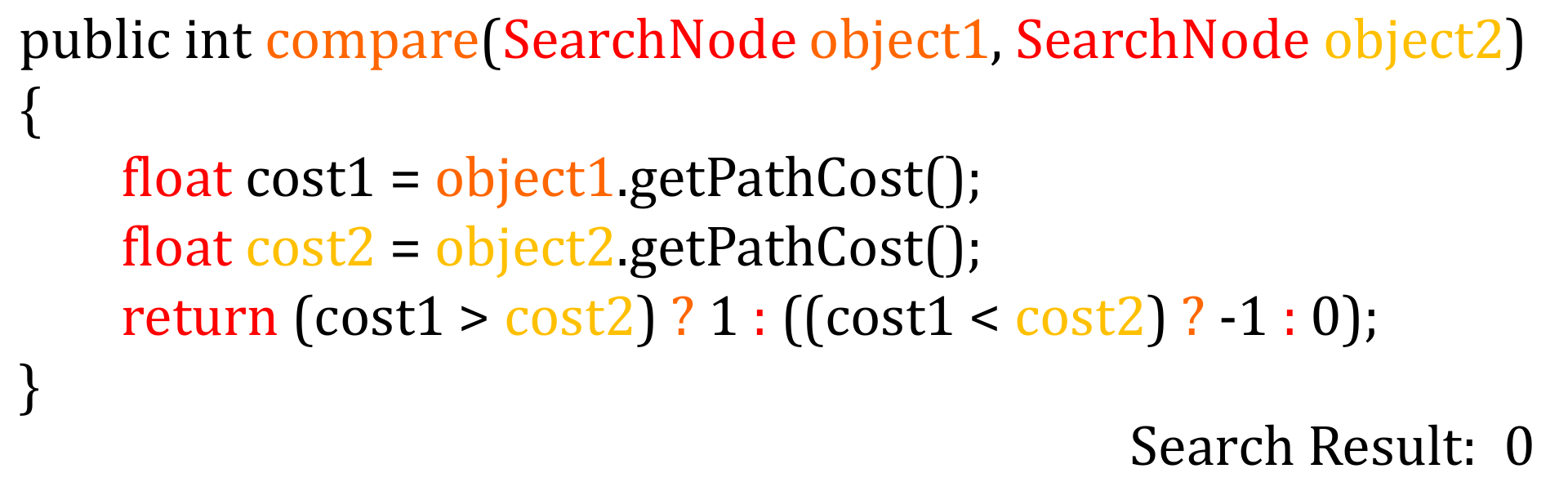}
      \label{fig:se-348281-GGNN}
    }
    \subfigure[Attribution score on ASTNN]{
      \includegraphics[width=0.48\columnwidth]{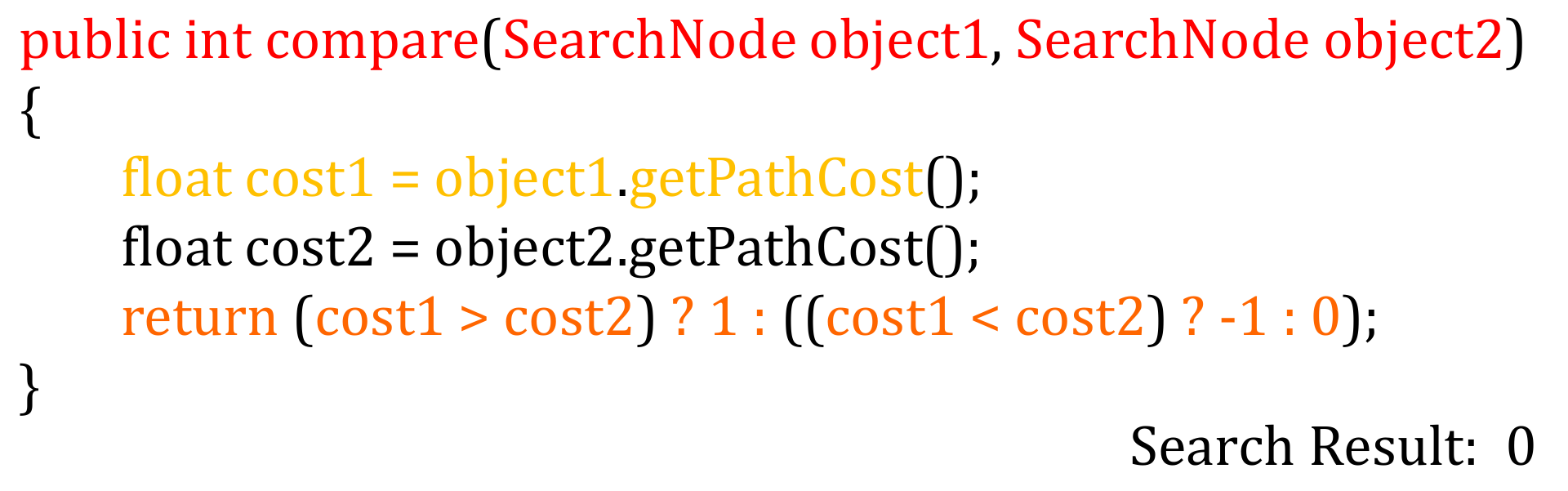}
      \label{fig:se-348281-ASTNN}
    }
  \caption{The example of code search. The query is "compares two search nodes by their path cost".}
  \label{fig:se-attr2} 
\end{figure*}
We present another two instances of code search task with different queries.
In Fig.~\ref{fig:se-attr1}, the selected natural language query aims to copy an array.
Only code2seq successfully ranks the matching program at the top 10.
Although most program representation models can capture the correlated token in the shared vocab like ``array'', ``copyInt'' and ``result'', few of them understand the functionality of the program that both of tokens, i.e. ``indexIterator'', and ``next'' are important to the copy process.
Similarly in Fig.~\ref{fig:se-attr2}, almost all models obtain relatively high attribution scores on ``compare'' and ``SearchNode'' under the query ``compares two search nodes by their path cost''.
However, only two token-based models and code2seq learn the subtoken of ``Path'' or the function name of ``getPathCost'', which is closely related to the purpose of the query. 
We can conclude that the token-based models benefit more from the splitting procedure to find similar subtokens and code2seq are more effective in learning semantics from natural language query.

\end{document}